\documentclass[12pt]{article}

\usepackage[margin=1.0in]{geometry}

\usepackage{cite}
\usepackage{booktabs}
\usepackage{amssymb,latexsym}

\usepackage{graphics,epsfig}

\usepackage{color}
\usepackage{xcolor}

\usepackage{amsmath,amsfonts}

\usepackage{mathtools}

\usepackage{mathrsfs}

\usepackage{slashed}

\usepackage{physics}
\usepackage{tensor}
\usepackage{graphicx}
\usepackage{float}
\usepackage{xcolor, colortbl}
\usepackage{hyperref}

\usepackage{authblk}

\usepackage{empheq}

\DeclareMathAlphabet{\mathpzc}{OT1}{pzc}{m}{it}

\usepackage{listings}
\usepackage{color}

\definecolor{dkgreen}{rgb}{0,0.6,0}
\definecolor{gray}{rgb}{0.5,0.5,0.5}
\definecolor{mauve}{rgb}{0.58,0,0.82}

\usepackage{feynmp}
\usepackage{feynmp-auto}

\let\a=\alpha \let\b=\beta \let\g=\gamma \let\d=\delta \let\e=\epsilon
\let\z=\zeta  \let\th=\theta  \let\k=\kappa
\let\l=\lambda \let\m=\mu \let\n=\nu \let\x=\xi \let\p=\pi 
\let\s=\sigma   \let\f=\phi  

\let\G=\Gamma \let\D=\Delta \let\Th=\Theta \let\L=\Lambda
\let\X=\Xi  
 \let\Y=\Psi

\let\la=\label \let\ci=\cite \let\re=\ref
  
\def\nn{\nonumber} \def\bd{\begin{document}} \def\ed{\end{document}}
\let\fr=\frac \let\bl=\bigl \let\br=\bigr
\let\Br=\Bigr \let\Bl=\Bigl
\let\bm=\bibitem
\let\na=\nabla
\def\tU{{\widetilde U}}
\let\pa=\partial \let\ov=\overline
\def\ie{{\it i.e.\ }}
\newcommand{\be}{\begin{equation}}
\newcommand{\ee}{\end{equation}}
\def\ba{\begin{array}}
\def\ea{\end{array}}
\def\ft#1#2{{\textstyle{{\scriptstyle #1}\over {\scriptstyle #2}}}}
\def\fft#1#2{{#1 \over #2}}
\def\F#1#2{{ F_{#1}^{(#2)} }}
\def\cF#1#2{{ {\cal F}_{#1}^{(#2)} }}

\def\R{{\bf R}}
\def\sst#1{{\scriptscriptstyle #1}}
\def\oneone{\rlap 1\mkern4mu{\rm l}}
\def\e7{E_{7(+7)}}
\def\td{\tilde}
\def\wtd{\widetilde}
\def\im{{\rm i}}
\def\bog{Bogomol'nyi\ }
\newcommand{\ho}[1]{$\, ^{#1}$}
\newcommand{\hoch}[1]{$\, ^{#1}$}
\newcommand{\bea}{\begin{eqnarray}}
\newcommand{\eea}{\end{eqnarray}}
\newcommand{\ra}{\rightarrow}
\newcommand{\lra}{\longrightarrow}
\newcommand{\Lra}{\Leftrightarrow}
\newcommand{\ap}{\alpha^\prime}
\newcommand{\bp}{\tilde \beta^\prime}
\newcommand{\cB}{{\cal B}}
\newcommand{\cO}{{\cal O}}
\newcommand{\vecx}{\vec{x}}
\newcommand{\vecy}{\vec{y}}
\newcommand{\vecp}{\vec{p}}
\newcommand{\vecq}{\vec{q}}
\newcommand{\NP}{Nucl. Phys. }

\newcommand{\cL}{{\cal L}}
\newcommand{\cA}{{\cal A}}
\newcommand{\cT}{{\cal T}}
\newcommand{\cR}{{\cal R}}
\newcommand{\cD}{{\cal D}}
\newcommand{\cH}{{\cal H}}

\newcommand{\SumInt}{\mathchoice
  {\rlap{$\displaystyle\sum$}\kern0.14em\int}
  {\rlap{$\textstyle\sum$}\kern0.14em\int}
  {\rlap{$\scriptstyle\sum$}\kern0.08em\int}
  {\rlap{$\scriptscriptstyle\sum$}\kern0.05em\int}
}

\def\Cb{\bar{C}}

\def\sst#1{{\scriptscriptstyle #1}}
\def\0{{\sst{(0)}}}
\def\1{{\sst{(1)}}}
\def\2{{\sst{(2)}}}
\def\3{{\sst{(3)}}}
\def\4{{\sst{(4)}}}
\def\5{{\sst{(5)}}}
\def\6{{\sst{(6)}}}
\def\7{{\sst{(7)}}}
\def\8{{\sst{(8)}}}
\def\9{{\sst{(9)}}}
\def\p{{\sst{(p)}}}
\def\q{{\sst{(q)}}}
\def\ve{\varepsilon}
\def\vf{\varphi}
\def\wg{\wedge}

\def\e{\epsilon}
\def\barl{\bar{l}}

\def \bi{\bibitem}
\def \la {\label}

\def \l {\lambda}
\def\foot{\footnote}
\def \tl  {{\tilde \l}}
\def \sql {{\sqrt \l}}
\def \adss {$AdS_5 \times S^5$\ }
\newcommand{\rf}[1]{(\ref{#1})}
\def \ov {\over}

\def\th{\theta}
\def\Th{\Theta}
\def\vth{\vartheta}
\def\btheta{{\bar\theta}}
\def\ttheta{{{\tilde\theta}}}
\def\bttheta{{{\bar\ttheta}}}
\def\vth{\vartheta}

\def\ra{\rightarrow}
\def\N{\nabla}
\def\F{{\cal F}}
\def\uM{\underline{M}}
\def\uA{\underline{A}}
\def\uN{\underline{N}}
\def\uP{\underline{P}}
\def\ua{\underline{a}}
\def\ub{\underline{b}}
\def\uc{\underline{c}}
\def\ud{\underline{d}}
\def\ue{\underline{e}}
\def\uf{\underline{f}}
\def\ui{\underline{i}}
\def\uj{\underline{j}}
\def\uk{\underline{k}}
\def\ul{\underline{l}}
\def\ual{\underline{\alpha}}
\def\ube{\underline{\beta}}
\def\um{\underline{m}}
\def\un{\underline{n}}
\def\up{\underline{p}}
\def\uq{\underline{q}}
\def\ur{\underline{r}}
\def\us{\underline{s}}
\def\umu{\underline{\mu}}
\def\unu{\underline{\nu}}
\def\ula{\underline{\l}}
\def\uka{\underline{\k}}
\def\usi{\underline{\s}}
\def\urh{\underline{\r}}
\def\cc{\circ}
\def\eqv{\equiv}

\def\ni{\noindent}

\def\Ep{E^{{}^{(+)}}}
\def\Em{E^{{}^{(-)}}}

\def\Mp{M^{{}^{(+)}}}
\def\Mm{M^{{}^{(-)}}}

\def \ha{{1\ov 2}}

\def\r{\rho}

\def\Y{{\rm Y}}
\def\X{{\rm X}}
\def\tY{\tilde{\rm Y}}
\def\tX{\tilde{\rm X}}
\def\dY{\dot{\rm Y}}
\def\dX{\dot{\rm X}}

\def \J {\mathcal{J}}
\def \del {\partial}

\def\dF{\dot{F}}
\def\dG{\dot{G}}
\def\df{\dot{f}}
\def \E {{\cal E}}
\def \J {{\cal J}}

\def\ms{\mathcal{S}}
\def\mj{\mathcal{J}}
\def\soj{\fr{\ms}{\mj}}
\def \R {{\bf R}}
\def \om {\omega}
\def \bE {\bar E}
\def \x {{\cal X}}

\def \bi{\bibitem}
\def \la {\label}

\def \l {\lambda}
\def\foot{\footnote}
\def \tl  {{\tilde \l}}
\def \sql {{\sqrt \l}}
\def \adss {$AdS_5 \times S^5$\ }
\def \ov {\over}

\def \varpi {{\rm w}}

\def\thb{\bar{\theta}}
\def\Thb{\bar{\Theta}}
\def\barp{\bar{p}}
\def\barq{\bar{q}}
\def\barc{\bar{c}}
\def\bard{\bar{d}}
\def\bare{\bar{e}}

\def\thb{\bar{\theta}}
\def\Thb{\bar{\Theta}}
\def\mb{\bar{\m}}
\def\ab{\bar{\a}}
\def\zb{\bar{z}}
\def\psib{\bar{\psi}}
\def\barl{\bar{l}}
\def\barp{\bar{p}}
\def\barq{\bar{q}}
\def\barc{\bar{c}}
\def\bard{\bar{d}}
\def\baru{\bar{u}}

\def\e{\epsilon}
\def\wb{\bar{w}}
\def\lb{\bar{\l}}
\def\Jb{\bar{J}}
\def\Nb{\bar{N}}
\def\Zb{\bar{Z}}
\def\pab{\bar{\pa}}

\def\At{\tilde{A}}
\def\Bt{\tilde{B}}
\def\Ct{\tilde{C}}
\def\Dt{\tilde{D}}
\def\Et{\tilde{E}}
\def\Ft{\tilde{F}}
\def\Gt{\tilde{G}}
\def\Ht{\tilde{H}}
\def\Kt{\tilde{K}}
\def\Mt{\tilde{M}}
\def\Nt{\tilde{N}}
\def\Rt{\tilde{R}}
\def\at{\tilde{a}}
\def\bt{\tilde{b}}
\def\ct{\tilde{c}}
\def\dt{\tilde{d}}
\def\et{\tilde{e}}
\def \zt{\tilde{z}}
\def \ztt{\tilde{\z}}
\def\Omt{\tilde{\Omega}}
\def \zetat{\tilde{\zeta}}
\def\htil{\tilde{h}}
\def\gt{\tilde{g}}
\def\nt{\tilde{n}}
\def\mut{\tilde{\mu}}
\def\nut{\tilde{\nu}}
\def\pht{\tilde{\f}}
\def\Phit{\tilde{\Phi}}
\def\vft{\tilde{\vf}}
\def\etat{\tilde{\eta}}

\def\rht{\tilde{\rho}}

\def\asth{\hat{*}}
\def\phh{\hat{\phi}}

\def\bA{{\bf A}}

\def\ola{\overleftarrow}
\def\ora{\overrightarrow}
\def\alt{\tilde{\a}}

\def\eh{\hat{e}}
\def\eph{\hat{\e}}
\def\ph{\hat{p}}
\def\alh{\hat{\a}}
\def\beh{\hat{\b}}
\def\gah{\hat{\g}}
\def\Fh{\hat{F}}
\def\muh{\hat{\m}}
\def\nuh{\hat{\n}}
\def\thh{\hat{\th}}
\def\rhh{\hat{\r}}
\def\dh{\hat{d}}
\def\ih{\hat{i}}
\def\jh{\hat{j}}
\def\hh{\hat{h}}
\def\nh{\hat{n}}
\def\gh{\hat{g}}
\def\kh{\hat{k}}
\def\deh{\hat{\d}}
\def\wh{\hat{w}}
\def\lah{\hat{\l}}
\def\Ah{\hat{A}}
\def\Gh{\hat{G}}
\def\Kh{\hat{K}}
\def\Nh{\hat{N}}
\def\Rh{\hat{R}}
\def\Ch{\hat{C}}
\def\Omh{\hat{\Omega}}

\def\xh{\hat{x}}

\def\ps{\rlap{\, /}\;\,p }
\def\ks{\rlap{\, /}\;\,k }

\def\gym{g_{YM}}

\def\adot{\dot{a}}
\def\bdot{\dot{b}}
\def\bpa{\bar{\pa}}

\def\pr{\prime}
\def\ssk{\medskip}
\def\clb{\color{blue}}
\def\clr{\color{red}}
\def\clg{\color{green}}
\def\clp{\color{purple}}
\def\clc{\color{cyan}}
\def\clm{\color{magenta}}
\def\cly{\color{yellow}}
\def\ah{\color{orange}}

\def\bfA{{\bf A}}
\def\bfB{{\bf B}}
\def\bfK{{\bf K}}
\def\bfU{{\bf U}}
\def\bfX{{\bf X}}
\def\bfY{{\bf Y}}
\def\bfZ{{\bf Z}}
\def\bfg{{\bf g}}
\def\bfn{{\bf n}}

\def\bsk{\bigskip}
\def\ssk{\medskip}

\def\Ec{{\cal E}}

\newcommand{\W}[1]{(W\,#1)}
\newcommand{\DT}[1]{\Delta^{(S)}_{T,#1}}   
\newcommand{\DP}[1]{\Delta^{(S)}_{P,#1}}   
\newcommand{\eps}{\epsilon}
\let\om=\omega

\newcommand{\DTm}{\Delta^{(S)}_{T}}
\newcommand{\DPm}{\Delta^{(S)}_{P}}

\newcommand{\Ro}{\mathcal{R}^{o}_{q}}
\newcommand{\TT}{\mathcal{T}}
\renewcommand{\SS}{\mathcal{S}}
\newcommand{\lR}{\ell_{\mathcal{R}}}
\newtheorem{remark}{Remark}

\vspace{ -3cm}
 \vspace{0.1cm}

\title{Kinetic-theory derivation and Bayesian analysis of the finite-temperature shifted CMB power spectrum}
 \date{}

\author{I. Y. Park\footnote{inyongpark05@gmail.com}}
\affil{\small{Department of Applied Mathematics and Engineering, Philander Smith University, Little Rock, AR 72202, USA.}}

\begin{document}
 \maketitle

\begin{abstract}

Finite-temperature quantum-gravitational corrections to the cosmological
constant introduce two additional density parameters, $\Omega_{\Lambda_2}$ and
$\Omega_{\Lambda_3}$, the associated terms of which scale as $a^{-4}$ and
$a^{-2}$, respectively. In previous work (Papers I and II),
these parameters were studied using machine-learning methods that identified a
best-fit point. Here we complete the picture with two complementary analyses.
First, we perform a Bayesian MCMC analysis using a modified version of
\textsc{class} against the Planck 2018 likelihoods. This tells us what the data
can constrain once all parameter uncertainties have been marginalized over. Both
parameters are consistent with zero, with $\Omega_{\Lambda_3}$ constrained only
from one side. Second, we build a CMB pipeline from scratch, following
Weinberg's kinetic-theory formulation and validating it against \textsc{class}.
This allows us to study the underlying physics. The response analysis carried
out with it shows that $\Omega_{\Lambda_3}$ is almost entirely absorbed by a
shift in $h$ over the well-measured acoustic multipoles. The part of the signal
that cannot be absorbed resides at low $\ell$, where cosmic variance dominates.
Meanwhile, the $\Omega_{\Lambda_2}$ signature is nearly degenerate with the
overall amplitude of the spectrum. The effects of both parameters on the CMB power spectrum
are small, since they represent higher-order thermal corrections to the
cosmological constant. The resulting constraints are correspondingly weak: the
induced shifts are largely absorbed by degeneracies with $h$ and with the
overall normalization of the spectrum. This is the reason why neither parameter
is detected.

\end{abstract}

\newpage

\section{Introduction }

The cosmological constant in the standard cosmological model is the simplest
explanation of the observed acceleration. Six parameters, one of which is a
constant vacuum energy, reproduce the Planck temperature and polarization
spectra and the galaxy clustering measured by eBOSS to sub-percent precision
\ci{Planck:2018vyg,Alam:2021inq}. The persistent disagreement between the value
of \(H_0\) inferred from the CMB under \(\Lambda\)CDM and the value measured
locally \ci{Riess:2016jrr} has led to a question of whether the vacuum term is
truly constant. One natural response is to treat the vacuum energy as something
that responds to the state of the universe rather than as a fixed input; this is
the approach taken in the running-vacuum program, where renormalization of the
vacuum energy in curved spacetime produces a density that evolves with the
expansion
\ci{Sola:2013gha,Moreno-Pulido:2020anb,SolaPeracaula:2022hpd,Moreno-Pulido:2022phq,Moreno-Pulido:2023ryo},
and which has recently been extended toward a unified description of dark energy
and inflation \ci{SolaPeracaula:2026trz}. The thermal history of the universe
provides a concrete and independent source of such evolution, since a quantum
field theory at finite temperature has a free energy that depends on temperature
\ci{Kapusta,Bellac,Laine:2016hma}. This is the direction pursued in
\ci{Park:2021ohu,Park:2021vro} and in the work that followed.

The cosmological constant receives corrections at finite temperature. In
\ci{Park:2025iog} (Paper I) it was shown that a one-loop quantum-gravitational
computation in a thermal background produces, in the high-temperature expansion,
terms scaling as $T^4$ and $T^2$; translated into the cosmological setting these
lead to two additional terms associated with density parameters, $\Omega_{\L_2}$
and $\Omega_{\L_3}$; the two terms scale as $a^{-4}$ and $a^{-2}$, respectively.
Their effect on the CMB power spectrum was examined with \textsc{class} through
brute-force parameter scans and quartic regression, and the extended model was
found to improve the fit by the standard information criteria. Paper II
\ci{Hatefi:2025lqe} pursued the same question with artificial neural networks,
and observed that the gross shape of the spectrum is set by the standard
parameters, with $\Omega_{\L_2}$ and $\Omega_{\L_3}$ refining the fine shape.
Both analyses are deterministic in character: they locate a best-fit point in
the extended parameter space.

In the present work we continue the endeavor with two further analyses that
address questions a best-fit point cannot. The first is a Bayesian MCMC
analysis, run with a modified \textsc{class} against the Planck 2018
likelihoods, which addresses what the data actually bound once every parameter
uncertainty is marginalized over rather than fixed at its optimum. The second is
a CMB power spectrum code built from scratch along the kinetic-theory lines of
Weinberg's \emph{Cosmology} \ci{Weinberg2008}, which tackles the questions of
how the two parameters imprint themselves on the spectrum and which standard
parameters can mimic that imprint. The three approaches --- the machine-learning
analyses of Papers I and II, the Bayesian analysis, and the response analysis of
the kinetic theory --- are complementary rather than competing, and we will
argue that their apparent tension is itself informative: a parameter can have a
well-defined best fit, a broad marginal posterior, and a clean physical
explanation for the spread, all at once.

A prior question must be settled first. Since the term associated with
$\Omega_{\L_2}$ scales as $a^{-4}$ and $\Omega_{\L_3}$ as $a^{-2}$, they share
their scale-factor dependence with radiation and with spatial curvature
respectively, and one may ask whether they are anything more than a relabeling
of $\Omega_R$ and $\Omega_K$. The question was addressed in
\ci{Park:2025iog,Hatefi:2025lqe}; we take it up again in
Sec.~\re{sec:redundancy} from a different direction, by way of a scalar field. A
scalar field admits an exact perfect-fluid representation for any potential, and
in the massless case its equation of state is that of radiation. What the
analysis in Sec.~\re{sec:redundancy} makes clear is that the origin of this
radiation-like term is evident at the level of the fundamental classical field
theory action, whereas the finite-temperature corrections considered here shift
the cosmological constant through quantum corrections. Sharing a power of $a$ is
not the same as being the same term. The distinction has a consequence which we
exploit throughout: $\Omega_{\L_3}$ enters the expansion rate alongside
$\Omega_K$, but it does not modify the perturbation equations; and
$\Omega_{\L_2}$ enters only through $H(a)$, unlike radiation, which also
perturbs.

That the terms are present is a theoretical statement. Detectability is set by
experimental precision, and it is here that the two analyses of this paper
converge on a single conclusion. The Bayesian analysis finds both parameters
consistent with zero, with $\Omega_{\L_3}$ bounded only from below. The response
analysis provides the reason: across the acoustic multipoles that Planck
measures best, the $\Omega_{\L_3}$ signature is very nearly reproducible by a
shift in $h$, and the portion of it that no standard parameter can imitate lies
at low $\ell$, where cosmic variance is irreducible; the $\Omega_{\L_2}$
signature is nearly flat in $\ell$ and therefore hides in the overall
normalization of the spectrum. Neither finding is an uninformative null result:
together they tell us which observations would be needed to do better.

\vspace{.2in}
The rest of the paper is organized as follows. In Section~\re{sec:setup} we
prepare the ground for both analyses. In Section~\re{sec:review}, we recall where
$\Omega_{\L_2}$ and $\Omega_{\L_3}$ come from and what Papers I and II
established, and in Sec.~\re{sec:redundancy} the redundancy question by the
scalar-field argument just described is settled. Section~\re{sec:theoryfile} then sorts out
a matter that is easy to get wrong and that the later sections depend on: four
different objects all get called ``the power spectrum'' --- the theoretical
curve as a function of the parameters, that curve evaluated at a particular
best-fit point, the same curve evaluated at the true parameters of Nature, and
the single realization that an experiment actually measures. The Planck data
products contain several of these, and for comparisons one should know which is
which.

Section~\re{sec:hydro} is the analytic part. We follow Weinberg's kinetic-theory
treatment of the acoustic era and carry it further, removing his approximations.
Section~\re{sec:exact} sets down the exact multipole formula that everything
else is measured against. Section~\re{sec:tc} obtains the tight-coupling
equations as the expansion of the photon hierarchy in $1/\om_c$, from which the
damping rate and the polarization coefficient $16/15$ follow rather than being
assumed. In Sections~\re{sec:debye} and~\re{sec:landau} we carry out the
large-$\ell$ reduction while keeping all three form factors, including the
polarization channel $\bar H$ that Weinberg drops, and Sec.~\re{sec:quadeval}
evaluates the resulting closed formula and compares its acoustic peak positions
with Weinberg's own. The intermediate steps are collected in
Appendix~\re{app:sec3}.

Section~\re{sec:bayesian} is the Bayesian analysis. Section~\re{sec:bayes_setup}
gives the sampled parameters, the priors, the likelihoods and the modifications
made to \textsc{class}; the following subsection reports how the chains were run
and how convergence was judged, which matters here because the new parameters
are poorly constrained and the chains mix slowly. Section~\re{sec:bayes_results}
gives the posteriors, and in Sec.~\re{sec:bayes_priors} we address the question
that a weakly constrained parameter raises: whether the width of a posterior
comes from the data or merely from the range the prior was given. The full
ten-dimensional posterior is shown there as well, in
Fig.~\re{fig:triangle_full}.

In Section~\re{sec:kinetic} we conduct the numerical kinetic theory analysis in
the Weinberg setup, and Sec.~\re{sec:pipeline} describes how the pipeline is
built. Its construction is governed by one design choice worth stating at the
outset: the photon Boltzmann hierarchy is integrated directly, with no
tight-coupling approximation and no switch from an approximate to an exact
system. Standard codes make such a switch, and its placement is tuned against a
fiducial cosmology. A change in the expansion history shifts the physics the
switch was tuned to, and the resulting error goes unnoticed --- it appears as a
change in the spectrum, not as a warning. Since our entire subject is a modified
expansion history, we preferred a slower code with no seam.
Section~\re{sec:pipeline_checks} establishes how far the resulting numbers can
be trusted: the smallest difference between two spectra that can be called
physical, the agreement with the Planck 2018 spectrum in absolute terms, and
three further checks that the features of the calculation emerge from the
dynamics rather than being imposed. The analysis of the spectral responses to
\(\Omega_{\L_2}\) and \(\Omega_{\L_3}\) (carried out in
Section~\ref{sec:responses}) indicates that for \(\Omega_{\L_3}\), \(h\) is
uniquely capable of absorbing the effect. Because the validity of this
conclusion relies entirely on the accuracy of the numerical \(h\)-response, we
establish the \(h\)-response in Section~\re{sec:validation} first (which allows
us to validate the code against a reliable, independent calculation): we compare
the pipeline with \textsc{class} on a standard parameter, the response to $h$.
The fact that it agrees there suggests that its findings on the non-standard
parameters should be taken seriously. Section~\re{sec:responses} computes the
two finite-temperature responses and closes with the Fisher analysis that turns
them into statements about detectability. The initial data for the code, and the
consistency of the gauge conditions used to set them, are treated in
Appendix~\re{app:ic}; the dictionary relating our variables to the more familiar
Ma--Bertschinger ones is given in Appendix~\re{sec:mb} (see \cite{Dodelson2021} 
for a review of the formalism).

Section~\re{sec:conclusion} collects the conclusions and states what the three
analyses, taken together, say about the prospects for a detection. Conventions
are gathered in Appendix~\re{app:conv}.


\section{Theoretical and computational preliminaries}
\la{sec:setup}

In this section we set the stage for the analyses that follow.
Section~\re{sec:review} has two purposes. The first is to recall, briefly, where
$\Omega_{\L_2}$ and $\Omega_{\L_3}$ come from and what Papers I and II
established about them. The second is to settle the redundancy question by an
argument that is independent of, though related to, the one given in
\cite{Park:2025iog,Hatefi:2025lqe}. In Section~\re{sec:theoryfile} we discuss
the rationale for the statistical framework adopted. We explain the nature of
the Planck data file used,
\texttt{COM\_PowerSpect\_CMB-base-plikHM-TTTEEE-\allowbreak lowl-lowE-lensing-\allowbreak minimum-theory\_R3.01.txt}
and justify the use of the plain Euclidean, as opposed to a weighted, distance
between two power spectra.

\subsection{Finite-temperature corrections to the cosmological constant}
\la{sec:review}

By extending the foliation-based approach to gravity quantization 
\cite{Park:2014tia,Park:2016zgt,Park:2014noa,Park:2018vci,Park:2019amz} 
(see \cite{Grassi:2024vkb} and \cite{Esposito:2023ymk} for recent works employing 
some ingredients of foliation-based quantization) to finite temperature, it was observed in \ci{Park:2025iog} 
that a one-loop quantum-gravitational computation in a thermal background shifts the cosmological constant, and that
the shift is a series in the temperature. In the high-temperature expansion the
leading behavior is $T^4$, followed by $T^2$; with $T \propto a^{-1}$ these
become contributions to the energy budget whose magnitudes run as $a^{-4}$ and
$a^{-2}$. Written as fractions of the critical density today they define
$\Omega_{\L_2}, \Omega_{\L_3}$, and the Friedmann equation of the extended model
reads
\bea
E^2(a) \;\eqv\; \left(\fr{H(a)}{H_0}\right)^2
= \Omega_M a^{-3} + \left(\Omega_R + \Omega_{\L_2}\right) a^{-4}
+ \left(\Omega_K + \Omega_{\L_3}\right) a^{-2} + \Omega_\L\,,
\la{eq:friedmann}
\eea
with $\Omega_\L$ fixed by the condition $E(1)=1$. The new components enter the
expansion rate and nothing else, and we do not perturb them.

In Paper I \ci{Park:2025iog} we examined the imprint of \rf{eq:friedmann} on the
CMB with \textsc{class}, through brute-force scans of the parameter space and
quartic regression, and found that the extended model improves upon $\L$CDM by
the standard information criteria. The question was revisited with artificial
neural networks in Paper II \ci{Hatefi:2025lqe}: the gross shape of the power
spectrum is set by the standard parameters, with $\Omega_{\L_2}$ and
$\Omega_{\L_3}$ refining the fine shape. That feature --- coarse shape from the
standard set and fine shape from thermal corrections --- constitutes the
physical core of the subsequent analysis.

\subsubsection*{The non-redundancy of
$\Omega_{\L_2}$ and $\Omega_{\L_3}$: another illustration}
\la{sec:redundancy}

Since the component associated with $\Omega_{\L_2}$ scales as $a^{-4}$ and
$\Omega_{\L_3}$ as $a^{-2}$, the same powers carried by radiation and by spatial
curvature, the question arises whether they are redundant. It was answered in
\ci{Park:2025iog,Hatefi:2025lqe}; we answer it again here from a different
direction, which has the advantage of making the distinction more prominent. We
consider two systems: Einstein--scalar and Einstein--Maxwell, for reasons that
will become clear. It is well known that their stress-energy tensors can be
written in hydrodynamic form \cite{EllisMaartensMacCallum2012}. A scalar field
$\phi(x)$ with action
\begin{equation}
I_{\phi} = -\int d^{4}x\sqrt{-\mathrm{Det}g}\left[\frac{1}{2} g^{\mu \nu}\partial_{\mu}\phi\,\partial_{\nu}\phi + V(\phi)\right],
\label{eq:scalar_action}
\end{equation}
has an energy-momentum tensor given by
\begin{equation}
T_{\phi}^{\mu \nu} = g^{\mu\rho}g^{\nu\sigma}\partial_{\rho}\phi\,\partial_{\sigma}\phi
- g^{\mu\nu}\left(\frac{1}{2}g^{\rho\sigma}\partial_{\rho}\phi\,\partial_{\sigma}\phi + V(\phi)\right).
\label{eq:scalar_stress_energy}
\end{equation}
As noted, e.g., in \cite{Weinberg2008}, this energy-momentum tensor can be
written in the form of a perfect fluid:\footnote{The scalar system at the
classical level above requires only a subset of the terms present in the general
form of the stress-energy tensor. (A general form of hydrodynamic stress-energy
tensor is given in \ci{EllisMaartensMacCallum2012}.) It is likely that the full
set will be required for the purpose of decomposing the quantum correction for
the scalar matter sector (similarly for a Maxwell system).}
\begin{equation}
T_{\phi}^{\mu \nu} = p g^{\mu\nu} + (p + \rho)u^{\mu}u^{\nu},
\label{eq:perfect_fluid_form}
\end{equation}
with
\begin{align}
\rho &= -\frac{1}{2} g^{\mu\nu}\partial_{\mu}\phi\,\partial_{\nu}\phi + V(\phi), \label{eq:weinberg_rho}\\
p &= -\frac{1}{2} g^{\mu\nu}\partial_{\mu}\phi\,\partial_{\nu}\phi - V(\phi), \label{eq:weinberg_p}\\
u^{\mu} &= -\left[-g^{\rho\sigma}\partial_{\rho}\phi\,\partial_{\sigma}\phi\right]^{-1/2}
g^{\mu\n}\partial_{\n}\phi. \label{eq:weinberg_u}
\end{align}
The velocity four-vector $u^{\mu}$ is chosen to be comoving with the scalar
field, so that in the local rest frame the spatial gradients of $\phi$ vanish.
In this comoving frame, the expressions reduce to the familiar forms
\begin{equation}
\rho = \frac{1}{2}\dot{\phi}^2 + V(\phi), \qquad 
p = \frac{1}{2}\dot{\phi}^2 - V(\phi),
\end{equation}
where the dot denotes differentiation with respect to proper time. This perfect
fluid interpretation is exact for any potential $V(\phi)$, including the massive
case $V(\phi) = \frac{1}{2}m^2\phi^2$, as well as more general potentials. One
may consider the massless case, in which the field can be interpreted as
radiation. Regardless of the mass, the finite-temperature corrections shift the
cosmological constant. This is genuinely different from the radiation-like term
that the scalar Lagrangian produces at the classical level.

Suppose one quantizes the system and obtains the quantum-level action, i.e., the
1PI effective action. For simplicity, let us first consider the zero-temperature
case. (See \cite{Park:2024kfn} for the finite-temperature extension.) The classical action is shifted by quantum corrections. In particular, the
cosmological constant term gets shifted. These corrections come from
vacuum-bubble diagrams. By functionally differentiating the quantum action, one
gets the quantum-corrected stress-energy tensor:
\bea
R_{\m\n}-\fr12 g_{\m\n}R=8\pi G T^{\, tot}_{\m\n}
\eea 
where $T^{\, tot}_{\m\n}$ denotes the total stress-energy tensor. It contains
the contributions from the matter sector stress-energy tensor
$T_{\phi}^{\mu \nu}$ given in \rf{eq:perfect_fluid_form}, and quantum
corrections, including {\em the shifts in the cosmological constant}. Consider
the massless scalar case. As just mentioned, the scalar radiation density term
is present (with certain quantum corrections belonging to that sector, of
course) in $T^{\, tot}_{\m\n}$. Since the total stress tensor contains the CC
and its shift by the loop corrections separately from the scalar radiation term,
the CC sector is genuinely different from the matter radiation sector. In other
words, the two sectors are not redundant. (Similarly for an Einstein--Maxwell
system: the CC sector is genuinely different from the classical radiation sector
associated with the photon field.)

In the previous papers \ci{Park:2025iog,Hatefi:2025lqe}, several different
systems were considered to illustrate the non-redundancy. We briefly summarize
the discussion here before we pinpoint the precise task that must be carried out
to {\em derive} the hydrodynamic system --- for which the non-redundancy must
ultimately be established --- from first principles. In fact one of the systems
considered was an Einstein--Maxwell system. However, the context in which this
system was considered is different from the Einstein-Maxwell analysis (or the
Einstein-scalar system) pondered above. As just stated, it is a hydrodynamics
system, as opposed to a fundamental QFT system such as a scalar or Maxwell
field, in which the non-redundancy between the finite-temperature shifted CC and
the matter radiation is to be established. The reason for considering a
fundamental QFT system above was that it provides the clearest path to show the
different origin of the quantum-induced CC terms, hence the non-redundancy. 

The employment of a Maxwell system in the previous works is in line with the
ultimate goal, establishing the non-redundancy of the two new parameters in the
hydrodynamic system. In Ch.~13 of \cite{WeinbergQFT1}, an isolated charge was
considered to show that consideration of the ladder diagrams leads to the
presence of an external field produced by the charge. In other words, by
starting with a fundamental QFT system, an effective Lagrangian term, i.e., a
newly appearing external field, was derived. For cosmology, three extensions are
needed. The first is curved space extension. The second is a finite-temperature
extension. As discussed in the previous works the vacuum loop contributions
produce the $T^4$- and $T^2$-scaling terms in the CC sector. The third is that
one would consider many source particles as in the kinetic theory setup. In
principle it should be possible to obtain the result analogous to Ch.~13 for the
kinetic theory setup of multiple particles.

Let us recall why carrying out the proofs, such as $\r=\fr13 p$, in the QFT
framework (as opposed to a kinetic theory analysis) does not justify the
redundancy argument. It is only meaningful for the purpose of identifying the
field as radiation but not beyond. This is because there are limitations to the
formalism. Firstly, such a formalism has not been carried out by including the
Einstein-Hilbert term, let alone all the other Standard Model fields, on a
curved background. In the presence of such a term, it will be necessary to set
the background metric to a specific geometry just as one sets the scalar field
$\f$ to $\f=0$ to get the onshell free energy. As far as we are aware, such an
analysis has not been carried out. Secondly, even if that direction can be
pursued it is a different line of research. Independently of it, computation of
the offshell quantum effective action (called the Helmholtz free energy in the
thermodynamic context --- offshell before any solution is substituted) is a
legitimate task.

Getting back to the QFT scalar system in the beginning, the moral is that
sharing a power of $a$ is not the same as being the same term. A massless scalar
contributes a radiation-like term to the stress-energy tensor; the
finite-temperature corrections considered here shift the cosmological constant.
The two arrive at $a^{-4}$ by different routes, and the difference survives into
the observables. It has, moreover, an operational consequence exploited in the
present work. Radiation perturbs; $\Omega_{\L_2}$ does not, entering only
through $H(a)$ in \rf{eq:friedmann}. $\Omega_{\L_3}$ rides alongside $\Omega_K$
in $E(a)$ but leaves the perturbation equations alone at the level of our
approximation. (For its precise meaning we refer to
\ci{Park:2025iog,Hatefi:2025lqe}.)

\subsection{Entries in the theory file, and which power spectrum they are}
\label{sec:theoryfile}

In our previous works, the target CMB power spectra were those in the file,\\
{\small\texttt{COM\_PowerSpect\_CMB-base-plikHM-TTTEEE-\allowbreak lowl-lowE-lensing-\allowbreak minimum-theory\_R3.01.txt}}.
To understand the meaning of these entries and their justification as the target
spectrum, it is useful to recall some of the relevant aspects of Bayesian
parameter inference.

The central theoretical object in the analysis of the cosmic microwave
background (CMB) is the angular power spectrum $C_\ell$, which encodes the
variance of the temperature fluctuations as a function of multipole moment
$\ell$. A Boltzmann solver computes $C_\ell$ as a function of the cosmological
parameters $\boldsymbol{\theta}$, and the Planck likelihood compares that
prediction against the measured sky. In other words, $C_\ell$ is the quantity
through which theory makes contact with observation. Because we shall be feeding
one particular Planck data product into the comparisons of the following
sections, it is worth being explicit about \emph{which} power spectrum that file
contains, since the symbol $C_\ell$ is used in the literature for several
conceptually distinct objects.

In the analysis of the cosmic microwave background, four distinct quantities
must be kept apart. The first is the ensemble-average spectrum
$C_\ell(\boldsymbol{\theta})$, which is the theoretical prediction viewed as a
function of the cosmological parameters $\boldsymbol{\theta}$. This is the
general mapping $\boldsymbol{\theta} \mapsto C_\ell(\boldsymbol{\theta})$ that
Boltzmann codes such as CLASS or CAMB implement: given a set of input
parameters, they return a predicted curve for the angular power spectrum as a
function of multipole $\ell$. The second quantity is the evaluation of that
function at the particular point $\boldsymbol{\theta}_{\mathrm{best}}$ returned
by an inference procedure. This gives
$C_\ell(\boldsymbol{\theta}_{\mathrm{best}})$, a single fixed curve in $\ell$.
The distinction between the function and its evaluation is the elementary one
familiar from $f(x)=x^2$ versus $f(3)$: the former is a rule, the latter a
number (or, here, a curve). Evaluating the same function at the true but
inaccessible parameters of Nature gives
$C_\ell(\boldsymbol{\theta}_{\mathrm{true}})$, a third distinct quantity. The
fourth quantity is $\hat{C}_\ell$, which denotes the single observed realization
that any real or ideal experiment actually measures. Unlike the theoretical
spectra, which are ensemble averages, $\hat{C}_\ell$ is a particular outcome
from our one universe.

The content of the first three quantities is captured by the ensemble average
\begin{equation}
  C_\ell(\boldsymbol{\theta})
  = \mathbb{E}\!\bigl[\hat{C}_\ell \,\big|\, \boldsymbol{\theta}\bigr],
  \label{eq:ensavg}
\end{equation}
which is the mean over infinitely many realizations of a universe with
parameters $\boldsymbol{\theta}$, free of instrumental noise, foregrounds,
sky-mask effects, and cosmic-variance fluctuations. The observed spectrum
$\hat{C}_\ell$ differs from the true theoretical spectrum by the cosmic-variance
fluctuation:
\begin{equation}
  \hat{C}_\ell
  = C_\ell(\boldsymbol{\theta}_{\mathrm{true}}) + \delta_\ell^{\mathrm{CV}} ,
  \label{eq:cosmicvar}
\end{equation}
where $\delta_\ell^{\mathrm{CV}}$ is the random deviation due to the fact that
we observe only one realization of the Universe. These distinctions clarify the
concept of cosmic variance and explain why best-fit parameters act as
estimators. They also justify using a Bayesian approach, instead of a
deterministic one, for parameter estimation.

Note that the gap in \eqref{eq:cosmicvar} cannot be closed by an improved
experiment. Three averages that arise in CMB theory are distinguished in
\cite{Weinberg2008} --- the average over all positions from which the radiation
could be observed, the average over all historical accidents that produced a
particular pattern of fluctuations, and the quantum-mechanical expectation value
over fluctuations generated during inflation --- which coincide under the
ergodic theorem given isotropy and homogeneity.\footnote{Since Planck performs
an angular average over the sky, and ergodicity lets that average serve as a
proxy for the position average, one might expect an ideal experiment (free of
noise, systematics and masking) to return $C_\ell(\boldsymbol{\theta})$ itself.
It does not. The angular average at each multipole is taken over only $2\ell+1$
modes, whereas the ensemble average of \eqref{eq:ensavg} corresponds to
infinitely many, which would require infinitely many independent realizations of
the Universe. No observer, however ideally equipped, has access to more than
$2\ell+1$ modes at multipole $\ell$, so the residual $\delta_\ell^{\mathrm{CV}}$
is irreducible. Ergodicity therefore justifies \emph{using}
$C_\ell(\boldsymbol{\theta})$ as the prediction for a single observer, but the
equality $\hat{C}_\ell = C_\ell(\boldsymbol{\theta}_{\mathrm{true}})$ holds only
in expectation, never for an individual measurement. It is thus more precise to
say that a Boltzmann code computes the spectrum an ideal experiment would
recover \emph{in ensemble average} than to say it computes what would be
observed.}

The same distinction governs the two parameter vectors. Here
$\boldsymbol{\theta}_{\mathrm{true}}$ is fixed by Nature but unobservable, while
$\boldsymbol{\theta}_{\mathrm{best}} = \arg\max_{\boldsymbol{\theta}} P(\boldsymbol{\theta} \mid \hat{C}_\ell)$
is an \emph{estimator} of it, constructed from the observed data (and coinciding
with the maximum-likelihood point when the priors are broad and flat). Two
sources of uncertainty separate them: cosmic variance, which displaces
$\boldsymbol{\theta}_{\mathrm{best}}$ from $\boldsymbol{\theta}_{\mathrm{true}}$
by a statistical fluctuation that no improvement of the instrument can reduce,
and, in a real experiment, instrumental noise and systematics. In expectation
over the ensemble the estimator is unbiased,
$\mathbb{E}[\boldsymbol{\theta}_{\mathrm{best}}]=\boldsymbol{\theta}_{\mathrm{true}}$,
but for any single observed Universe the equality fails; the spread of
$\Delta\boldsymbol{\theta}\equiv\boldsymbol{\theta}_{\mathrm{best}}-\boldsymbol{\theta}_{\mathrm{true}}$
is characterized by the posterior width, i.e.\ the parameter uncertainties
reported by Cobaya. Correspondingly,
$C_\ell(\boldsymbol{\theta}_{\mathrm{best}})$ is not an experimental outcome but
the ensemble-mean prediction of the cosmology that best fits $\hat{C}_\ell$; it
depends on the particular realization we happen to observe, and coincides with
$C_\ell(\boldsymbol{\theta}_{\mathrm{true}})$ only in the limit of infinite data
and vanishing cosmic variance. (Note that in Bayesian inference
$\boldsymbol{\theta}$ is itself a random variable with posterior
$P(\boldsymbol{\theta}\mid\hat{C}_\ell)$, but that is a separate layer; the
ensemble in \eqref{eq:ensavg} is always over realizations at fixed
$\boldsymbol{\theta}$.)

\subsubsection*{The minimum-theory file}

With these distinctions made, we can now state which Planck data file we are
using:
{\small\texttt{COM\_PowerSpect\_CMB-base-plikHM-TTTEEE-\allowbreak lowl-lowE-lensing-\allowbreak minimum-theory\_R3.01.txt}},
which came with the Planck 2018 release, gives the best-fit theoretical spectra
for the $\Lambda$CDM model. These spectra are computed at the parameter values
$\boldsymbol{\theta}_{\mathrm{best}}$ that minimize the negative log-likelihood
$-\ln \mathcal{P}(\boldsymbol{\theta}\mid\mathrm{data})$ for the standard
likelihood combination called \texttt{Planck TT,TE,EE+lowE+lensing}. The file
contains quantities defined as
\begin{equation}
    \mathcal{D}_\ell^{XY} \;=\; \frac{\ell(\ell+1)}{2\pi}\,
    C_\ell^{XY}\!\bigl(\boldsymbol{\theta}_{\mathrm{best}}\bigr),
    \qquad XY \in \{TT,\, TE,\, EE\},
\end{equation}
with units of $\mu\mathrm{K}^2$. (The file also carries $XY \in \{BB,\, dd \}$.)
Here, $C_\ell^{XY}(\boldsymbol{\theta}_{\mathrm{best}})$ is the power spectrum
computed by a Boltzmann solver (CAMB, in Planck's case) at the best-fit point.
In the terminology introduced earlier, this file provides a tabulation of
$C_\ell(\boldsymbol{\theta}_{\mathrm{best}})$ --- that is, the theoretical
average over many realizations of the Universe --- not the measured power
spectrum $\hat{C}_\ell$ from a single observation.\footnote{The Planck Legacy
Archive documentation~\cite{PLA_wiki_CMB} says these are the best-fit
$\Lambda$CDM spectra from the baseline TT,TE,EE+lowE+lensing analysis. It also
notes that before comparing them with the actual coadded CMB spectra, one must
divide them by the square of the best-fit calibration parameter
\texttt{calPlanck}. See
\url{https://wiki.cosmos.esa.int/planck-legacy-archive/index.php/CMB_spectrum_\%26_Likelihood_Code}.
The best-fit \texttt{calPlanck} value itself is given in the companion file
\texttt{COM\_PowerSpect\_CMB-base-plikHM-TTTEEE-\allowbreak lowl-lowE-lensing-\allowbreak minimum\_R3.01.txt}
as entry 25, $\texttt{calPlanck}=1.000442$, so that the correction is
$\texttt{calPlanck}^2=1.000884$.}

Three files from the release are in play here, and it is important not to
confuse their roles. Two of them belong to the theory side. The companion file
\texttt{...-minimum\_R3.01.txt} records the \emph{input}, namely the best-fit
values of all cosmological, nuisance and foreground parameters ($\omega_b$,
$\omega_c$, $\theta_{\mathrm{MC}}$, $\tau$, $A_s$, $n_s$, together with the
calibration and foreground parameters) at the point minimizing
$-\ln\mathcal{P}(\boldsymbol{\theta}\mid\mathrm{data})$, along with the total
$-\ln(\mathrm{Like})$ and $\chi^2$. The file quoted above,
\texttt{...-minimum-theory\_R3.01.txt}, records the \emph{output}, the spectra
$\mathcal{D}_\ell^{XY}$ in columns $\ell$, $\mathcal{D}_\ell^{TT}$,
$\mathcal{D}_\ell^{TE}$, $\mathcal{D}_\ell^{EE}$, $\mathcal{D}_\ell^{BB}$,
$\mathcal{D}_\ell^{dd}$. The third file,
\texttt{COM\_PowerSpect\_CMB-TT-full\_R3.01.txt}, is of a different kind
altogether: it tabulates the \emph{measured} temperature spectrum with its
uncertainties, and so is a single realization --- the $\hat{C}_\ell$ of the
distinction drawn above --- not an ensemble average. The first two files answer
``what did the fit assume, and what curve did it produce''; the third is the
thing the fit was fitting to. The first two are related as the input and the
output of a single Boltzmann computation: the cosmological subset of the
parameters recorded in \texttt{...-minimum\_R3.01.txt} is what Planck handed to
CAMB, and the spectra recorded in \texttt{...-minimum-theory\_R3.01.txt} are
what CAMB returned,
\begin{equation}
    \underbrace{\boldsymbol{\theta}_{\mathrm{best}}}_{\texttt{...-minimum}}
    \;\xrightarrow{\;\text{CAMB}\;}\;
    \underbrace{C_\ell^{XY}\!\bigl(\boldsymbol{\theta}_{\mathrm{best}}\bigr)}
              _{\texttt{...-minimum-theory}} .
\end{equation}
(The remaining entries of the first file --- the foreground and nuisance
parameters, together with the calibration $\texttt{calPlanck}$ --- take no part
in this step; they enter only when the predicted spectra are confronted with the
data.) One more important point about $\boldsymbol{\theta}_{\mathrm{best}}$: it
is the maximum a posteriori (MAP) point, \emph{not} the posterior mean
\begin{equation}
    \langle \boldsymbol{\theta} \rangle_{\mathrm{post}}
    \;=\; \int \mathrm{d}\boldsymbol{\theta}\;\boldsymbol{\theta}\;
    P(\boldsymbol{\theta} \mid \mathrm{data}),
\end{equation}
which is what gets reported in the marginalized parameter tables of
Ref.~\cite{Planck:2018vyg}. These two summaries of the posterior are generally
different. And since $C_\ell^{XY}$ depends on $\boldsymbol{\theta}$ in a
nonlinear way, the corresponding spectra also differ:
\begin{equation}
    C_\ell^{XY}\!\bigl(\boldsymbol{\theta}_{\mathrm{best}}\bigr)
    \;\neq\;
    C_\ell^{XY}\!\bigl(\langle\boldsymbol{\theta}\rangle_{\mathrm{post}}\bigr)
    \;\neq\;
    \bigl\langle C_\ell^{XY}(\boldsymbol{\theta}) \bigr\rangle_{\mathrm{post}} .
\end{equation}
So the theory file should \emph{not} be treated as the spectrum evaluated at the
tabulated marginalized parameters.

\subsubsection*{Reproducing the best-fit spectrum with a nonstandard model}

Now that we have clearly established that
$C_\ell(\boldsymbol{\theta}_{\mathrm{best}})$ is a theoretical average over many
possible universes, not an actual measurement, we can ask a new question. Could
a more complicated model --- one with extra parameters like $\Omega_{\Lambda_2}$
and $\Omega_{\Lambda_3}$ --- reproduce this theoretical curve? In other words,
does there exist some set of parameters $\boldsymbol{\theta}'$ in the extended
model such that
\begin{equation}\label{eq:matching}
  C_\ell^{\mathrm{exten}}(\boldsymbol{\theta}')
  \approx C_\ell^{\Lambda\mathrm{CDM}}(\boldsymbol{\theta}_{\mathrm{best}})
\end{equation}
This is a well-posed question, but it is important not to confuse it with a
different one: whether the extended model fits the actual data.
Equation~\eqref{eq:matching} is about comparing two theories --- both sides are
ensemble averages, and no observed data enter. The statistical question, by
contrast, is about comparing theories to data: does the extended model give a
higher likelihood than the standard model, i.e., is
$\mathcal{L}(\hat{C}_\ell\mid\boldsymbol{\theta}') \geq \mathcal{L}(\hat{C}_\ell\mid\boldsymbol{\theta}_{\mathrm{best}})$?
These two questions are not the same, because there are three different things
at play: the actual observed spectrum $\hat{C}_\ell$, the best-fit $\Lambda$CDM
theoretical curve
$C_\ell^{\Lambda\mathrm{CDM}}(\boldsymbol{\theta}_{\mathrm{best}})$, and the
best-fit extended-model curve
$C_\ell^{\mathrm{exten}}(\boldsymbol{\theta}'_{\mathrm{best}})$. The first is
not equal to the second, because of cosmic variance (as in equation
\eqref{eq:cosmicvar}). So even if two theoretical curves match each other
perfectly, that does not mean the extended model will fit the real data equally
well. The likelihood depends on many things --- the full covariance structure,
nuisance parameters, foregrounds, and additional data sets --- not just on the
average spectrum by itself.

If such a $\boldsymbol{\theta}'$ does exist, then the extended model is said to
be \emph{spectrally degenerate} with $\Lambda$CDM. This is possible because the
mapping from parameters to spectra
$\boldsymbol{\theta}\mapsto C_\ell(\boldsymbol{\theta})$ is not one-to-one:
shifts in the late-time expansion history, matter density, sound horizon,
angular-diameter distance, and reionization amplitude can partly cancel each
other out, and the extra parameters $\Omega_{\Lambda_2}$ and
$\Omega_{\Lambda_3}$ may provide exactly that kind of compensation (as they
indeed do in the degeneracy directions we identify later). However, spectral
degeneracy does not imply degeneracy at the level of the likelihood. It is
therefore legitimate to use $C_\ell(\boldsymbol{\theta}_{\mathrm{best}})$ as a
target for testing whether the nonstandard model can mimic $\Lambda$CDM, for
locating degeneracy directions, and for initializing parameter scans. It should
not, however, be used as a substitute for a likelihood analysis against the
actual Planck data.

\section{The acoustic era: Weinberg's treatment carried further}
\la{sec:hydro}

In this section we develop the analytical machinery used for the CMB analysis in
Sec.~\re{sec:kinetic} by updating an analytic method by Weinberg
\ci{Weinberg2008}. His formulation rests on a sequence of simplifying
assumptions: a tight-coupling treatment of the photon-baryon plasma before
recombination, WKB approximations of the acoustic oscillation and its damping,
and sudden decoupling at last scattering.\footnote{In Paper I, Weinberg's CMB
power spectrum curve was checked and reproduced by employing an Mathematica
analysis.} We improve his original formulation to make it more accurate.
Weinberg's analysis ends in a closed-form expression for the temperature
multipoles, his eq. (7.2.41). In that approach every ingredient remained
identifiable---allowing one to point directly to the terms responsible for Silk
damping or the Doppler contribution. Our treatment keeps this identifiability
but removes the approximations (one at a time), allowing us to achieve much
higher accuracy without losing track of the individual physical processes.

The endpoint of this section is \rf{eq:final}, our final analytic result. It has
the same structure as Weinberg's (7.2.41) but carries one additional ingredient,
the parameter $\eps_P$, which measures the residual polarization correction to
the acoustic source over a photon mean free time; setting $\eps_P\to0$ renders
\rf{eq:final} identical to \W{7.2.41}, and retaining it is the accuracy
improvement that the closed form gains here. Several of the intermediate results
are of independent interest, in that they follow directly from the Boltzmann
hierarchy with no reference to fluid equations, viscosity coefficients, or heat
conduction: the damping rate \rf{eq:Gamma}, in particular, reproduces \W{6.4.25}
with its $16/15$ descending from the polarization ladder and the slaved shear
rather than from a transport coefficient assigned to an imperfect fluid. What is
gained thereby is not only accuracy but also traceability --- each coefficient
retains an identifiable kinetic origin. The numerical approach of
Sec.~\ref{sec:kinetic} works the same way, so it preserves the same
traceability: the form factors $F$, $G$ and $H$ are computed separately there
too, so a feature of a parameter response can be attributed to the term that
produced it.

Because the derivation is long, we have kept in this section only the key steps.
The definitions of Weinberg's kinetic variables, the Boltzmann hierarchy they
satisfy, and the intermediate algebra of several of the reductions below are
collected in Appendix~\re{app:sec3}, to which we refer at each point where the
material has been set aside.

\subsection{The exact multipole formula}
\la{sec:exact}

Weinberg's starting point is kinetic rather than hydrodynamic. The photon
distribution is carried by a dimensionless intensity matrix
$J_{ij}(\mathbf{q},\hat p,t)$, whose trace at the observer is the fractional
temperature fluctuation. Expanding $J_{ij}$ in partial waves gives temperature
and polarization amplitudes $\DT{\ell}$, $\DP{\ell}$ that satisfy a coupled
Boltzmann hierarchy. Those definitions, the hierarchy, the photon fluid moments
and the photon--baryon momentum balance are all recorded in
Appendix~\re{app:sec3}.

Defining the temperature fluctuation as
\begin{equation}
\Delta T(\hat{\mathbf{n}})\equiv T(\hat{\mathbf{n}})-T_0
=\sum_{\ell m}a_{\ell m}\,Y_\ell^{m}(\hat{\mathbf{n}})\,,
\qquad
T_0\equiv\frac{1}{4\pi}\int \dd^2\hat{\mathbf{n}}\;T(\hat{\mathbf{n}})\,
\label{eq:W2-6-1}
\end{equation}
where $a_{\ell m}$, denoted by $a^S_{T,\ell m}$ below, is the scalar
partial-wave amplitude of the observed fractional temperature fluctuation, the
goal is the angular power spectrum
\begin{equation}
C_\ell=\frac{1}{4\pi}\int \dd^2\hat{\mathbf{n}}\,\dd^2\hat{\mathbf{n}}'\,
P_\ell(\hat{\mathbf{n}}\cdot\hat{\mathbf{n}}')\,
\bigl\langle\Delta T(\hat{\mathbf{n}})\,\Delta T(\hat{\mathbf{n}}')\bigr\rangle\,.
\label{eq:W2-6-5}
\end{equation}
The temperature fluctuation can be written as
\begin{equation}
\frac{\Delta T(\hat{\mathbf{n}})}{T_0}
=\frac14\,J_{ii}(\mathbf{x}=0,\,-\hat{\mathbf{n}},\,t_0)\,,
\label{eq:W7-1-1}
\end{equation}
where the dimensionless fractional density matrix $J_{ij}$ is introduced through
the photon distribution matrix for small perturbations,
\begin{equation}
n_{ij}(\mathbf{x},\mathbf{p},t)
=\bar n_\gamma\!\bigl(a(t)\,p^0(\mathbf{x},\mathbf{p},t)\bigr)
\left[\,\frac12\!\left(g_{ij}(\mathbf{x},t)
-\frac{g_{ik}(\mathbf{x},t)\,g_{jl}(\mathbf{x},t)\,p_k p_l}
{[\,p^0(\mathbf{x},\mathbf{p},t)\,]^{2}}\right)\right]
+\delta n_{ij}(\mathbf{x},\mathbf{p},t),
\label{eq:W6-1-6}
\end{equation}
with
\begin{equation}
p^0(\mathbf{x},\mathbf{p},t)\equiv\sqrt{g^{ij}(\mathbf{x},t)\,p_i p_j}\,,
\qquad
\bar n_\gamma(p)\equiv\frac{1}{(2\pi)^3}
\left[\exp\!\left(\frac{p}{k_B\,a(t)\,\bar T(t)}\right)-1\right]^{-1}.
\label{eq:W6-1-8}
\end{equation}
The matrix $J_{ij}$ is then defined by
\begin{equation}
a^{4}(t)\,\bar\rho_\gamma(t)\,J_{ij}(\mathbf{x},\hat{\mathbf{p}},t)
\equiv a^{2}(t)\int_0^{\infty}
\delta n_{\gamma\,ij}(\mathbf{x},p\hat{\mathbf{p}},t)\,4\pi p^{3}\,\dd p\,,
\label{eq:W6-1-13}
\end{equation}
where
$\bar\rho_\gamma(t)\equiv a^{-4}(t)\int 4\pi p^{3}\,\bar n_\gamma(p)\,\dd p$. It
is shown in \ci{Weinberg2008} that $a^S_{T,\ell m}$ can be expressed as
\begin{equation}
\begin{split}
a^S_{T,\ell m}=4\pi\,i^\ell\,T_0\int \dd^3q\;\alpha(q)\,
Y^{m*}_\ell(\hat q)\int_{t_1}^{t_0}\dd t\,
\Bigl[&j_\ell\bigl(qr(t)\bigr)F(q,t)
+j'_\ell\bigl(qr(t)\bigr)G(q,t)\\
&+j''_\ell\bigl(qr(t)\bigr)H(q,t)\Bigr].
\end{split}
\label{eq:alm}
\end{equation}
Propagating the line-of-sight solution \ci{Seljak:1996is}\ci{Weinberg2008} to the
observer gives the temperature power spectrum as
\be
C^S_{TT,\ell}=16\pi^2 T_0^2\int_0^\infty q^2\,\dd q\,
\Bigl|\int_{t_1}^{t_0}\dd t\,
\bigl[\,j_\ell\bigl(qr(t)\bigr)F(q,t)+j'_\ell\bigl(qr(t)\bigr)G(q,t)
+j''_\ell\bigl(qr(t)\bigr)H(q,t)\bigr]\Bigr|^2 ,
\la{eq:7210}
\ee
via $\langle\alpha(q)\alpha^*(q')\rangle=\delta^3(q-q')$, where
$r(t)\equiv \int_t^{t_0}\dd t'/a(t')$ is the unperturbed radial coordinate. This
is exact within linear perturbation theory. The three form factors are
\bea
F(q,t)&=&e^{-\tau(t)}\Bigl\{
\om_c(t)\Bigl[\fr34\Phi+\fr3{16}\Pi
-\fr12 a^2\ddot B_q-\fr12 a\dot a\dot B_q\Bigr]
-\fr12\fr{\dd}{\dd t}\Bigl[A_q+a^2\ddot B_q+a\dot a\dot B_q\Bigr]\Bigr\},
\la{eq:Fqt}\\
G(q,t)&=&-q\,\om_c(t)\,e^{-\tau(t)}
\Bigl[\fr{\delta u_{Bq}(t)}{a(t)}+\fr{a(t)\dot B_q(t)}{2}\Bigr],
\la{eq:Gqt}\\
H(q,t)&=&\fr3{16}\,\om_c(t)\,e^{-\tau(t)}\,\Pi(q,t),
\la{eq:Hqt}
\eea
where $\om_c(t)$ is the collision rate of a photon with electrons in the
baryonic plasma, $A_q$ and $B_q$ are the scalar modes of the metric
perturbation,
\bea
\delta g_{ij}(\mathbf{x},t) = a^2(t) \sum_n \int d^3q \, \alpha_n(\mathbf{q}) \left[ A_{nq}(t) \delta_{ij} - q_i q_j B_{nq}(t) \right] e^{i \mathbf{q} \cdot \mathbf{x}},
\eea
and $\delta u_{Bq}(t)$ is the velocity potential of the baryonic plasma. The
last term of $F$, a total time derivative, is the integrated Sachs--Wolfe (ISW)
contribution.\footnote{Since the ISW term is a total time derivative, it can be naturally integrated by parts. The rearrangement is algebraically exact but
numerically disastrous: it replaces the small time derivative of the metric
combination $A_q+a^2\ddot B_q+a\dot a\dot B_q$ by the undifferentiated
combination, which is of order unity, multiplied by the unbounded factor $q/a$.
A small quantity is thereby written as the difference of two large ones, which
are then integrated against different Bessel weights, $j_\ell$ and $j'_\ell$. In
a numerical evaluation the resulting $C_\ell$ grows as $\ell^2$ instead of
damping. The pipeline of Sec.~\re{sec:kinetic} therefore keeps the ISW term in
the printed form \rf{eq:Fqt} and takes its time derivative directly.} The two
sources $\Phi$ and $\Pi$ are combinations of the lower multipoles,
\bea
\Phi = \fr16\left(2\DT{0} - \DP{0} - \DT{2} - \DP{2}\right), \qquad
\Pi = \DP{0} + \DT{2} + \DP{2}.
\la{eq:sources}
\eea
The source $\Pi$ will be seen to control both the damping and the polarization
correction to the temperature spectrum. One consequence of the hierarchy is
worth displaying here, because the whole of the improvement below follows from
it. The identity $3\Phi=\DT{0}-\ft12\Pi$, derived in Appendix~\re{app:sec3},
turns the scattering source of $F$ in \rf{eq:Fqt} into
\begin{equation}
\omega_c\Bigl[\tfrac34\Phi+\tfrac3{16}\Pi\Bigr]
=\omega_c\Bigl[\tfrac14\DT{0}+\tfrac1{16}\Pi\Bigr]:
\label{eq:Fsource}
\end{equation}
the intrinsic $\delta T/\bar T=\tfrac14\delta_{\gamma}$ plus an explicit
polarization correction $\Pi/16$. Together with \rf{eq:Hqt}, this means that the
polarization source $\Pi$ enters the temperature spectrum in \emph{two} places.
Both are kept below, and keeping them is what produces the $\eps_P$ term of
\rf{eq:final}.

\subsection{Improved approximation schemes}

The analysis above is exact within linear perturbation theory. In
\ci{Weinberg2008} the power spectrum $C^S_{TT,\ell}$ is then computed with a
series of approximations. One of the tasks of the present work is to calculate
the spectrum more accurately: in this section by improving Weinberg's
hydrodynamic approximation, and in Sec.~\re{sec:kinetic} by evaluating his
kinetic formulation without those approximations. Starting from the exact
kinetic multipole formula we introduce the tight-coupling ($1/\omega_c$)
expansion, the large-$\ell$ Debye reduction, and a WKB treatment of the acoustic
mode, and assemble them into a closed formula that generalizes Weinberg's
(7.2.41). Every \emph{numerical} value quoted in this part is the evaluation of
that closed analytic formula, \rf{eq:final}, obtained \emph{after} the
tight-coupling, sudden-decoupling and Debye approximations have been made. They
are conceptually distinct from, and are not to be compared line-by-line with,
the full-hierarchy numbers in Sec.~\re{sec:kinetic}, which are computed without
any such approximation.

\subsubsection{Tight coupling as the $1/\om_c$ expansion of the hierarchy \label{sec:tc}
}

The expansion carried out in this subsection is not new in itself, and it is
worth saying at the outset where it stands in relation to the standard
treatments. An expansion in the inverse collision rate is the oldest
approximation in the subject: it goes back to Peebles and Yu
\ci{Peebles:1970ag}, who used it to reduce the photon--baryon system to a damped
acoustic oscillator, and the viscosity coefficient that carries the polarization
correction --- the $16/15$ recovered in \rf{eq:Gamma} below --- was obtained by
Kaiser \ci{Kaiser:1983nrc}. In the form used by every modern Boltzmann code it
is the first-order tight-coupling scheme of Ma and Bertschinger
\ci{MaBertschinger1995}, in which the photon--baryon slip and the photon shear
are solved for to first order in the Thomson mean free time and substituted back
into the $\ell\le1$ equations; second-order versions were later given by
Cyr-Racine and Sigurdson \cite{Cyr-Racine:2010qdb} and by Pitrou
\ci{Pitrou:2010ai}, and one of these is what \textsc{class} \ci{Blas:2011rf}
integrates at early times. The analytic use made of the same expansion ---
carrying it through to a closed form for the spectrum rather than to a system to
be integrated --- is due to Hu and Sugiyama \ci{Hu:1994uz,Hu:1995en}.

Two differences from that literature should be kept in view. The first is one of
purpose. In the numerical codes the expansion is a device for evading the
stiffness of the exact equations before recombination, and it is abandoned at a
switch point once the collision rate has dropped far enough; the accuracy that
matters there is the accuracy of the solution up to the switch. Here it is used
analytically, to exhibit which kinetic quantity supplies which coefficient of
the closed form \rf{eq:final}, and the numerical calculation of
Sec.~\re{sec:kinetic} makes no such expansion and no such switch at all. The
second is one of variables: the references above work in conformal time with Ma
and Bertschinger's $\delta$, $\theta$, $\sigma$ and the metric pair $h$, $\eta$,
whereas we follow Weinberg's cosmic-time variables throughout, so that the
powers of $a$ and the placement of the collision terms differ term by term even
where the physics is identical. The dictionary between the two is set out in
Appendix~\re{sec:mb}.

Deep in the pre-recombination era, the collision rate $\omega_c$ is much larger
than both the acoustic frequency and the Hubble expansion rate,
\[\omega_c \gg \omega_0 \gg H,\]
where $\omega_0$ is the acoustic frequency defined below. This hierarchy
naturally suggests an expansion in the small parameter
\[\epsilon \sim \frac{q/a}{\omega_c}\sim \frac{\omega_0}{\omega_c}= \omega_0 t_\gamma \ll 1,\qquad t_\gamma \equiv \frac{1}{\omega_c}.\]
This hierarchy is the organizing principle of this section. Tight coupling is
not an approximation imposed on an otherwise independent fluid description.
Rather, as we will see, it emerges as the leading-order term in a systematic
expansion of the underlying kinetic hierarchy, with damping appearing at the
next order. We therefore expand Eqs. \eqref{eq:hierP}--\eqref{eq:hierT} and
\eqref{eq:momclean} in powers of $\epsilon$. For the fast (acoustic) modes, we
can neglect the metric source terms $\dot A_q$ and $\dot B_q$, which drive the
slow mode discussed below, as well as all terms suppressed by the small ratio
$H/\omega_0$. At leading order, the free-streaming couplings in the polarization
hierarchy are of \(\mathcal{O}(\epsilon)\) relative to the collision terms,
effectively locking --- or ``slaving'', in standard terminology --- the
polarization ladder to $\Pi$. At the same order the photon quadrupole is slaved
to the velocity gradient, and the photon--baryon slip is slaved to the
acceleration; both are $\mathcal{O}(\eps)$. Feeding these into the $\ell \leq 1$
equations and \rf{eq:momclean}, and letting all fast-mode quantities vary as
$\exp(-i\!\int\! \om\, \dd t)$ with $\om$ slowly varying, yields the WKB
dispersion relation, whose real part reproduces the acoustic frequency $\om_0$
and whose imaginary part gives the damping rate $\G$ of \rf{eq:Gamma}.

\vspace{.1in}

\ni {\bf The polarization ladder and $\Pi=\frac52\DT{2}$}

In the tight-coupling limit, Thomson scattering maintains the photon
distribution very close to isotropy, but the small quadrupole anisotropy
$\Delta_{T,2}^{(S)}$ is the primary source of both the temperature and
polarization anisotropies. At leading order the free-streaming couplings in
\eqref{eq:hierP} are of $\mathcal{O}(\eps)$ relative to the collision terms.
Dropping the free-streaming term entirely, \eqref{eq:hierP} becomes
\begin{equation}
\Delta_{P,\ell}^{(S)} = \frac{1}{2}\Pi
\left( \delta_{\ell 0} + \frac{\delta_{\ell 2}}{5} \right) + \mathcal{O}(\epsilon),
\end{equation}
so
\begin{equation}
\DP{0}=\tfrac12\Pi+\mathcal{O}(\eps),\qquad
\DP{2}=\tfrac1{10}\Pi+\mathcal{O}(\eps),\qquad
\DP{\ell}=\mathcal{O}(\eps)\ \ \text{otherwise}.
\label{eq:ladder}
\end{equation}
Inserting \eqref{eq:ladder} into the definition $\Pi=\DP{0}+\DT{2}+\DP{2}$ of
\eqref{eq:sources},
\begin{equation}
\Pi=\tfrac12\Pi+\tfrac1{10}\Pi+\DT{2}
\quad\Longrightarrow\quad
\;\Pi=\tfrac52\,\DT{2}\;\,.
\label{eq:Pi}
\end{equation}
This is the well-known polarization enhancement: the source $\Pi$ that generates
polarization is not simply the bare temperature quadrupole $\Delta_{T,2}^{(S)}$,
but is amplified by the polarization it itself creates. This coefficient
underlies the $16/15$ of eq.~(8) of \ci{Kaiser:1983nrc} and the $16/15$ of
\rf{eq:Gamma} below.

\vspace{.1in}
\ni {\bf Slaved shear}

The $\ell=2$ line of \eqref{eq:hierT}, with $\dot\Delta_{T,2}=O(\eps)$ relative,
$\DT{3}=O(\eps^2)$, and $\Pi$ from \eqref{eq:Pi}, gives
\begin{equation}
\omega_c\Bigl(\DT{2}-\tfrac1{10}\Pi\Bigr)
=\tfrac34\,\omega_c\,\DT{2}
=\frac{2q}{5a}\,\DT{1}+O(\eps^2)
\quad\Longrightarrow\quad
\;\DT{2}=\frac{8q}{15\,a\,\omega_c}\,\DT{1}\;\,.
\label{eq:shear}
\end{equation}
This is the photon shear (anisotropic stress, cf.\
$q^2\pi^S_{\gamma q}=\bar\rho_\gamma\DT{2}$) slaved to the dipole, i.e.\ kinetic
shear viscosity. The numerical coefficient is where the treatment of Thomson
scattering matters, and it is what puts the $16/15$ into the damping rate
\rf{eq:Gamma}; the comparison with the no-polarization and isotropic-scattering
treatments is given in Appendix~\re{app:sec3}.

\vspace{.1in}
\ni {\bf Slip}

The $\ell=1$ line of \eqref{eq:hierT}, combined with the fluid moments
\eqref{eq:moments}, relates the time derivative of the dipole to the momentum
exchanged between photons and baryons by scattering. That exchange is
proportional to the \emph{difference} of the two velocities, as momentum
conservation requires, and in the tight-coupling regime the difference --- the
slip --- is of order $\epsilon$. Solving for it gives \rf{eq:slip} of
Appendix~\re{app:sec3}: the slip too is slaved to the photon multipoles and
suppressed by the large collision rate, so photons and baryons move almost
together and any residual relative velocity is quickly erased by scattering.

\vspace{.1in}
\ni {\bf WKB dispersion relation and the damping rate}
\label{sec:dispersion}

The fast modes oscillate on a timescale far shorter than the Hubble time, so we
treat them by the WKB ansatz: each fast-mode quantity is taken to vary as
\begin{equation}
\Delta_{T,0}^{\text{fast}}(t) \propto \exp\!\left(-i \int \omega(t')\, dt'\right),
\end{equation}
with $\omega(t)$ an instantaneous frequency that drifts slowly,
$|\dot\omega/\omega|\ll\omega$, as the universe expands and the sound speed of
the photon-baryon fluid changes. The phase accumulated is then the time integral
of $\omega$ rather than $\omega t$. With this ansatz the $\ell=0$ line of
\eqref{eq:hierT} collapses, by the identity behind \rf{eq:Fsource} and after
dropping the metric terms for the fast mode, to\footnote{The intermediate step
is given in Appendix~\re{app:sec3}.}
\begin{equation}
-\,i\omega\,\DT{0}+k\,\DT{1}=0
\qquad\Longrightarrow\qquad
\DT{1}=\frac{i\omega}{k}\,\DT{0},
\qquad
\delta u_{\gamma q}=-\frac{3}{4k}\,\DT{1}
=-\frac{3i\omega}{4k^2}\,\DT{0},
\label{eq:e0}
\end{equation}
where $k\equiv q/a$. Inserting \eqref{eq:e0} into the slip \eqref{eq:slip} and
dropping the $O(\eps^2)$ shear term expresses the slip through the monopole,
\begin{equation}
\delta u_{Bq}-\delta u_{\gamma q}
=\frac{R}{4(1+R)\,\omega_c}\,\DT{0}+O(\eps^2),
\label{eq:slipval}
\end{equation}
and feeding \eqref{eq:e0}, \eqref{eq:shear} and \eqref{eq:slipval} into the
momentum balance \eqref{eq:momclean} gives the dispersion relation
\begin{equation}
\omega^2(1+R)
=\frac{k^2}{3}
-\,i\,\omega\,\frac{k^2}{\omega_c}
\Bigl[\frac{16}{45}+\frac{R^2}{3(1+R)}\Bigr].
\label{eq:disp}
\end{equation}
The intermediate algebra is given in Appendix~\re{app:sec3}. Writing
\be
\omega=\omega_0-i\Gamma
\ee
and expanding to first order,
\begin{empheq}{align}
\omega_0=\frac{k}{\sqrt{3(1+R)}}\,,
\qquad
\Gamma=\frac{k^2\,t_\gamma}{6(1+R)}
\Bigl[\frac{16}{15}+\frac{R^2}{1+R}\Bigr],
\qquad k=\frac{q}{a},\quad t_\gamma=\frac1{\omega_c},
\label{eq:Gamma}
\end{empheq}
which is precisely \W{6.4.25}---here obtained with no reference to fluid
equations, viscosity coefficients, or heat conduction: the $16/15$ descends from
the ladder \eqref{eq:Pi} and the slaved shear \eqref{eq:shear}, the $R^2/(1+R)$
from the slip \eqref{eq:slipval}. The standard two-timescale (WKB) analysis of
the same $\ell\le1$ system supplies the slowly varying amplitude,
$\DT{0}\propto(1+R)^{-1/4}$, exactly as in \W{6.4.21}; we do not repeat it. Thus
the fast part of the monopole through recombination is
\begin{gather}
\Delta^{(S)\,\mathrm{fast}}_{T,0}(q,t)
=-A(t)\cos\Theta(t),
\label{eq:fast}\\
A(t)=\frac{3\Ro}{5}\,(1+R)^{-1/4}\,e^{-\int_0^t\Gamma\,\dd t'}\,\SS(\kappa),
\qquad
\Theta(t)=\int_0^t\frac{q\,\dd t'}{a\sqrt{3(1+R)}}+\Delta(\kappa),
\nonumber
\end{gather}
with the normalization, and the functions $\SS(\kappa)$, $\Delta(\kappa)$ of the
rescaled wave number
\begin{equation}
\kappa\equiv\frac{\sqrt2\,q}{a_{\rm EQ}H_{\rm EQ}}
=\frac{q\,d_T}{a_L},
\label{eq:kappa}
\end{equation}
fixed by matching to the slow-mode analysis.

\vspace{.2in}
\ni {\bf The slow mode and the transfer functions}
\label{sec:slow}

The slow modes are the counterpart of the fast ones: they are the particular
solutions driven by the metric perturbations, they evolve on the Hubble
timescale rather than the acoustic one, and they are the gravitational growth
modes that eventually become large-scale structure. Their amplitude is
controlled by a transfer function $\TT(\kappa)$, which measures how much a mode
lost by entering the horizon while radiation pressure still resisted collapse;
modes that entered during matter domination are unsuppressed, $\TT\to1$. A
fuller account, with the slow-mode solutions for $\delta_{Dq}$ and $\psi_q$, is
given in Appendix~\re{app:sec3}.

At zeroth order in $\epsilon$ the hierarchy truncates: all photon multipoles
with $\ell \ge 2$, and all polarization multipoles $\Delta_{P,\ell}$, are of
order $\epsilon$ and therefore drop out. The remaining equations for
$\ell \le 1$, together with the baryon momentum equation \eqref{eq:momclean},
the dark-matter and neutrino equations, and the Einstein field equations of
Weinberg's Section 5.3, are \emph{identical} to the hydrodynamic system solved
in his \S\S 6.2--6.5. This is the precise sense in which the kinetic approach
contains Weinberg's fluid treatment: tight coupling is not an additional
assumption imposed on a fluid picture; it is simply the leading-order term in
the $1/\omega_c$ expansion of the full Boltzmann hierarchy
\eqref{eq:hierP}--\eqref{eq:hierT}. The damping rate $\Gamma$ from
\eqref{eq:Gamma} then appears as the first correction beyond that leading order.
The solution of the slow system, interpolating between the long- and
short-wavelength regimes, is parameterized by three functions of the rescaled
wave number: the amplitude $\SS(\kappa)$, the growth suppression $\TT(\kappa)$,
and the phase shift $\Delta(\kappa)$. Their explicit fits, their limits at small
and large $\kappa$, and the resulting expressions for $\psi_q(t_L)$,
$\delta_{\gamma q}(t_L)$ and $\delta u_{\gamma q}(t_L)$ are collected in
Appendix~\re{app:sec3}. In the two limits, $\TT,\SS\to1$ and
$\Delta\to2\kappa/\sqrt3$ for $\kappa\ll1$, while $\TT$ falls off
logarithmically, $\SS\to5$ and $\Delta\to0$ for $\kappa\gg1$.

\subsubsection{Large-$\ell$ reduction keeping all three form factors}
\label{sec:debye}

The exact multipole formula \eqref{eq:7210} is now reduced at large $\ell$,
using the Debye asymptotic expansion of the spherical Bessel functions, with all
three form factors retained --- including the polarization channel $\bar H$,
which Weinberg drops. The reduction itself assumes nothing about the plasma: the
tight-coupling solutions of the previous subsection enter only afterwards, when
the visibility-averaged form factors are evaluated. This completes the analytic,
approximation-based route.\footnote{Its numerical outputs (the closed
$\beta$-quadrature evaluated below) must not be conflated with the
full-hierarchy numbers of Sec.~\re{sec:kinetic}.}

\vspace{.1in}
\ni {\bf Thin-shell factorization}

Because the visibility function $P(t)$ is sharply peaked at the time of last
scattering $t_L$, and the optical depth factor $e^{-\tau(t)}$ is a smoothed step
function that rises from zero to unity around $t_L$, the time integral in
\rf{eq:7210} localizes to $t \simeq t_L$. Localization holds term by term for
everything that carries the visibility function $P(t)$. The one exception is the
integrated Sachs--Wolfe (ISW) piece of \eqref{eq:Fqt}, which is multiplied by
the bare $e^{-\tau}$ rather than by $P$, and which therefore collects
contributions from the whole period after last scattering. The analytic route of
this section separates it out and drops it, as \W{7.2.41} does; the numerical
pipeline of Section~\re{sec:kinetic} makes no such approximation and keeps it.
Consequently, we may write
\bea
&&\int_{t_1}^{t_0} dt \, \Big[ j_\ell(q r(t)) F(q,t) + j'_\ell(q r(t)) G(q,t) + j''_\ell(q r(t)) H(q,t) \Big] \nn\\
\;&\longrightarrow&\;
j_\ell(q r_L) \, \bar F(q) + j'_\ell(q r_L) \, \bar G(q) + j''_\ell(q r_L) \, \bar H(q),
\qquad r_L \equiv r(t_L),
\label{eq:thin}
\eea
where $\bar F(q)$, $\bar G(q)$, and $\bar H(q)$ are the visibility-weighted time
integrals of the sources. The finite width of the visibility function $P(t)$ is
retained \emph{inside} these form factors, through a Gaussian average over the
moment of last scattering, exactly as in Weinberg's procedure
(7.2.25)--(7.2.27). This averaging accounts for the fact that different photons
last scattered at slightly different times, leading to a suppression of the
rapidly oscillating fast modes.

The residual time dependence of the spherical Bessel phase across the
last-scattering shell is governed by
$\frac{\dd}{\dd t}\bigl[\nu(\tan b - b)\bigr] = -\,q\sin b/a$, with
$\nu \equiv \ell + 1/2$ and $\cos b \equiv \nu/(qr(t))$; the derivation is in
Appendix~\re{app:sec3}. This variation is suppressed relative to the source
phase $\omega_0 = q/[a\sqrt{3(1+R)}]$ by a factor $\sqrt{3(1+R_L)}\,\sin b$,
which vanishes at the dominant edge $\beta \to 1$ of the $\beta$-integral below.
The cross term it would produce is of the same order as the corrections already
dropped in the Landau-damping treatment, and it affects mainly the Doppler
contribution, itself suppressed by $\beta^2-1$ in the final integral.

\vspace{.1in}
\ni {\bf Debye asymptotics and the generalized \W{7.2.17}}

For $\ell\gg1$ the spherical Bessel functions are replaced by their uniform
asymptotic (Debye) forms, and each $\rho$-derivative acts at leading order on
the phase rather than on the amplitude, so that $j'_\ell$ and $j''_\ell$ carry
factors of $-\sin b$ and $-\sin^2 b$ relative to $j_\ell$. The first of these is
\W{7.2.15}; the second is its immediate extension, needed because $\bar H$ is
kept as well. Inserting these together with the thin-shell factorization
\eqref{eq:thin} into \eqref{eq:7210}, and averaging the rapidly oscillating
phase, leaves $\bigl(\bar F-\sin^2 b\,\bar H\bigr)^2+\sin^2 b\,\bar G^2$ inside
the integral. The temperature and polarization form factors therefore interfere
coherently --- both multiply $\cos\varphi$, and they combine with a relative
minus sign --- while the Doppler form factor multiplies $\sin\varphi$ and adds
in quadrature, reflecting the fact that the Doppler effect does not correlate
with the intrinsic temperature or polarization fluctuations on the same angular
scale. The steps are written out in Appendix~\re{app:sec3}.

Changing variables from $q$ to $\beta\equiv1/\cos b=qr_L/\nu$, so that
$q=\nu\beta/r_L$, $\dd q=\nu\,\dd\beta/r_L$, and
\begin{equation}
\cos^2 b=\frac{1}{\beta^2},\qquad
\sin b=\frac{\sqrt{\beta^2-1}}{\beta},\qquad
\frac{\cos^2 b}{\sin b}=\frac{1}{\beta\sqrt{\beta^2-1}} 
\end{equation}
eq. \eqref{eq:7210} becomes, with $\ell(\ell+1)\simeq\nu^2$ at this order,
\bea
\frac{\ell(\ell+1)\,C^S_{TT,\ell}}{2\pi}
=\frac{4\pi T_0^2\,\ell^3}{r_L^3}
\int_1^\infty\frac{\beta\,\dd\beta}{\sqrt{\beta^2-1}}\,
\Biggl\{
\Bigl[\bar F\Bigl(\tfrac{\ell\beta}{r_L}\Bigr)
-\frac{\beta^2-1}{\beta^2}\,\bar H\Bigl(\tfrac{\ell\beta}{r_L}\Bigr)\Bigr]^2
+\frac{\beta^2-1}{\beta^2}\,\bar G^2\Bigl(\tfrac{\ell\beta}{r_L}\Bigr)
\Biggr\}.
\label{eq:gen7217}
\eea
Setting $\bar H \to 0$ recovers Weinberg's equation (7.2.17). Equation
\eqref{eq:gen7217} is the first main result of this section: it gives the
large-$\ell$ temperature power spectrum as a one-dimensional integral over the
kinetic form factors, with the $j''_\ell$ (polarization) channel fully retained.

What has been used to reach it is only the exact line-of-sight solution and the
Debye asymptotics of the spherical Bessel functions. Nothing has been assumed
about the plasma dynamics --- not the equation of state, the nature of the dark
matter, the recombination history, or the evolution of the metric perturbations.
The transfer functions $\TT(\kappa)$ and $\SS(\kappa)$, the acoustic phase
$\Theta(t)$, the damping length $d_D$ and the polarization source $\Pi$ are
placeholders for whatever the underlying physics provides. The angular
projection from Fourier space to multipoles is therefore computed once and for
all, and all model dependence sits in $\bar F$, $\bar G$ and $\bar H$. We
evaluate those form factors below.

\subsubsection{Assembly and improved accuracy}

Equation \eqref{eq:7210} is exact, but it cannot be evaluated analytically,
since the form factors $F(q,t)$, $G(q,t)$ and $H(q,t)$ are known only
numerically as solutions of the full Boltzmann hierarchy. Weinberg's (7.2.41) is
what \eqref{eq:7210} becomes once the time integral is done with the
tight-coupling solutions, the visibility average, sudden decoupling and the
Debye asymptotics. We follow the same route, with one difference. At that last
stage Weinberg discards the polarization source: the $\Pi/16$ of \rf{eq:Fsource}
and the whole of $H$ in \rf{eq:Hqt} are both set to zero, which is why his
(7.2.17) carries no $j''_\ell$ channel and why $\Pi$ survives in his result only
indirectly, through the $16/15$ of the damping rate. We instead retain $\Pi$ in
both of the places where it acts as a source. Both survive the visibility
average, and they are what produce the $\eps_P$ terms of \rf{eq:final}.

\vspace*{.1in}
\ni {\bf Landau damping from the visibility width}
\label{sec:landau}

Following Weinberg's (7.2.25), we model the visibility function \(P(t)\) as a
Gaussian of width \(\sigma_t\) centered at $t_L$. Averaging the rapidly
oscillating factors of the fast-mode solution over the last-scattering shell,
with the acoustic phase expanded as
\(\Theta(t) \simeq \Theta_L + \omega_L (t - t_L)\) and
\(\omega_L \equiv \omega_0(t_L) = q/[a_L \sqrt{3(1+R_L)}]\), gives his
(7.2.26)--(7.2.27): both \(\langle \cos \Theta \rangle_P\) and
\(\langle \sin \Theta \rangle_P\) reduce to their values at $t_L$ multiplied by
\(e^{-\omega_L^2 \sigma_t^2 / 2}\). The factor
\(e^{-\omega_L^2 \sigma_t^2 / 2}\) is the Landau damping factor: the finite
width of the visibility function averages out the rapid oscillations of the fast
modes and suppresses them at high wavenumber.

Let us combine this with the Silk damping. Both exponents --- the accumulated
Silk damping \(\int_0^{t_L}\Gamma\,dt\) of the fast-mode solution and the Landau
factor \(\tfrac12\omega_L^2\sigma_t^2\) from the visibility width --- carry the
same wavenumber dependence: \eqref{eq:Gamma} gives
\(\Gamma = (q^2/a^2)\, t_\gamma [16/15 + R^2/(1+R)] / [6(1+R)]\), while the
definition of \(\omega_L\) above gives \(\omega_L^2 = q^2 / [3 a_L^2 (1+R_L)]\).
In the sum the factor \(q^2\) therefore comes out, leaving a coefficient that
depends only on the background history, and this is what permits a single
damping length \(d_D\) to be defined as in (7.2.28)--(7.2.31):
\begin{equation}
\int_0^{t_L} \Gamma \, dt + \frac{1}{2} \omega_L^2 \sigma_t^2
= \frac{q^2 d_D^2}{a_L^2},
\qquad
d_D^2 = d_{\rm Silk}^2 + d_{\rm Landau}^2,
\end{equation}
with
\begin{equation}
d_{\rm Silk}^2
= a_L^2 \int_0^{t_L} \frac{t_\gamma}{6a^2(1+R)}
\left[ \frac{16}{15} + \frac{R^2}{1+R} \right] dt,
\qquad
d_{\rm Landau}^2 = \frac{\sigma_t^2}{6(1+R_L)}.
\label{eq:dD}
\end{equation}
The \(16/15\) inside $d_{\rm Silk}$ is the same kinetic coefficient that appears
in the damping rate \eqref{eq:Gamma}, and it traces back through
\eqref{eq:shear} and \eqref{eq:Pi} to the polarization ladder rather than to any
assigned transport coefficient.

\vspace*{.1in}
\ni {\bf The visibility-averaged form factors}

We can now evaluate $\bar F$, $\bar G$, $\bar H$ of the thin-shell factorization
\rf{eq:thin} in the tight-coupling solution, treating the sudden-decoupling
limit with the Gaussian correction above. The three evaluations are carried out step by step in
Appendix~\re{app:sec3}; we quote the results and what is new in each.

For $\bar F$, the slow pieces of $\delta_{\gamma q}(t_L)$ and $\psi_q(t_L)$
combine into $3\Ro\,\TT(\kappa)R_L/5$ and the fast piece picks up the Landau
damping factor, exactly as in Weinberg. The new ingredient is the explicit
polarization source $P(t)\,\Pi/16$ of \rf{eq:Fsource}, whose $1/\omega_c$
cancels against the $\omega_c$ inside the visibility function
$P=\om_c e^{-\tau}$ and which is therefore finite and localized at last
scattering. It enters the bracket as a $\sin\Theta_L$ term:
\begin{equation}
\bar F(q)=\frac{\Ro}{5}\Bigl\{
3\,\TT(\kappa)\,R_L
-(1+R_L)^{-1/4}\,\SS(\kappa)\,e^{-q^2d_D^2/a_L^2}
\Bigl[\cos\Theta_L+\frac{\omega_L\bar t_\gamma}{3}\,\sin\Theta_L\Bigr]
\Bigr\},
\label{eq:Fbar}
\end{equation}
with $\Theta_L$ as in \rf{eq:fast} and $\bar t_\gamma\equiv1/\om_c(t_L)$.

For $\bar H$, the same polarization source is averaged with the coefficient
$3/16$ of \rf{eq:Hqt} instead of $1/16$:
\begin{equation}
\bar H(q)=\frac{3}{16}\bigl\langle\Pi\bigr\rangle_P
=-\frac{\Ro}{5}\cdot \omega_L\bar t_\gamma\,
(1+R_L)^{-1/4}\,\SS(\kappa)\,e^{-q^2d_D^2/a_L^2}\,\sin\Theta_L .
\label{eq:Hbar}
\end{equation}
The ratio $\bar H/\bar F_{\rm pol}=3$ is fixed entirely by those two source
coefficients, since both average the same underlying $\Pi$. It is in that sense
a prediction of the hierarchy rather than of the plasma dynamics.

For $\bar G$, the slow parts of $\delta u_{\gamma q}$ and $\psi_q$ cancel
identically --- Weinberg's observation below (7.2.24) --- because both carry the
same $\TT(\kappa)$ with matched coefficients. What survives is a pure Doppler
oscillation,
\begin{equation}
\bar G(q)=-\frac{\sqrt3\,\Ro}{5\,(1+R_L)^{3/4}}\,
e^{-q^2d_D^2/a_L^2}\,\SS(\kappa)\,\sin\Theta_L ,
\label{eq:Gbar}
\end{equation}
which is his (7.2.37) unchanged.

\vspace{.2in}

Assembling \rf{eq:Fbar}--\rf{eq:Gbar} into \rf{eq:gen7217} gives the closed form
\begin{align}
\fr{\ell(\ell+1)\,C^S_{TT,\ell}}{2\pi}
&=\fr{4\pi T_0^2N^2e^{-2\tau_{\rm reio}}}{25}
\int_1^\infty \dd\beta\,
\Bigl(\fr{\beta\ell}{\lR}\Bigr)^{n_S-1}
\nonumber\\[1mm]
&\times\Biggl\{
\fr{1}{\beta^2\sqrt{\beta^2-1}}
\biggl[3\,\TT\!\Bigl(\ft{\beta\ell}{\ell_T}\Bigr)R_L
-(1+R_L)^{-1/4}\,\SS\!\Bigl(\ft{\beta\ell}{\ell_T}\Bigr)\,
e^{-\beta^2\ell^2/\ell_D^2}
\nonumber\\[-0.5mm]
&\hskip9em\times
\Bigl(\cos\Theta_{\ell\beta}
+\tfrac43\eps_P\,\Bigl[1-\ft{3(\beta^2-1)}{\beta^2}\Bigr]\sin\Theta_{\ell\beta}
\Bigr)\biggr]^2
\nonumber\\[1mm]
&\hskip2.2em
+\fr{3\sqrt{\beta^2-1}}{\beta^4\,(1+R_L)^{3/2}}\,
e^{-2\beta^2\ell^2/\ell_D^2}\,
\SS^2\!\Bigl(\ft{\beta\ell}{\ell_T}\Bigr)\,
\sin^2\Theta_{\ell\beta}
\Biggr\},
\la{eq:final}
\\[1mm]
\Theta_{\ell\beta}&\eqv\fr{\beta\ell}{\ell_H}
+\Delta\!\Bigl(\fr{\beta\ell}{\ell_T}\Bigr),
\qquad
\eps_P=\fr{\beta\,\ell\,\bar t_\gamma}{4\,d_A\sqrt{3(1+R_L)}} .
\nonumber
\end{align}
This is the closed-form endpoint of the analytic route: the Boltzmann hierarchy
enters through the transfer functions $\TT(\kappa)$ and $\SS(\kappa)$ and
through the damping rate $\Gamma$ (which determines $\ell_D$), but the time
integral of \eqref{eq:7210} has been done once and for all under the
sudden-decoupling approximation. (Sec.~\re{sec:kinetic} takes the other route, a
more direct numerical evaluation of \rf{eq:7210}.) Every ingredient is
physically identifiable. The first bracket carries the monopole contribution and
its acoustic oscillation, damped by $e^{-\beta^2\ell^2/\ell_D^2}$. The second
term is the Doppler contribution, damped twice as fast in the exponent and
phase-shifted by a quarter period relative to the monopole term. The transfer
functions $\TT$ and $\SS$ encode the pre-recombination evolution of the slow and
fast modes, and $\Theta_{\ell\beta}$ is the acoustic phase accumulated by last
scattering.

Setting $\eps_P \to 0$, \eqref{eq:final} reduces verbatim to Weinberg's
(7.2.41). The $\eps_P$ terms are the polarization correction, retained here
exactly rather than estimated. The factor $\bigl[1 - 3(\beta^2-1)/\beta^2\bigr]$
multiplying $\eps_P$ arises from the coherent interference between the
polarization source in $\bar F$ (coefficient $1/16$) and the $j''_\ell$ term
$\bar H$ (coefficient $3/16$), noted below \eqref{eq:avg}. This is the main
improvement over Weinberg's treatment.

\subsection{Evaluation of the closed formula}
\label{sec:quadeval}

Evaluating \eqref{eq:final} with the substitution and sampling prescribed above,
at Weinberg's parameters --- $\ell_T=97.60$, $\ell_H=96.15$, $\ell_D=1598$,
$\ell_R=708$, $R_L=0.6234$ and $n_S=0.95820$, the values given in Ch. 7 of
\cite{Weinberg2008} --- places the first five acoustic peaks at $\ell=242$,
$564$, $842$, $1154$ and $1444$, against the values $220$, $541$, $821$, $1134$,
$1425$ of Weinberg's Table~7.1. What is being tested here is only the internal
consistency of \eqref{eq:final}. The scales above are taken from Weinberg rather
than computed independently, so the question being asked is whether our improved
closed formula returns his peak positions when it is fed his own inputs. The
offset is $+19$ to $+23$ multipoles --- nearly constant in absolute terms rather
than proportional to $\ell$, as an error in the acoustic scale $\ell_H$ would
be. A peak sits where $\Theta_{\ell\beta}$ passes a multiple of $\pi$. The
$\beta$ integral in \eqref{eq:final} is concentrated at its lower limit, since
the monopole term carries the integrable singularity $(\beta^2-1)^{-1/2}$ there
while the damping factor $e^{-\beta^2\ell^2/\ell_D^2}$ falls off as $\beta$
grows, so the peak positions are set by the edge $\beta\to1$. Differentiating
the definition of $\Theta_{\ell\beta}$ given with \eqref{eq:final} and putting
$\beta=1$,
\begin{equation}
\frac{\partial\Theta_{\ell\beta}}{\partial\ell}\bigg|_{\beta=1}
=\frac{1}{\ell_H}+\frac{\Delta'(\kappa)}{\ell_T},
\qquad \kappa=\frac{\ell}{\ell_T} ,
\end{equation}
in which the second term is negligible compared with the first.\footnote{The
second term is a few percent of the first across the range the peaks occupy:
$\ell_T=97.60$ and $\ell_H=96.15$ are nearly equal, while $\Delta$ has left its
small-$\kappa$ linear regime by $\kappa\simeq2$ and decays from $0.33$ to $0.07$
over $2\lesssim\kappa\lesssim15$, so that $|\Delta'|\,\ell_H/\ell_T$ falls from
$4\%$ to $0.5\%$ across that range.} An offset of this size therefore
corresponds to an additive phase deficit of $\delta\ell/\ell_H\simeq0.22$ rad,
about three percent of a cycle. We checked whether the offset was a numerical
artifact by doubling the sampling density --- from twelve to twenty-four points
per half period of $\Theta_{\ell\beta}$. The peak positions did not move.
Therefore the offset is built into the formula, not introduced by the numerical
integration.

We have not traced the offset to a single cause, so we list the candidates
rather than choose among them. The most likely is the exponent in the fit for
the phase transfer function \(\Delta(\kappa)\). In the original printing of
\cite{Weinberg2008} this fit is raised to the power \(1/4\); the erratum drops
the exponent altogether, and that form is what we use in this section and in
\eqref{eq:final}. Both versions fail the analytic small-\(\kappa\) limit
\W{6.5.10}. Taking the square root of the same fit passes that limit, and it
also brings all five peaks to within one multipole of Weinberg's Table~7.1. So
it is possible that the erratum swapped one wrong exponent for another, and that
the intended power is \(1/2\), not \(1\). We do not adopt the \(1/2\) exponent
here. Our check of the fit against Weinberg's Table~6.1 used the printed form,
and that check would need to be redone before any exponent is changed.
Appendix~\ref{app:sec3} lays out all three possibilities. This candidate is
plausible as the origin of the offset because \(\Delta\) enters
\(\Theta_{\ell\beta}\) additively: if one shifts it, all five peaks will move by
nearly the same amount, which is exactly what we see. A second possibility is
that the offset simply measures the accumulated cost of the approximation chain
--- tight coupling, WKB phase, sudden decoupling, Debye asymptotics --- which is
what \eqref{eq:final} exists to exhibit. A third is that the reference values
themselves derive from a full numerical computation rather than from \W{7.2.41},
in which case the comparison is not like-for-like. One candidate can be ruled
out. Switching on the polarization term $\eps_P$ of \eqref{eq:final}---set by
the mean free time $\bar t_\gamma$ at last scattering, whose natural scale is
$\bar t_\gamma\sim d_{\rm Landau}$ so that $\eps_P\sim10^{-2}$ at the first
peak---moves the peaks by between one and eight multipoles, and in the direction
that \emph{increases} the offset rather than reducing it.

None of this affects the results of Sec.~\re{sec:kinetic}. There the exact
formula \eqref{eq:7210} is evaluated directly: the hierarchy is integrated
without a tight-coupling expansion, and the transfer functions $\TT$, $\SS$ and
$\Delta$ --- and with them this entire class of uncertainty --- do not appear at
all. The comparison with Planck in \S\ref{sec:planck} is performed there. 
What this subsection establishes is the internal consistency of the closed
form, and with it the identifiability of the coefficients that
Sec.~\re{sec:kinetic} goes on to test numerically.

\section{Bayesian analysis}
\la{sec:bayesian}

In Papers~I and II, our goal was to find the best-fit values. In this section we
ask a different question: given the Planck data, what values of
$\Omega_{\Lambda_2}$ and $\Omega_{\Lambda_3}$ are still allowed once we account
for our ignorance of all other parameters? These two questions have different
answers, and the gap between them tells us something useful.

\subsection{Setup}
\la{sec:bayes_setup}

We added the two new parameters of Eq.~\eqref{eq:friedmann} to
\textsc{class}~\cite{Blas:2011rf}; the modified code is publicly available
\cite{mod_CLASS}. This required changing only the background
expansion: $\Omega_{\Lambda_2}$ and $\Omega_{\Lambda_3}$ enter the function
$E(a)$ and nothing else, for the reasons explained in Sec.~\ref{sec:redundancy}.
We then used \textsc{cobaya}~\cite{Torrado:2020dgo} to sample the modified code
against the Planck 2018 likelihood suite~\cite{Aghanim:2019ame}: the high-$\ell$
\texttt{plik\_lite} TTTEEE spectrum, the low-$\ell$ TT and EE likelihoods, and
the lensing reconstruction. We did not combine these with any external data. The
analysis is Planck-only by design, because the question is what the CMB alone
can tell us. On the theory side, we ran \textsc{class} with $\ell_{\max}=2500$,
one massive neutrino species, $N_{\rm ur}=2.0328$ massless species, and the
\texttt{hmcode} non-linear prescription.

The sampled parameters are the six standard $\Lambda$CDM parameters, plus
spatial curvature $\Omega_K$, plus the two new parameters, plus the Planck
calibration $A_{\rm planck}$ --- ten in total. The last of these is an
instrumental nuisance parameter: it is the overall calibration of the Planck
maps, not to be confused with the primordial amplitude $A_s$. It is the only
parameter with a Gaussian prior rather than a flat one, $1.000\pm0.0025$,
inherited from the likelihood.\footnote{It appears as $y_{\rm cal}$ in the
figures, following the \textsc{getdist} convention.} The remaining priors were
flat and deliberately wide: $h\in[0.55,0.85]$, $\Omega_K\in[-0.05,0.05]$, and
$\Omega_{\Lambda_3}\in[-0.1,0.1]$, with proposal width $5\times10^{-4}$.

One numerical issue required care. The parameter $\Omega_{\Lambda_2}$ is
expected to be of order $10^{-7}$,\footnote{When a heavy field 
is integrated out, the resulting effective interactions generate a leading off-shell 
contribution that scales as $T^6/m^2$ \cite{Matsumoto:1999kd}. This would modify the 
Friedmann equation by introducing an additional term proportional to 
$\Omega_{\Lambda_4}\, a^{-6}$. Although such higher-order corrections would also arise 
via higher-order loops, the magnitude of $\Omega_{\Lambda_4}$ must be sufficiently 
suppressed to remain compatible with observational data, much like the stringent bounds 
required for $\Omega_{\Lambda_2}$.
 
} which is six orders of magnitude or more
below every other sampled parameter. Feeding a number this small straight into
the sampler makes the proposal covariance ill-conditioned; early runs confirmed
this. We therefore did not sample $\Omega_{\Lambda_2}$ directly. Instead, we
sampled a rescaled variable $\tilde\Omega_{\Lambda_2}\in[-10,10]$, which is of
order unity, with unit proposal width, and supplied
$\Omega_{\Lambda_2}=\tilde\Omega_{\Lambda_2}\times10^{-7}$ to the theory code,
while quoting the physical value $\Omega_{\Lambda_2}$ in the results.
$\Omega_{\Lambda_3}$, being of order $10^{-2}$, required no rescaling and was
sampled directly.

We report three chains below. It will be convenient to name them. \emph{Run~A}
is the production chain. It uses the configuration just described, was sampled
with eight parallel chains on the Purdue Anvil cluster, and is the source of
every number in Table~\ref{tab:mcmc}. \emph{Run~B} is an independent chain over
the same extended model. It was run on local hardware as a single chain, and
differs from run~A in a single respect: $\Omega_{\Lambda_3}$ was given the wider
prior $[-0.5,0.5]$. Its purpose is to provide a check that does not rely on
run~A's internal diagnostics, and to answer the prior-sensitivity question of
Sec.~\ref{sec:bayes_priors}. The \emph{baseline run} is the eight-parameter
chain obtained by deleting $\Omega_{\Lambda_2}$ and $\Omega_{\Lambda_3}$ and
changing nothing else. We include it because interpreting a null result requires
knowing what the same data and the same code return when the new parameters are
simply absent. Its likelihoods, \textsc{class} settings, shared-parameter
priors, and Gelman--Rubin stopping criterion $R-1<0.05$ are identical to
run~A's. The two differ only in sampler-efficiency settings, which affect the
speed at which a chain explores the posterior, not the posterior it explores.

Before these runs, we carried out a longer exploratory stage, also on local
hardware. In that stage we varied the priors and the sampling scheme for the two
new parameters --- in particular, whether they were sampled directly or through
a rescaled variable. The goal was twofold: to diagnose the conditioning problem
described above, and to find prior ranges wide enough that they would not
truncate the posterior. That was achieved for $\Omega_{\Lambda_2}$, but not for
$\Omega_{\Lambda_3}$, whose posterior is cut off by a prior edge in both
production runs for reasons that turn out to be structural rather than a matter
of choosing a wider range; Sec.~\ref{sec:bayes_priors} takes this up. Once those
exploratory runs had served their purpose, we stopped them, and no quantitative
result below depends on them.

\subsection{Convergence}

Convergence warrants a dedicated discussion for two key reasons: the extensive
process required to achieve it, and the fact that insights gained along the way
directly influence how the final results should be interpreted. Run~A used eight
MPI chains on a full 128-core node. It was resumed across several multi-day jobs
over three weeks of wall-clock time, and stopped when \textsc{cobaya}'s
Gelman--Rubin statistic reached $R-1=0.049$, below the requested threshold of
$0.05$. Reaching that point was not straightforward. Four-chain runs approached
the threshold repeatedly but failed to hold it. They reached $R-1=0.077$ and
$0.069$ at two separate stages, then rebounded to $0.16$ and $0.12$ as sampling
continued. The cause was that their chains kept wandering between two apparently
separate regions of $\Omega_{\Lambda_3}$. Only with eight chains did the
statistic reach the threshold at all, and even then not smoothly: it stood at
$0.24$ after the first eight-chain job, fell to $0.064$ after the second, and
crossed at $0.049$ in the third, after some $309{,}000$ accepted steps in total.

Two features of the converged chain deserve a comment. First, the
acceptance rate settled at $82\%$, far above the $20$--$30\%$ that a well-tuned
Metropolis proposal targets. A high acceptance rate means the chain takes small
steps, so successive samples are strongly correlated. Of roughly $309{,}000$
accepted samples, the effective number is only about $1{,}400$. This is ample
for the means and the $68\%$ limits quoted below, but the chain does not have
enough samples to constrain the extremes. That is why we do not quote $99\%$
intervals. Second, the convergence statistic for the confidence limits,
$(R-1)_{\rm cl}$, was evaluable only at the final check, where it returned
$0.18$ against a threshold of $0.2$. This statistic is the binding criterion for
a quoted limit, and it is the least well determined diagnostic of the run.

That history is worth recording also because of what the four-chain runs
appeared to show. Seen from inside such a run, the $\Omega_{\Lambda_3}$
posterior looks bimodal: two peaks, one on either side of zero, separated by a
dip. This picture seems to imply that two physically distinct solutions fit the
data equally well. It is an appealing picture, but it is an incorrect one. Two
things were producing it. First, the chains had not yet communicated their way
across the parameter space, so the gap between the populated regions was a gap
in the sampling rather than in the likelihood. Second, the prior in use at that
stage, $\Omega_{\Lambda_3}\in[-0.05,0.05]$, was narrow enough that the posterior
filled it to both edges. A broad distribution truncated on both sides is easily
mistaken for a peaked one on each. Widening the prior and mixing the chains each
removed part of the effect; together they removed all of it. Neither run~A nor
run~B shows a second peak. (We record this as a caution: along a weakly
constrained direction, chains that have not yet mixed can show peaks that do not
exist, especially when a prior edge truncates the posterior.)

Run~B was held to a looser standard than run~A, so we use it only for coarse
comparisons. We list its diagnostics here alongside run~A's. It is a single
chain of roughly $62{,}000$ accepted samples, of which the first thirty percent
are discarded as burn-in, with an acceptance rate of $51\%$. It stopped when
$R-1$ reached $0.091$ against a requested threshold of $0.1$, with
$(R-1)_{\rm cl}=0.19$ against the same threshold of $0.2$ used for run~A. Its
effective sample count is about $140$, close to an order of magnitude coarser
than run~A. We therefore use it for agreement at the level of posterior means
and widths, and for the qualitative question of Sec.~\ref{sec:bayes_priors}, not
for the third digit of any interval.

Convergence statistics are internal diagnostics, however, so an independent
check is worth having. Run~B serves that purpose: for $\omega_b$,
$\omega_{\rm cdm}$, $A_s$, $n_s$, $\tau_{\rm reio}$ and $\Omega_K$, it agrees
with run~A well within the quoted uncertainties. Its $\Omega_{\Lambda_2}$
posterior has the same width ($4.67$ against $4.60$, in units of $10^{-7}$), and
it shows the same non-detection of both new parameters. The two runs differ, as
discussed in Sec.~\ref{sec:bayes_priors}, in where the upper limit on
$\Omega_{\Lambda_3}$ falls --- and that difference turns out to be informative.

\subsection{Results}
\la{sec:bayes_results}

\begin{table}[h]
\centering
\begin{tabular}{lcc}
\toprule
Parameter & Baseline & Extended model (run~A)\\
\midrule
$\omega_b$          & $0.022457\pm0.000165$ & $0.022460\pm0.000157$\\
$\omega_{\rm cdm}$  & $0.11864\pm0.00142$   & $0.11869\pm0.00148$\\
$h$                 & $0.6527\pm0.0212$     & $0.6436^{+0.035}_{-0.065}$\\
$10^{9}A_s$         & $2.0723\pm0.0351$     & $2.0730\pm0.0349$\\
$n_s$               & $0.96824\pm0.00470$   & $0.96813\pm0.00548$\\
$\tau_{\rm reio}$   & $0.04987\pm0.00807$   & $0.04991\pm0.00804$\\
$\Omega_K$          & $-0.00727\pm0.00567$  & $-0.00792\pm0.00652$\\
$A_{\rm planck}$    & $1.00000\pm0.00247$   & $1.00000\pm0.00247$\\
\midrule
$10^{7}\Omega_{\L_2}$ & ---                 & $-0.20\pm4.60$\\
$\Omega_{\L_3}$       & ---                 & $>-0.0092$ \ (68\%)\\
\midrule
$\chi^2_{\rm CMB}$    & $1016.3\pm3.6$      & $1016.8\pm3.7$\\
\bottomrule
\end{tabular}
\caption{Marginalized posterior means and $68\%$ intervals. The baseline column
is the eight-parameter run with $\Omega_{\L_2}$ and $\Omega_{\L_3}$ absent; the
extended column adds them. Likelihoods, \textsc{class} settings and the priors
on all shared parameters are identical between the two. The $\Omega_{\L_3}$
entry is a one-sided limit, for the reason given in Sec.~\re{sec:bayes_priors}.}
\la{tab:mcmc}
\end{table}

Table~\ref{tab:mcmc} collects the marginalized constraints from run~A, with the
baseline run alongside for comparison. The first thing the table shows is that
the two new parameters neither degrade the fit nor improve it. Every standard
parameter is recovered within a small fraction of its own uncertainty, and the
fit quality is unchanged: $\chi^2_{\rm CMB}=1016.3\pm3.6$ without them against
$1016.8\pm3.7$ with them. The agreement also holds separately for each
likelihood: $586.5$ against $586.6$ for high-$\ell$ TTTEEE, $396.7$ against
$396.7$ for low-$\ell$ EE, $22.5$ against $22.6$ for low-$\ell$ TT, and $10.6$
against $10.8$ for lensing. This means the two extra degrees of freedom have not
redistributed the fit between datasets either. In other words, the two new
parameters are not being used to trade off one dataset against another --- each
likelihood is fit just as well as it was without them.\footnote{These are
posterior-mean values rather than best-fit minima, so the difference between
them is not a $\Delta\chi^2$ in the formal likelihood-ratio sense. The usual
likelihood-ratio machinery compares the single best-fit point of each model,
whereas here we are comparing the centers of the two posterior distributions.
However, the two distributions are $1016.3 \pm 3.6$ and $1016.8 \pm 3.7$: their
centers differ by $0.5$, while each distribution is roughly $3.6$ wide. They
overlap so thoroughly that no sensible choice of summary statistic would
separate them.} Run~B gives $1017.1 \pm 3.9$, consistent with both, confirming
that the result does not depend on the specific run either.

\begin{figure}[H]
  \centering
  \includegraphics[width=0.8\columnwidth]{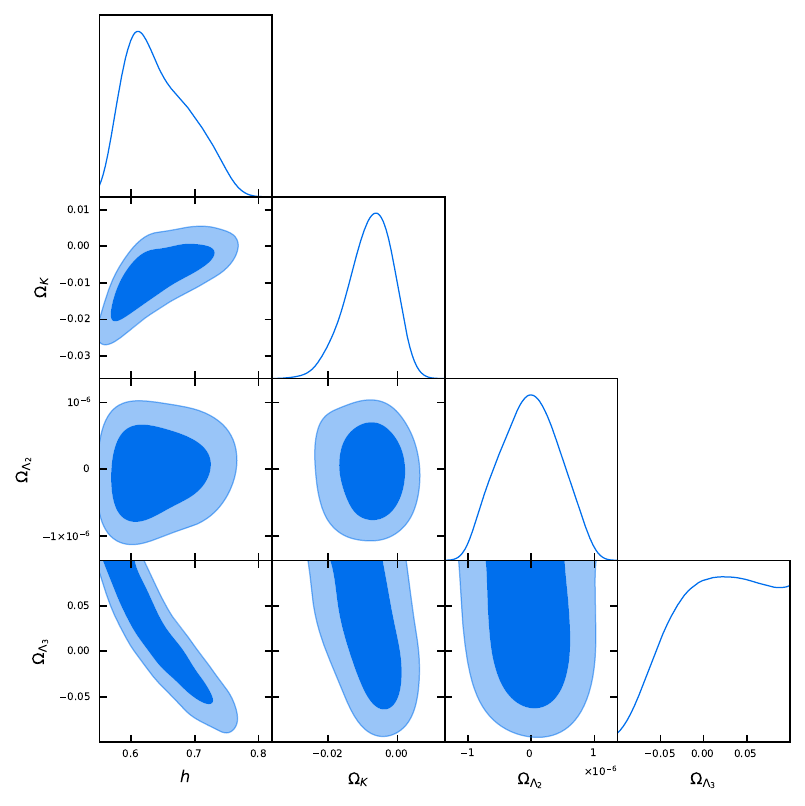}
\caption{Marginalized one- and two-dimensional posteriors from run~A, with
$68\%$ and $95\%$ contours, for the four parameters that carry the discussion;
the full ten-dimensional posterior is Fig.~\re{fig:triangle_full}. The
$h$--$\Omega_{\L_3}$ panel shows a narrow diagonal band: the two parameters are
tightly anticorrelated that $h$ can slide along the band almost without any cost
to the fit. In other words, there is a nearly flat direction in parameter space
--- a combination of $h$ and $\Omega_{\L_3}$ that the data barely constrains ---
so the fit can trade one against the other freely. This has two visible
consequences: the allowed range for $h$ alone becomes much wider, with
$\sigma(h)$ more than doubling relative to the baseline run, and the $h$
posterior becomes skewed toward low values. The $\Omega_{\L_2}$ posterior is
peaked and symmetric about zero, falling away on both sides within its prior
range of $\pm10^{-6}$. The $\Omega_{\L_3}$ posterior rises out of the negative
region and then flattens into the upper edge of its prior at $+0.1$, which is
the origin of the one-sided limit in Table~\re{tab:mcmc} and the subject of
Sec.~\re{sec:bayes_priors}.}
  \la{fig:triangle}
\end{figure}

The one standard parameter that shifts is $h$. Its uncertainty more than
doubles, from $\pm0.021$ to $^{+0.035}_{-0.065}$, and it acquires a marked
asymmetry toward low values. The data themselves are unchanged; what changed is
the geometry of parameter space. $\Omega_{\Lambda_3}$ opened a direction along
which $h$ can slide almost without penalty. This anticorrelation is the narrow
diagonal band visible in Fig.~\ref{fig:triangle}. It is also the degeneracy that
Sec.~\ref{sec:fisher} derives from the differential response of the spectrum, by
differentiating rather than by sampling. The two calculations converge here, our
main internal cross-check.

Spatial curvature comes out slightly negative, $\Omega_K=-0.0079\pm0.0065$.
Taken at face value, this is a pull of about $1.2\sigma$ toward a closed
universe, but the $95\%$ interval, $[-0.0211,+0.0039]$, contains zero, so the
preference is not statistically significant. Planck-only analyses are known to
show this mild preference~\cite{Planck:2018vyg}, and external data remove
it~\cite{Alam:2021inq}. We therefore draw no conclusion from it. We include
$\Omega_K$ only so that it can absorb any curvature-like signal that would
otherwise be misattributed to $\Omega_{\Lambda_3}$; it is not included in order
to measure curvature itself.

The two new parameters give results of different kinds. $\Omega_{\Lambda_2}$ is
a clean two-sided null: $(-0.20\pm4.60)\times10^{-7}$, symmetric about zero,
with a peak that falls away on both sides. At the best-fit $h$,
$\Omega_R\simeq9\times10^{-5}$ when we count photons and the massless neutrino
background. So the data allow a thermal correction to the vacuum energy of at
most about half a percent of the radiation density, and no more. This is an
empirical finding, not a prior expectation imposed on the fit, as
Sec.~\ref{sec:bayes_priors} shows. Why is $\Omega_{\Lambda_2}$ bounded? One
might think the bound comes from the acoustic peaks. It does not. It comes from
the overall amplitude of the spectrum, which is already fixed by other means.
The evidence is in the full posterior of Fig.~\ref{fig:triangle_full}. There,
$A_{\rm planck}$ shows no correlation with any other parameter, including
$\Omega_{\Lambda_2}$. This matters because $A_{\rm planck}$ rescales the whole
predicted spectrum. If $\Omega_{\Lambda_2}$ acted only as an overall amplitude,
the two would be correlated. They are not, because the calibration prior of
$\pm0.0025$ is far too tight to let $A_{\rm planck}$ move. The amplitude is
therefore pinned externally --- by $A_s$, by $\tau_{\rm reio}$ through the
low-$\ell$ polarization, and by that same calibration prior. Nothing in the
acoustic peaks is responsible for the bound. Sec.~\ref{sec:responses} reaches
the same conclusion by a different route, differentiating the spectrum rather
than sampling the posterior.

$\Omega_{\Lambda_3}$ is a different case. It is bounded from below,
$\Omega_{\Lambda_3}>-0.0092$ at $68\%$, but not from above. The posterior rises
out of the negative region, crosses zero, and does not come back down before the
prior range ends. This does not mean the data are indifferent to large positive
$\Omega_{\Lambda_3}$. It means only that the prior range ends before the
posterior starts to fall off. The next subsection takes this up directly. What
can be said for certain is that zero lies comfortably inside the allowed region:
neither parameter is detected.

\begin{figure}[H]
  \centering
  \includegraphics[width=\columnwidth]{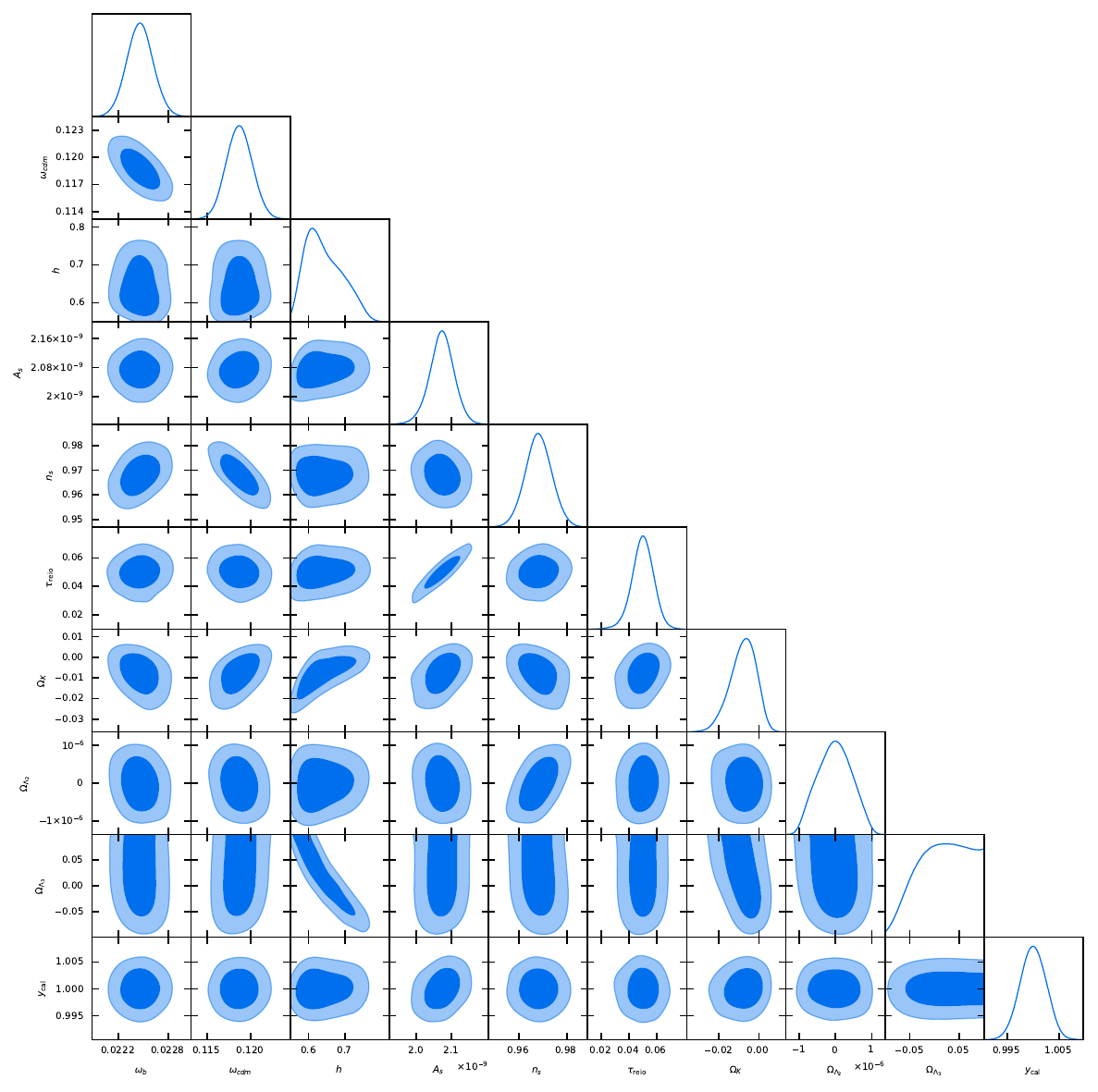}
\caption{Marginalized one- and two-dimensional posteriors for all ten sampled
parameters of the extended model, from the converged eight-chain run, with
$68\%$ and $95\%$ contours. The six standard $\L$CDM parameters retain their
usual shapes and correlations --- in particular the $A_s$--$\tau_{\rm reio}$
degeneracy characteristic of temperature-dominated data --- so the two
additional parameters distort no part of the baseline picture.}
  \la{fig:triangle_full}
\end{figure}

For completeness we close with the full ten-dimensional posterior of the same
run, Fig.~\re{fig:triangle_full}, of which Fig.~\re{fig:triangle} is the
four-parameter corner used above. One panel deserves separate comment. In the
\(\Omega_K\)--\(h\) panel the two parameters are positively correlated, whereas
the \(h\)--\(\Omega_{\L_3}\) panel of Fig.~\ref{fig:triangle} shows a tight
negative correlation, and \(\Omega_K\) is bounded several times more tightly
than \(\Omega_{\L_3}\), whose posterior runs to its prior edge. Neither fact can
originate in the expansion history, where by \eqref{eq:friedmann} the two
parameters appear only through the sum \(\Omega_K+\Omega_{\L_3}\) and are
interchangeable. What distinguishes them lies outside \(E(a)\), and it is what
Sec.~\ref{sec:responses} sets out: spatial curvature bends the
comoving-to-radial map, and it modifies the perturbation equations, while a
smooth \(a^{-2}\) component in a flat universe does neither. Of the two it is
the bending that the data grip here. It is of order \(1.7\,\Omega_K\) at last
scattering, larger than the \(E(a)\) effect and opposite to it in sign,
whereas the curvature terms in the perturbation equations scale as $K/q^2$ and
are negligible on every scale where Planck carries information.\footnote{In a
curved background the plane waves are replaced by eigenfunctions of the
curved-space Laplacian, and curvature enters the perturbation equations only in
combinations such as $q^2-3K$ and $q^2-(\ell^2-1)K$ \ci{Hu:1997mn}. With
$K=-\Omega_K H_0^2$ and $q\simeq\ell/r_L$, the relative correction is
$K/q^2\simeq\Omega_K(H_0r_L)^2/\ell^2\approx10\,\Omega_K/\ell^2$, about
$10^{-4}$ at $\ell=30$ for the $\Omega_K$ of Table~\ref{tab:mcmc}.} This
effect is what makes $\Omega_K$ behave differently from $\Omega_{\L_3}$ in the
data, and it reverses the sense of the correlation with $h$. It does not by
itself close the posterior, since a shift of $d_A$ is exactly what $h$ can
compensate; the tighter bound on $\Omega_K$ needs the further effects described
in Sec.~\ref{sec:responses}: curvature also acts on CMB lensing and the
late-time ISW, and its steeper degeneracy with $h$ runs into the limits on $h$
sooner. This makes concrete the
argument of Sec.~\ref{sec:bayes_setup} for keeping $\Omega_K$ free. Had it been
fixed to zero, we would have silently absorbed part of the $\Omega_{\L_3}$
constraint into an assumption, and then reported the rest as if it were a
measurement. Keeping $\Omega_K$ free lets the data decide how the shared
$a^{-2}$ term is split between the two parameters.

\subsection{Prior dependence}
\la{sec:bayes_priors}

The reported constraints on $\Omega_{\Lambda_2}$ and $\Omega_{\Lambda_3}$ above
need a check. The reason is that a flat prior is not the same as no prior. It is
itself a definite statement of belief --- that the parameter lies somewhere
within the chosen interval and nowhere else --- and when the data constrain that
parameter only weakly, the posterior simply reproduce the prior's range. In that
case the width we report measures our choice of interval, not the information in
the data, and one would be mistaken to read it as a measurement. A flat prior is
therefore trustworthy only when the data are strong enough to override it. Both
of our new parameters fall into this regime. $\Omega_{\Lambda_2}$ is a parameter
of order $10^{-7}$ whose posterior may be broad enough that the prior range,
rather than the data, could set its width. $\Omega_{\Lambda_3}$, meanwhile, is
quoted as a one-sided limit, which means its reported bound is shaped by where
the prior was cut off rather than by a genuine two-sided constraint from the
data. Both numbers therefore require a check that an ordinary, well-constrained
parameter estimate would not.

\textbf{$\Omega_{\Lambda_2}$: the data set the width.} The shape of its own
posterior answers the question. If the data were too weak to constrain
$\Omega_{\Lambda_2}$, then the posterior would simply mirror the prior and look
uniform on $\tilde\Omega_{\Lambda_2}\in[-10,10]$. Such a uniform distribution
has standard deviation $10/\sqrt3=5.77$ and places its $68\%$ and $95\%$ limits
at $1.18\sigma$ and $1.65\sigma$. If instead the likelihood were doing the work
--- that is, if the data were strong enough to dominate the prior --- the
posterior would be shaped by the likelihood rather than by the prior's
boundaries. For a Gaussian likelihood, the posterior would then be a Gaussian,
and a Gaussian has a characteristic signature: by definition, $68\%$ of its
probability lies within $1.00\sigma$ of the mean and $95\%$ within $1.96\sigma$.
These what we should see if the data are setting the width.

Run~A returns $\sigma=4.60$ --- twenty percent narrower than the uniform case
--- with $68\%$ limits at $1.08\sigma$ and $1.05\sigma$, and $95\%$ limits at
$1.90\sigma$ and $1.91\sigma$. Every one of these numbers sits at the Gaussian
end of the range. The conclusion is therefore that the likelihood, not the
prior, is setting the width. Run~B, five times shorter and on different
hardware, reproduces the same width at $68\%$, confirming that the result is not
an artifact of the run. The prior only matters in the far tails, where the
$99\%$ limits come within three percent of the edge --- one reason we do not
quote them. The constraint $|\Omega_{\Lambda_2}|\lesssim5\times10^{-7}$ is
therefore a statement about the data, with one qualification: $\Omega_R$ is held
fixed at its FIRAS value \ci{Fixsen:2009ug} throughout, so the bound is
conditional on the radiation density being known to that precision. We return to
this in the Outlook in the conclusion.

\textbf{$\Omega_{\Lambda_3}$: the prior edge, not the data, sets the limit.}
Here the answer is different and more interesting. Run~A cannot settle the
question on its own, because its prior $\Omega_{\Lambda_3}\in[-0.1,0.1]$ ends
exactly where the posterior does --- so the posterior appears to run into the
prior edge, and we cannot tell whether it would have continued had the prior
allowed it. Run~B used the same $h$ prior but let $\Omega_{\Lambda_3}$ range
over $[-0.5,0.5]$, five times wider. There the posterior peaks and falls off:
$\Omega_{\Lambda_3}=0.035\pm0.062$, with a $95\%$ interval $[-0.078,+0.152]$
well inside the prior range. So the one-sided limit in Table~\ref{tab:mcmc} is
not the data refusing to bound $\Omega_{\Lambda_3}$ from above. It is the prior
ending first.

That would be easy to fix, except for what actually does the bounding in run~B.
The $95\%$ lower limit on $h$ there is $0.55000$ --- the prior minimum, to five
figures. The chain is pressed against the floor of the $h$ range, and $h$ is
anticorrelated with $\Omega_{\Lambda_3}$ along a narrow band. That floor is what
stops $\Omega_{\Lambda_3}$ from climbing further. Widening the
$\Omega_{\Lambda_3}$ prior did not measure $\Omega_{\Lambda_3}$; it moved the
binding constraint from one prior edge to another. In other words, the
degeneracy between $h$ and $\Omega_{\Lambda_3}$ is so tight that whichever prior
first cuts off that degenerate direction becomes the thing that sets the limit
--- and in run~B it is $h$'s prior floor, not the data.

\textbf{The upshot.} The correct statement is stronger than a
bound and weaker than a measurement. The $h$--$\Omega_{\Lambda_3}$ degeneracy is
tight enough that the upper end of the $\Omega_{\Lambda_3}$ posterior is
controlled by whichever prior first cuts off the degenerate direction: its own
at $\pm0.1$ in run~A, or $h$'s at $0.55$ in run~B. No CMB-only analysis of this
model can produce a prior-independent upper limit on $\Omega_{\Lambda_3}$, and
quoting one would amount to quoting a choice of prior. What survives every
choice is the lower bound, the non-detection, and the fact that zero remains
comfortably allowed. This is the sampling-side counterpart of the response calculation in
Sec.~\ref{sec:fisher}, which finds the same degeneracy by a completely different
route and measures how nearly exact it is over the multipoles where Planck is
precise. Together, the two results say that breaking the degeneracy requires
information from outside the CMB temperature spectrum: either an external
measurement of $h$, which would turn the degenerate direction into a real
constraint, or the low-$\ell$ measurements, where the absorbable part of the
$\Omega_{\Lambda_3}$ signature fails and cosmic variance --- not instrumental
precision --- sets the floor.

\section{Power spectrum from Weinberg's kinetic theory}
\la{sec:kinetic}

The Bayesian analysis in Section~\ref{sec:bayesian} showed that the data cannot
determine the two new parameters, and it pointed to the Hubble constant $h$ as
the reason why the parameter $\Omega_{\Lambda_3}$ is so poorly constrained. In
this section we examine why that happens. We compute how each parameter affects
the temperature spectrum, one multipole $\ell$ at a time, and then we ask ---
using the real Planck TT covariance --- which combinations of the standard
cosmological parameters can mimic the effect of changing $\Omega_{\Lambda_3}$.
To answer this, we evaluate the exact multipole formula \eqref{eq:7210}
directly, without using any of the approximations from Section~\ref{sec:hydro}.
In other words we do not use the tight-coupling expansion, we do not assume that
decoupling happens instantaneously, we keep the polarization channel $H$, and we
compute the integrated Sachs--Wolfe (ISW) term from \eqref{eq:Fqt} instead of
discarding it. Consequently, Silk damping, Landau damping, and the ISW effect
emerge naturally from the kinetic equations rather than being inserted by hand.
Every number quoted below comes from this full-hierarchy calculation.

The closed formula \eqref{eq:final} loses accuracy compared to the exact
expression \eqref{eq:7210} in three specific ways, and our pipeline is designed
to remove all three. The first is the finite width of the visibility function
$P(t)=\omega_c e^{-\tau(t)}$. The sudden-decoupling approximation replaces
$P(t)$ by a sharp spike at $t_L$, but in reality the form factors oscillate in
$q$ across the width of $P(t)$. Averaging an oscillation over a finite window
suppresses its amplitude --- this suppression is Landau damping, a real physical
effect, not a numerical artifact. Weinberg estimates this suppression
analytically; the present approach instead keeps the full time integral of
\eqref{eq:7210} and lets the suppression emerge naturally from the integration
itself. (Silk damping, by contrast, is already present in the damping rate
$\Gamma$ before any averaging, and it appears in \eqref{eq:final} through
$\ell_D$; the two mechanisms separate in the computed spectrum.) The second
approximation is the WKB treatment of the acoustic phase $\Theta_{\ell\beta}$,
which is accurate only to order $\epsilon$ by construction. The pipeline
replaces this with a direct integration of the full hierarchy. The third is the
omission of the integrated Sachs--Wolfe term. Because it carries the bare
$e^{-\tau(t)}$ rather than the visibility function, it does not localize at
$t_L$, and it was therefore dropped when \eqref{eq:final} was assembled in
Section~\re{sec:debye}, as it is in \W{7.2.41}; the pipeline computes it from
\eqref{eq:Fqt} along with the rest of the source. The pipeline's inclusion of
the ISW term is the most consequential of the three: the ISW effect is often
thought to touch only the lowest multipoles, but it in fact reshapes the whole
rise to the first acoustic peak, as Section~\re{sec:planck} quantifies. These
improvements are not free: keeping the time integral means the form factors must
be computed on a grid in both $q$ and $t$, rather than at a single time, and the
projection integral now contains a Bessel function that oscillates, and the
numerical integration must sample it finely enough to capture those
oscillations. It takes several minutes per spectrum, compared to a single
quadrature for \eqref{eq:final} and seconds for \textsc{class}.\footnote{The
main benefit of our method is that it eliminates a risky transition point.
Standard codes evolve an approximate tight-coupling system early and switch to
the full hierarchy near recombination. This switch is calibrated to work
perfectly for a ``normal,'' standard universe. However, because we examine specifically an altered universe with a modified history, that built-in
switch point is no longer accurate. This means the software can introduce a
hidden error during the switch --- altering the final results without ever
triggering an error message or warning.} Our pipeline therefore integrates the
full hierarchy throughout. The increased computation time is acceptable because
the analysis below only needs differences between spectra at nearby parameters,
not a full survey of parameter space.

Our new method is not meant to replace the standard codes, such as
\textsc{class} \ci{Blas:2011rf}, which is about a thousand times faster and more
accurate. In fact, we used \textsc{class} for our main Bayesian analysis of
Sec.~\re{sec:bayesian} because testing all the options requires a hundred
thousand calculations, which would take too long with our slower code. Instead,
the two tools work together and are complementary. While fast codes yield a
final, summarized number, if a result looks strange, one cannot look backward to
find out why. Our method here is slower, but it allows us to look inside the
calculation and see the exact step-by-step cause and effect.

There are four properties that make the present method well-suited for
understanding the physics rather than just producing numbers. First, we can
attribute each feature in the spectrum to the physical term that produced it.
Each cosmological parameter enters at a single identified place --- through the
expansion history $E(a)$ in equation \eqref{eq:friedmann} --- and then
propagates through the form factors $F$, $G$, and $H$, which are computed
separately and can be switched on and off independently. This means that a
feature in the derivative $\partial C_\ell / \partial \Omega_{\Lambda_i}$ can be
traced directly to the term that produced it, rather than being inferred from
the total sum. This is what allows us to make the statements in
Section~\ref{sec:responses}: that the $\Omega_{\Lambda_2}$ signature is purely
an amplitude effect, and that the unabsorbable part of the $\Omega_{\Lambda_3}$
signature is the low-$\ell$ ISW response. A general-purpose code only returns
the sum, making such identification impossible.

The second property is independence. Our pipeline and \textsc{class} share
almost no machinery: they use different perturbation variables, a different way
of fixing the residual gauge freedom in synchronous gauge (see
Appendix~\re{app:ic}), a line-of-sight integral evaluated directly against
\eqref{eq:7210} rather than through a source-function interpolation, and no
tight-coupling expansion at all. Agreement between them on the $h$-response at
the level of Section~\ref{sec:validation} is therefore evidence about the
physics itself, rather than merely a consistency check of one code against a
variant of itself. The same independence is valid for the Fisher analysis of
Section~\ref{sec:fisher} as well: the analysis reproduces the
$h$--$\Omega_{\Lambda_3}$ anticorrelation of the chains without sampling --- one
calculation differentiates a spectrum and the other explores a likelihood, yet
they agree.

Third, the numerical calculation confirms the analytic results of
Section~\ref{sec:hydro}. Those results were obtained under approximations ---
the tight-coupling expansion, the WKB treatment of the acoustic mode, and a
Gaussian visibility function --- and they contain specific coefficients that
the derivation predicts. The clearest is the $16/15$ in the damping rate
$\Gamma$ of \eqref{eq:Gamma}, which arises from the polarization ladder and the
slaved shear. The pipeline makes none of those approximations and contains no
damping rate or viscosity coefficient: it integrates the full hierarchy and
nothing more. Yet the damping it produces matches $\Gamma$, with its $16/15$,
to better than $1\%$, and the Landau damping length it implies agrees with
Weinberg's (7.2.31) to $0.42\%$ (Section~\ref{sec:further}). These agreements
therefore test the analytic derivation itself, not merely one implementation of
it. A comparison with \textsc{class} could not do this: agreement between two
codes tests the codes, not the analytic steps of Section~\ref{sec:hydro}.

Fourth, the present analysis allows us to quantify the cost of each
approximation. Sections~\ref{sec:hydro} and~\ref{sec:kinetic} both evaluate the
same expression \eqref{eq:7210}, one analytically and one numerically. By
switching each approximation back on one at a time, we can read off exactly what
it costs: the sudden-decoupling limit, the WKB treatment of the acoustic phase,
and the conventional lower cutoff on the projection integral. That last one, for
instance, turns out to discard $18\%$ of the exact Bessel normalization at
$\ell=2$ (see Section~\re{sec:further}). A code that never makes these
approximations cannot say what they cost, because it has no point of comparison.

Before we proceed, two remarks about the new parameters themselves are in order.
First, both $\Omega_{\Lambda_2}$ and $\Omega_{\Lambda_3}$ enter the calculation
in only one place: the expansion history $E(a)$ of Eq.~\eqref{eq:friedmann}.
Second, both are small. The data allow $\Omega_{\Lambda_2}$ at the level of
$10^{-7}$ --- about a tenth of a percent of the radiation density --- and
$\Omega_{\Lambda_3}$ at the level of $10^{-2}$. This smallness is what makes the
analysis below tractable in the form we present it. Over the range the data
allow, each parameter produces a linear response, so a single derivative
captures how each acts. And each adds its own degenerate direction to parameter
space --- $\Omega_{\Lambda_2}$ with the overall amplitude, $\Omega_{\Lambda_3}$
with $h$ --- without changing the degeneracies that already exist among the
standard parameters.

The last statement should be read as an empirical claim, not a theorem. 
A new parameter large enough to make the effective combination depend
on $\ell$ would break an existing degeneracy. What we find is that, at the
magnitudes the data permit, neither of our parameters does so. The
parameter $\Omega_{\Lambda_2}$ enters $E(a)$ exactly as an addition to the background
radiation density would, and differs from genuine extra radiation only in
carrying no perturbations of its own. That difference is too small to measure
at the fraction of a percent of $\Omega_R$ that the data allow, so in practice
$\Omega_{\Lambda_2}$ is degenerate with a small change of $\Omega_R$, or
equivalently of $N_{\rm eff}$. Since $\Omega_R$ is held at its FIRAS value in
our chains, the degeneracy shows up instead as one with the overall amplitude
(Sec.~\ref{sec:responses}).

\vspace{.2in}

The rest of this section proceeds as follows. In Section~\ref{sec:pipeline} we
describe the four stages of the pipeline and fix the numerical settings and the
two baseline parameter sets used throughout. In
Section~\ref{sec:pipeline_checks} we examine how far the resulting numbers can
be trusted: Section~\ref{sec:noisefloor} measures the numerical noise floor, the
smallest difference between two spectra that can be taken as physical;
Section~\ref{sec:planck} measures the absolute accuracy against the Planck 2018
best-fit spectrum; and in Section~\ref{sec:further} we collect three further
checks --- that Silk and Landau damping emerge from the dynamics rather than
being imposed, that the two cuts on the projection integral are quantified, and
what surface of last scattering the calculation produces for itself.
Section~\ref{sec:validation} then validates the code's response to a parameter
change, using \(h\), against \textsc{class}, where an independent calculation
exists. In Section~\ref{sec:responses} we compute how the spectrum responds to
\(\Omega_{\L_2}\) and \(\Omega_{\L_3}\), and then ask which of the standard
parameters can mimic each response, answering that question with a Fisher
analysis carried out with the real Planck covariance.

\subsection{Construction of the pipeline}
\la{sec:pipeline}

The pipeline is built in four stages, with each stage checked against the one
before it. It is written from scratch in Python, rather than adapted from an
existing Boltzmann code.

\medskip\noindent\emph{Stage 1: background and recombination history.} We
integrate the Friedmann equation \eqref{eq:friedmann} to obtain the expansion
history and the distance $r_L$ to last scattering. At the same time, we solve
for the ionization history using an effective three-level hydrogen model (the
Peebles equation, similar to the treatment in \textsc{recfast}) and a Saha
treatment for helium. We compute the angular diameter distance in two ways: once
with radiation included in the integral, and once using Weinberg's
radiation-free approximation from equation (7.2.43); the two differ by $0.5\%$.
Production spectra use Weinberg's radiation-free form throughout: the resulting
$r_L$ is computed once in Stage~1, stored with the source caches, and read from
there by the projector, and the same $r_L$ also sets the wavenumber grid, the
Bessel sampling in Stage~4, and the analytic evaluation of
Section~\re{sec:quadeval}. The two routes of this paper therefore share one
distance convention, and the $0.5\%$ difference is an explicit design choice
rather than an uncounted discrepancy; Section~\re{sec:validation} accounts for
its contribution to the acoustic-scale offset. We do not take last scattering
from a table; instead, we define it as the peak of the visibility function
$P(t)=\omega_c e^{-\tau}$. This means that $z_L$, $t_L$, and the width of the
last-scattering shell $\sigma_t$ emerge as outputs of the integration, rather
than being taken from the values listed in Weinberg (7.2.46). For the parameters
used below, this gives $z_L \simeq 1076$, which is close to but slightly
different from the \textsc{recfast} value $z_{\rm rec} \simeq 1089$. The
difference is definitional --- visibility peak versus the fitting-formula
surface --- and we return to it when we validate the recombination history
independently through the Landau damping length.

\medskip\noindent\emph{Stage 2: synchronous-gauge perturbations.} For each
comoving wavenumber $q$, we integrate the coupled system of baryons, cold dark
matter, neutrinos, and the metric from deep in the tightly coupled era
($q/aH \ll 1$, super-horizon) to the present. The metric evolves through
Weinberg's equations (6.1.70)--(6.1.73), and neutrinos are evolved as a
truncated collisionless hierarchy following Weinberg (6.1.53). The initial
conditions are set to the adiabatic growing mode, imposed as
$A_q(a_i) = q^2 B_q(a_i) = 0$. In this setup, the matter mode fixes the metric
\emph{rates} through equations (6.1.72)--(6.1.73), while the metric
\emph{values} are left as the residual gauge freedom. These are therefore the
correct growing-mode conditions, not arbitrary hand-set zeros (we check the
consistency of this choice in Appendix~\re{app:ic}). The output of this stage is
the full time evolution of the metric and matter perturbations, which then serve
as the source for the photon hierarchy.

\medskip\noindent\emph{Stage 3: the full photon-plus-polarization hierarchy and
the form factors.} We evolve the temperature and polarization multipoles
$\Delta_{T,\ell}$ and $\Delta_{P,\ell}$ directly from Weinberg's equations
(6.1.27) and (6.1.28), using the scalar and polarization sources $\Phi$ and
$\Pi$ from (6.1.29) and (6.1.30). Baryons carry their own velocity $\delta u_B$:
the $\ell=1$ scattering term is proportional to the photon--baryon slip
$(\delta u_\gamma - \delta u_B)$, and momentum conservation from Weinberg
(6.1.5) then fixes
$\dot{\delta u}_B = (\omega_c/R)(\delta u_\gamma - \delta u_B)$ with no further
input. From this stage we read off the form factors $F(q,t)$, $G(q,t)$, $H(q,t)$
of equation \eqref{eq:7210} --- these are Weinberg's (7.2.7)--(7.2.9), quoted in
\eqref{eq:Fqt}, \eqref{eq:Gqt}, \eqref{eq:Hqt} --- on a grid of $372$ modes (see
below for how the 372 modes arise).
Each form factor is built from the solver's own metric variables, without
substituting the matter-dominated relations. The combination
$W \equiv q^2 \dot B_q = 3\dot A_q - 2\psi_q$ from Weinberg (6.1.73), and its
time derivative $d(aW)/dt$, are evaluated from the solver's right-hand side and
a centered finite difference on the dense output, so that $F$ and $G$ carry the
exact radiation-era corrections rather than only their matter-dominated limits.
Two features of these form factors govern the implementation. First, the
visibility function $P = \omega_c e^{-\tau}$ multiplies all four terms in the
bracket of \eqref{eq:Fqt}, including the metric pair; only the total-derivative
(ISW) term carries the bare $e^{-\tau}$, so the ISW term is integrated to the
present, while the other three are confined to the last-scattering shell.
Second, by the identity \eqref{eq:phiid}, the scattering source reduces to
$\omega_c [ \frac{1}{4}\Delta_{T,0} + \frac{1}{16}\Pi ]$ --- the intrinsic
temperature perturbation $\delta T/\bar T = \frac{1}{4}\delta_\gamma$ together
with a polarization correction --- so the anisotropic stress $\Pi$ enters the
temperature spectrum \emph{twice}: once here, in the scattering source, and
again through $H$ in \eqref{eq:Hqt}. Both contributions are retained.

\medskip\noindent\emph{Stage 4: line-of-sight projection.} Equation \rf{eq:7210}
is evaluated with exact spherical Bessel functions rather than their Debye
asymptotics, and the visibility function is retained at its full width, so that
Landau damping emerges from the integration itself rather than from an inserted
estimate. Because the sources are smooth in $q$ while the Bessel factor
$j_\ell(qr_L)$ oscillates on the far finer scale $2\pi/r_L$, the projector fits
a smooth interpolating curve through the sources on the $372$-mode grid, and then evaluates that curve wherever the
Bessel factor needs sampling --- a fixed number of points per half-oscillation.
Interpolating the sources in this way is the same strategy the standard codes
use.

\medskip\noindent\emph{Production settings.} For reference, the numerical
controls\footnote{By a numerical control we mean a setting in the code that has
nothing to do with the physics: how finely a grid is spaced, how many steps an
integral is divided into, where a sum over multipoles is cut off, how tightly
the differential-equation solver is asked to converge.} are held at the
following values throughout, and the convergence of the first four is
established in Section~\re{sec:noisefloor}: the sources are tabulated on
$n_t = 400$ steps in $\ln a$; the photon and polarization hierarchies are
truncated at $\ell_{\max}^\gamma = 10$ and the neutrino hierarchy at
$\ell_{\max}^\nu = 8$; the ODE solver runs at a relative tolerance of $10^{-6}$;
the projector samples six points per half-oscillation of the Bessel factor; and
the wavenumber grid carries $372$ modes. Above $q = 0.003\,\mathrm{Mpc}^{-1}$
the modes are spaced linearly with step $\Delta q = 0.0015\,\mathrm{Mpc}^{-1}$
up to $q_{\max} = 0.5\,\mathrm{Mpc}^{-1}$; the code caps the grid at the smaller
of $0.5\,\mathrm{Mpc}^{-1}$ and
$12\times 700/r_L \simeq 0.6\,\mathrm{Mpc}^{-1}$, where $700$ is the highest
multipole projected. The linear part therefore has
$1 + \lfloor (0.5 - 0.003)/0.0015 \rfloor = 1 + 331 = 332$ modes, the last at
$q = 0.4995\,\mathrm{Mpc}^{-1}$. Below $q = 0.003\,\mathrm{Mpc}^{-1}$, $40$
geometrically spaced modes reach down to
$q_{\min} = 6\times10^{-5}\,\mathrm{Mpc}^{-1}$, giving $332 + 40 = 372$ modes in
total. The two cuts on the projection integral itself are $\ell$-dependent;
their production values, and the reasons for them, are given in
Section~\re{sec:further}.

\medskip\noindent\emph{Parameter sets.} Two baselines are used below. Weinberg's
reference set, from his (7.2.44)--(7.2.45), is
\begin{align}
\Omega_M h^2 = 0.13299,\quad \Omega_B h^2 = 0.02238,\quad h = 0.71992,\quad n_S = 0.95820,
\nonumber\\
T_0 = 2.725~{\rm K},\quad k_R = 0.05~{\rm Mpc}^{-1},\quad N^2 = 1.736\times10^{-10},\quad e^{-2\tau_{\rm reio}} = 0.80209,
\label{eq:weinbergparams}
\end{align}
the last of which corresponds to $\tau_{\rm reio} = 0.11027$. The Planck 2018
set, the best fit of the \texttt{base\_plikHM\_TTTEEE\_lowl\_lowE\_lensing}
chain, is
\begin{align}
\Omega_M h^2 = 0.14237,\quad \Omega_B h^2 = 0.02237,\quad h = 0.6736,\quad n_S = 0.9649,
\nonumber\\
T_0 = 2.7255~{\rm K},\quad k_R = 0.05~{\rm Mpc}^{-1},\quad \tau_{\rm reio} = 0.0544.
\label{eq:planckparams}
\end{align}
Both are spatially flat, and both are evaluated at
$\Omega_{\Lambda_2} = \Omega_{\Lambda_3} = 0$ wherever they serve as a baseline;
the constant $\Omega_\Lambda$ is then fixed by closure, so that $E(1) = 1$
exactly with the finite-temperature components included. In the Planck case the
amplitude $N^2$ is left at Weinberg's value, since it is absorbed by the fitted
$\alpha$ of Section~\re{sec:planck} and never enters a shape comparison. The
noise floor of Section~\re{sec:noisefloor} and the quadrature cuts of
Section~\re{sec:further} are measured at \eqref{eq:planckparams}; the damping
check of that same subsection is measured at \eqref{eq:weinbergparams}.

\subsection{Checks of the pipeline}
\la{sec:pipeline_checks}

Before using the present method to compute finite-temperature responses, one
needs to know how far the numbers it generates can be trusted. The two
subsections that follow establish that: the numerical noise floor, which is the
smallest change in $C_\ell$ that can be regarded as physical rather than
numerical, and the absolute accuracy against the Planck 2018 spectrum.
Section~\re{sec:further} then collects the three further checks listed above,
which are of a different kind: they do not measure accuracy but confirm that
features of the calculation emerge from the dynamics rather than having been
inserted.

\subsubsection{Numerical noise floor}
\label{sec:noisefloor}

The pipeline reproduces the observed spectrum to $1.8\%$ RMS, as shown in
Sec.~\ref{sec:planck}. Taken at face value, that number would seem to invalidate
everything this subsection and Sec.~\ref{sec:responses} later claim: a numerical
noise floor of $0.004\%$ --- roughly four hundred times finer --- and responses
to the finite-temperature parameters measured at the level of a tenth of a
percent. The two numbers are not, however, measuring the same thing. The $1.8\%$
is the accuracy of the \emph{absolute} spectrum, and it is dominated by
systematic effects common to every spectrum the pipeline produces --- missing
lensing, the recombination history, the residual low-$\ell$ deficit, quadrature
choices. Whereas a systematic shared by two spectra cancels when the two are
differenced, the same is not true for numerical noise. So what limits a response
is not the absolute accuracy but the numerical noise floor: the level below
which a change in $C_\ell$ is an artifact of finite resolution rather than real
physics. That is the floor we analyze in this subsection.

Measuring a floor this way is standard practice for Boltzmann codes, though it
is normally carried out by comparing two independent implementations rather than
one code against itself. In \cite{Seljak:2003th} three independent linear
perturbation codes were compared and, once the dominant sources of numerical
instability had been removed, brought into agreement at the $10^{-3}$ level
across most of the range $\ell \lesssim 3000$; later in
\cite{Lesgourgues:2011rg} the comparison between \textsc{class} and
\textsc{camb} was pushed to $0.01\%$ for lensed spectra at a fixed recombination
history, and the common limit of the two codes was used to calibrate precision
settings against a stated likelihood error. Closest in method to what follows is
Ref.~\cite{Hamann:2009yy}, in which the internal precision parameters that
control the error of \textsc{camb} were identified, its default settings were
shown to bias Planck-era parameter estimates by several tenths of a standard
deviation, and those parameters were then optimized for precision at acceptable
cost --- the same logic that makes the measurement below worth carrying out,
namely that a numerical error smaller than the effect one is trying to measure
is the condition for the measurement to be meaningful. Our situation differs in
one respect: no second independent implementation of Weinberg's kinetic
formulation exists, so the benchmark here is the same code at finer settings
rather than another code. What we measure is therefore self-consistency under
refinement, not agreement between implementations. The absolute accuracy, which
is the question a code comparison answers, is taken up separately in
Section~\re{sec:planck} against the Planck 2018 spectrum and in
Section~\re{sec:validation} against \textsc{class} at the level of the
$h$-response.

All of the measurements in this subsection are made at the Planck 2018
parameters \eqref{eq:planckparams}, with
$\Omega_{\Lambda_2} = \Omega_{\Lambda_3} = 0$. That is deliberate: the floor is
the error bar on the finite-temperature responses of Section~\re{sec:responses},
and those responses are measured as departures from this same baseline, so the
floor must be measured there and not at some other point in parameter space.
Each numerical control has a production value, chosen as a compromise between
accuracy and running time. If a production value is fine enough, making it finer
should leave the spectrum unchanged. Whatever change remains is numerical noise,
not physics. We measure this noise one control at a time, using four controls;
let us call this list $\mathbf{n}$,
\begin{align}
\mathbf{n} \;=\; \big(n_t,\; \ell_{\max}^\gamma,\; \epsilon_{\rm tol},\; n_J\big),
\label{eq:controls}
\end{align}

\begin{itemize}
\item $n_t$ --- the number of time steps on which the sources are tabulated;
\item $\ell_{\max}^\gamma$ --- the multipole at which the photon and
polarization hierarchies are truncated;
\item $\epsilon_{\rm tol}$ --- the tolerance of the ODE solver;
\item $n_J$ --- the number of sample points per half-oscillation of the
spherical Bessel function in the projector.
\end{itemize}

\noindent A fifth control, the wavenumber grid spacing $\Delta q$, is special
and is handled separately below. Let $\mathbf{n}^{\rm prod}$ be the production
settings, and let $\mathbf{n}^{\rm ref}_i$ be the same list with only the $i$-th
entry tightened. We tighten the step counts by a factor of two, raise the
multipole cutoff from $10$ to $16$, and tighten the ODE tolerance by a factor of
ten. Recomputing the full spectrum for each of the four tightened settings gives
four spectra to compare with the production spectrum. We combine the four
fractional changes in quadrature:

\begin{align}
\left.\frac{\Delta C_\ell}{C_\ell}\right|_{\rm numerical}
\;\equiv\;
\left[\;\sum_{i=1}^{4}\;
\left\langle
\left(\frac{C_\ell(\mathbf{n}^{\rm ref}_i)-C_\ell(\mathbf{n}^{\rm prod})}
           {C_\ell(\mathbf{n}^{\rm prod})}\right)^{\!2}
\right\rangle_{\ell\ge30}
\;\right]^{1/2}
\;=\;0.004\% ~{\rm RMS}\quad(\ell\ge30).
\label{eq:floor}
\end{align}

\noindent Here $\langle\cdot\rangle_{\ell\ge30}$ is the average over the ten
sampled multipoles with $\ell\ge30$ listed in Table~\ref{tab:planckshape}, so
the bracket is a root-mean-square. The result is a noise floor of $0.004\%$ RMS.

Of the four controls, one accounts for essentially the entire noise floor, while
the other three barely register. Refining the time grid moves the spectrum by
\(0.004\%\) RMS; its single largest deviation, \(0.007\%\), occurs at
\(\ell=700\). Raising the multipole cutoff moves it by \(0.001\%\). Tightening
the ODE tolerance and doubling the Bessel sampling each move it by under
\(0.001\%\) --- that is, by an amount our precision cannot resolve. The floor
therefore has a single cause rather than many small errors added together: the
time derivative in the ISW term of \eqref{eq:Fqt}, which the code takes as a
finite difference on the time grid. An independent test supports this
identification: comparing \(n_t=400\) with \(n_t=1600\) --- a fourfold
refinement instead of a doubling --- and carrying the change through to
\(C_\ell\) yields \(2.8\times10^{-5}\), or \(0.0028\%\), the same order as the
\(0.004\%\) found above for the time grid.

The wavenumber grid deserves separate remarks. Its role in the comparison
differs from that of the other four controls: they were measured at their
production values, while the grid was improved until its effect became
negligible. To see why, let us start with the original spacing,
$0.003\,\mathrm{Mpc}^{-1}$. At that spacing the grid was the only control that
moved the spectrum appreciably. The effect was $0.101\%$ RMS overall, but it was
not spread evenly across multipoles. It appeared as a localized error of about
$0.3\%$ between $\ell \simeq 50$ and $\ell \simeq 80$, on the rise to the first
acoustic peak. This is where the projection kernel is most sensitive to how
finely the acoustic oscillation in the sources is sampled, so a coarse grid
shows up there first. We removed the error by halving the spacing to
$0.0015\,\mathrm{Mpc}^{-1}$ and doubling the low-$q$ tail. This is the
production grid, with $372$ modes. We then re-measured the four controls of
Eq.~\eqref{eq:controls} on the refined grid; those re-measured values are what
appear in Eq.~\eqref{eq:floor}. The grid thus does not enter
Eq.~\eqref{eq:floor} as a term of its own. It was refined until its contribution
to the comparison vanished, rather than being measured at its production
setting. One question remains open. To show that the production grid is itself
converged, we would have to halve the spacing once more, and we have not done
so. We expect the residual to lie below the floor quoted here, for the following
reason: the error the refinement removed was localized to $\ell \simeq 50$--$80$
and absent everywhere else, and a residual of the same character would be
confined to the same narrow range.

The floor of $0.004\%$ is roughly uniform in $\ell$ and lies a factor of about
four hundred below the absolute accuracy of the code --- the gap that makes a
differential measurement possible. It is quoted for $\ell \ge 30$, and with the
two cuts on the projection integral of Section~\re{sec:further} already at their
production values. At the lowest multipoles it is those cuts, not the controls
listed here, that set the accuracy. Any response to a finite-temperature
parameter change that lies above this floor is therefore physical.

\subsubsection{Comparison with Planck 2018}
\label{sec:planck}

We evaluate the pipeline at the Planck 2018 best-fit parameters, taken from the
base-plikHM-TTTEEE-lowl-lowE-lensing chain, and compare against the
corresponding best-fit \emph{theory} spectrum. At $\ell \gtrsim 30$, the
spectrum depends on the primordial amplitude and the reionization optical depth
only through the combination $N^2 e^{-2\tau_{\rm reio}}$. This creates a
normalization mismatch: Weinberg writes the spectrum in terms of
$|\mathcal{R}_q^o|^2$, Planck quotes $A_s$, and converting between them would
require Weinberg's mode-function conventions, which was never needed and never
fixed in the present calculation; we thus do not convert. Instead we fit a
single overall amplitude $\alpha$ by least squares over $\ell \ge 30$, and
compare the \emph{shape}.\footnote{The amplitude is free by construction here,
not by choice. Weinberg's $N^2 = 1.736\times10^{-10}$ was fitted to WMAP at his
own $\tau_{\rm reio} = 0.110$, whereas Planck quotes $A_s$ in a different
Fourier normalization. Converting between the two requires Weinberg's
mode-function conventions. Guessing the conversion would therefore inject a
constant factor into every comparison without our noticing. Fitting $\alpha$
instead leaves the shape untouched.} The shape is what the code computes from
first principles, so it is the meaningful thing to test. Reionization is not
modeled. At $\ell \gtrsim 30$ its only effect is the factor
$e^{-2\tau_{\rm reio}}$, which $\alpha$ absorbs. At the lowest multipoles it
does more than rescale, and it is therefore one candidate for the residual
deficit noted below. Table~\ref{tab:planckshape} gives representative values.

\begin{table}[H]
\centering
\begin{tabular}{rrrrr}
\toprule
$\ell$ & $\alpha\,\mathcal{D}_\ell^{\rm kin}$ & $\mathcal{D}_\ell^{\rm Planck}$ & ratio & dev. \\
\midrule
2   &  902.2 & 1016.7 & 0.8874 & $-11.26\%$ \\
3   &  877.0 &  963.7 & 0.9100 & $-9.00\%$ \\
5   &  826.9 &  874.5 & 0.9456 & $-5.44\%$ \\
10  &  805.8 &  817.2 & 0.9861 & $-1.39\%$ \\
20  &  901.8 &  905.2 & 0.9963 & $-0.37\%$ \\
\midrule
30  & 1057.2 & 1054.7 & 1.0024 & $+0.24\%$ \\
50  & 1428.5 & 1421.3 & 1.0051 & $+0.51\%$ \\
80  & 2128.9 & 2118.3 & 1.0050 & $+0.50\%$ \\
100 & 2722.9 & 2697.9 & 1.0093 & $+0.93\%$ \\
150 & 4435.2 & 4382.3 & 1.0121 & $+1.21\%$ \\
220 & 5750.4 & 5730.1 & 1.0035 & $+0.35\%$ \\
300 & 3991.0 & 4077.7 & 0.9788 & $-2.12\%$ \\
400 & 1699.5 & 1746.7 & 0.9730 & $-2.70\%$ \\
550 & 2543.3 & 2572.9 & 0.9885 & $-1.15\%$ \\
700 & 1892.2 & 1868.1 & 1.0129 & $+1.29\%$ \\
\bottomrule
\end{tabular}
\caption{Computed spectrum against the Planck 2018 best-fit theory spectrum, in
$\mathcal{D}_\ell = \ell(\ell+1)C_\ell/2\pi$ with units of $\mu$K$^2$, with a
single overall amplitude $\alpha = 0.9646$ fitted by least squares on
$\ell \ge 30$ and everything else computed from first principles. The first
block lists the low multipoles, where a residual deficit remains; the second
samples the acoustic range $\ell \ge 30$. The multipoles listed here are a
sample: Figure~\ref{fig:planck} shows the comparison on a denser grid, on which
the largest deviation is $-4.0\%$ near $\ell \simeq 360$, between the tabulated
entries at $\ell = 300$ and $\ell = 400$.}
\label{tab:planckshape}
\end{table}

\begin{figure}[H]
  \centering
  \includegraphics[width=0.8\columnwidth]{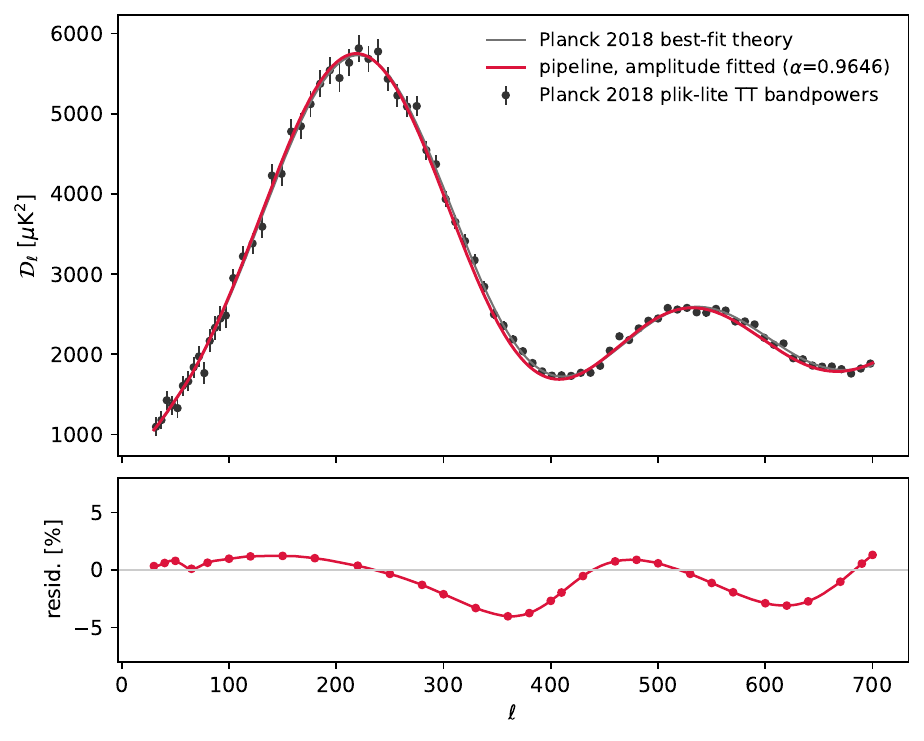}
\caption{The computed spectrum against the Planck 2018 best-fit theory spectrum
(grey) and the \texttt{plik\_lite} TT bandpowers (points, with $1\sigma$
errors). The amplitude is not fitted here: it is the single value
$\alpha = 0.9646$ of Table~\ref{tab:planckshape}, applied unchanged across the
whole range, with everything else computed from first principles at the Planck
2018 parameters. \emph{Bottom:} fractional residual against the theory spectrum.
The residual is oscillatory rather than random, reaching $-4.0\%$ near
$\ell\simeq360$ and changing sign near each acoustic extremum --- the signature
of a small displacement in $\ell$, which dominates the residual rather than an
error in amplitude or in shape.}
  \label{fig:planck}
\end{figure}

Over the tabulated multipoles with $\ell \ge 30$, the RMS of the fractional
deviation in the last column --- which, with $\alpha$ fitted, measures only the
difference in shape --- is $1.34\%$. Figure~\ref{fig:planck} shows the same
comparison on a denser grid of thirty-one multipoles, where the RMS is $1.8\%$
and the largest single deviation is $-4.0\%$, near $\ell \simeq 360$. The
residual is not random. It is oscillatory, changing sign near each acoustic
extremum. That pattern is the signature of a small displacement in $\ell$, and
it is this displacement --- not an error in amplitude or in shape --- that
dominates the residual. A single overall rescaling of $\ell$ by $0.8\%$ reduces
the RMS to $0.9\%$. A rescaling of this kind, in which every $\ell$ is
multiplied by the same factor, is the natural form for the correction: at fixed
physical densities the acoustic pattern is set by the angle $\theta_*=r_s/d_A$,
and a small error in either the sound horizon or the distance to last scattering
moves every feature by the same fraction of $\ell$ --- the geometric degeneracy
of Refs.~\ci{Efstathiou:1998xx,Zaldarriaga:1996xe}. The $0.8\%$ is the rescaling
we found by fitting the residual, chosen because it removes the displacement
most effectively. A single rescaling is a first approximation only, since the
per-feature offsets found in the comparison with \textsc{class} in
Sec.~\ref{sec:validation} range from $0.7\%$ to $1.3\%$. They are the same
acoustic-scale offset, traced there to the distance and recombination
conventions of Stage~1.

One further difference between the two spectra is that the Planck best-fit
spectrum is lensed while our pipeline computes the unlensed spectrum; we have
not implemented lensing. To gauge the size of this effect, we compared the
lensed and unlensed \textsc{class} spectra at the same parameters. Over
$30 \le \ell \le 700$, the effect on $TT$ nowhere exceeds about $1\%$, reaching
that level at the first and second troughs, and is $+0.39\%$ at
$\ell \simeq 360$. Correcting the reference for lensing lowers the largest
deviation from $-4.0\%$ to $-3.7\%$, and lowers the RMS that remains after the
$0.8\%$ rescaling of $\ell$ from $0.9\%$ to $0.6\%$. Lensing is therefore a
small component on top of the acoustic-scale offset, not an explanation for it.

The ISW term plays a much larger role than expected. With it switched off and
nothing else changed, the RMS shape deviation on the tabulated multipoles rises
from $1.34\%$ to $22.6\%$. This quantifies the remark made in the introduction
to this section: the ISW effect is often thought to affect only the lowest
multipoles, but it in fact reshapes the whole rise to the first acoustic peak.
Its influence peaks near $\ell=100$, where it raises $C_\ell$ by $78\%$. It
remains large at the first peak itself, contributing $+40\%$ at $\ell=220$. It
then falls away through $\ell\simeq350$, changes sign near $\ell=400$, and
contributes no more than a few percent beyond that. Below $\ell=10$ the
comparison leaves a residual deficit, about $11\%$ at $\ell=2$. Part of this is
the $0.6\%$ of the Bessel normalization from Eq.~\eqref{eq:besselid} that is
still outside the mode grid; we have not isolated the rest. At $\ell=2$, the
cosmic variance is $\sqrt{2/5}=63\%$, so this is a matter of numerical
bookkeeping, not of physics.

\subsubsection{Three further checks}
\label{sec:further}

Three further checks made while the pipeline was built confirm that features of
the calculation emerge from the dynamics, or are under control, rather than
having been inserted by hand. They do not measure its accuracy, which is the
task of the two preceding subsections. We summarize the three checks without
presenting the technical details.

\medskip\noindent\emph{Damping is emergent.} The code contains no damping
rate and no viscosity coefficient; it integrates the full photon and
polarization hierarchy and nothing more. Damping nevertheless appears in its
output, and at the rate derived in Section~\ref{sec:hydro}. To isolate it, we
compare with a tight-coupling solver from which damping has been removed. At
last scattering the amplitude of the $\delta_\gamma$ oscillation, relative to
that undamped solution, has fallen to $0.073$ by $q = 0.45\,\mathrm{Mpc}^{-1}$,
and the acoustic energy decays at the rate $\Gamma$ of \eqref{eq:Gamma} to
better than $1\%$. The numerical code thus confirms the $16/15$ derived in
Section~\ref{sec:hydro} without having been given it. The agreement is also
sharp enough to discriminate: replacing $16/15$ by $8/9$ (no polarization) or
by $4/5$ (isotropic scattering), the alternatives tabulated in
Appendix~\ref{app:sec3}, would change $\Gamma$ by more than $10\%$.
Landau damping can be checked separately, because it depends on the
recombination history only through the width $\sigma_t$ of the visibility
function. Inserting the width our recombination calculation produces,
$\sigma_t = 0.01511$~Mpc, into Weinberg's (7.2.31) gives
$d_{\rm Landau} = 0.004829$~Mpc, against his $0.004809$~Mpc --- agreement at
$0.42\%$. This is therefore also an independent check of the Stage-1
recombination.

\medskip\noindent\emph{The two cuts on the projection integral are quantified.}
The projection integral over \(q\) must in practice be truncated, and the
standard lower cut is \(q_{\rm lo} = 0.85\,\nu/r_L\) with
\(\nu = \ell + 1/2\). This cut is usually justified by a WKB
turning-point argument: below \(q_{\rm lo}\), the Bessel function
\(j_\ell(qr)\) is supposed to be exponentially suppressed and contribute
nothing. The trouble is that WKB is only valid at large \(\ell\). At small
\(\ell\) there is no WKB regime, so the argument does not justify the cut there. How much does the cut discard? An exact identity answers this directly:
\begin{align}
\int_0^\infty \frac{dq}{q}\, j_\ell(qr)^2 = \frac{1}{2\ell(\ell+1)}.
\label{eq:besselid}
\end{align}
Comparing the truncated integral to this exact value shows that the cut
throws away \(17.7\%\) at \(\ell = 2\), but less than \(0.4\%\) from
\(\ell = 20\) upward. The loss is therefore severe at low \(\ell\) and
negligible at high \(\ell\), which is why the two regimes must be treated differently. The production run does exactly this. It applies the standard cut
\(\beta_{\rm lo} = 0.85\) only for \(\ell \ge 20\); at \(\ell = 2\) this
recovers \(0.9936\) of the exact normalization, against the \(82.3\%\)
that the naive cut would retain. The upper cut is likewise made
\(\ell\)-dependent: \(\beta_{\max} = 40\) below \(\ell = 60\) and
\(\beta_{\max} = 10\) at and above it. The reason is that the
source-weighted integrand, unlike the bare kernel \(j_\ell^2\), still
carries weight in the \(j_\ell\) tail at low \(\ell\), so the integral
must be pushed further out in \(q\) to capture it. At high \(\ell\) the
integrand decays faster and the smaller cutoff suffices.

\medskip\noindent\emph{The last-scattering surface is an output.} The visibility
peak sits at $z_L = 1076.18$ at Weinberg's parameters and $1076.35$ at Planck's,
about $15$ below the Hu--Sugiyama fitting value \ci{Hu:1995en} that both
Planck's $z_*$ and Weinberg's $1090$ quote. The offset is definitional rather
than an error in the recombination history: the peak moves by only $0.16$
between the two parameter sets, while the fitting formula sits about $15$ above
both.

\subsection{Validation of the $h$-response against \textsc{class}}
\label{sec:validation}

This subsection and the next form a pair, and it is worth saying at the outset
how they fit together. In Section~\ref{sec:responses} we compute the responses
of the spectrum to \(\Omega_{\L_2}\) and \(\Omega_{\L_3}\), then analyze which
standard parameters can absorb them; for \(\Omega_{\L_3}\) the answer turns out
to be \(h\), and essentially nothing else. That answer is only as good as the
code's own \(h\)-response, so we establish the \(h\)-response first, here, where
an independent calculation exists to check it against. The order is therefore:
show that the derivative machinery is correct on a parameter that can also be
computed independently, then apply the same machinery to the two parameters for
which no independent calculation exists. In both subsections the derivatives are
computed in the same way, on the same grid and at the same settings: the noise
floor value from Section~\ref{sec:noisefloor} serves as the measure of
uncertainty, or the error bar, for the calculations.

Everything below is a derivative of the spectrum with respect to a parameter.
What must be validated is therefore not the spectrum itself but its response to
a parameter change. We validate that response by comparing
$\partial\ln C_\ell/\partial h$ against \textsc{class}~3.3.4 under matched
conventions. The density parameters are held fixed: $\Omega_M h^2 = 0.14237$ and
$\Omega_B h^2 = 0.02237$. The Hubble constant is stepped to
$h = 0.6736 \pm 0.0070$, and a central difference is taken. This is the
fixed-physical-density convention. The remaining settings are $\tau = 0.0544$
fixed, flat geometry, $T_0 = 2.7255$ K, and three massless neutrino species
($N_{\rm ur} = 3.0$, the convention of the background module, \emph{not}
$3.046$).\footnote{This massless-neutrino convention is used throughout the
code. It differs from the $N_{\rm ur} = 2.0328$ plus one massive species used in
the companion Cobaya/\textsc{class} chains of Sec.~\ref{sec:bayesian}. The
difference is immaterial to the $h$-response validated here, but should be
matched before any absolute cross-comparison of the two analyses.} We also set
$Y_{\rm He} = 0.24$ as hard-coded in the recombination module, use
\textsc{recfast} recombination, and compute unlensed TT to $\ell = 700$. The
fractional derivative $(C_\ell^{+} - C_\ell^{-}) / (2\,\Delta h\, C_\ell^{0})$
is formed identically on both sides. It cancels the free overall amplitude
$N^2 e^{-2\tau}$ of our code and the $A_s$ normalization of \textsc{class}, so
no amplitude conversion is needed.

The following three results are obtained, in decreasing order of strength.

\emph{(i) The acoustic-scale channel.} With the physical densities fixed, the
sound horizon is fixed, so the $h$-response is dominated by the shift of the
angular acoustic scale. The pipeline gives
$d\ln\theta_s/dh = -d\ln r_L/dh = +0.2863$ from the cached distances; the
\textsc{class} derived parameter $100\,\theta_s$ gives $+0.2879$. The dominant
physical channel of the $h$-derivative agrees to $0.55\%$, with no free
parameters and no amplitude fitting.

\begin{figure}[H]
  \centering
  \includegraphics[width=\columnwidth]{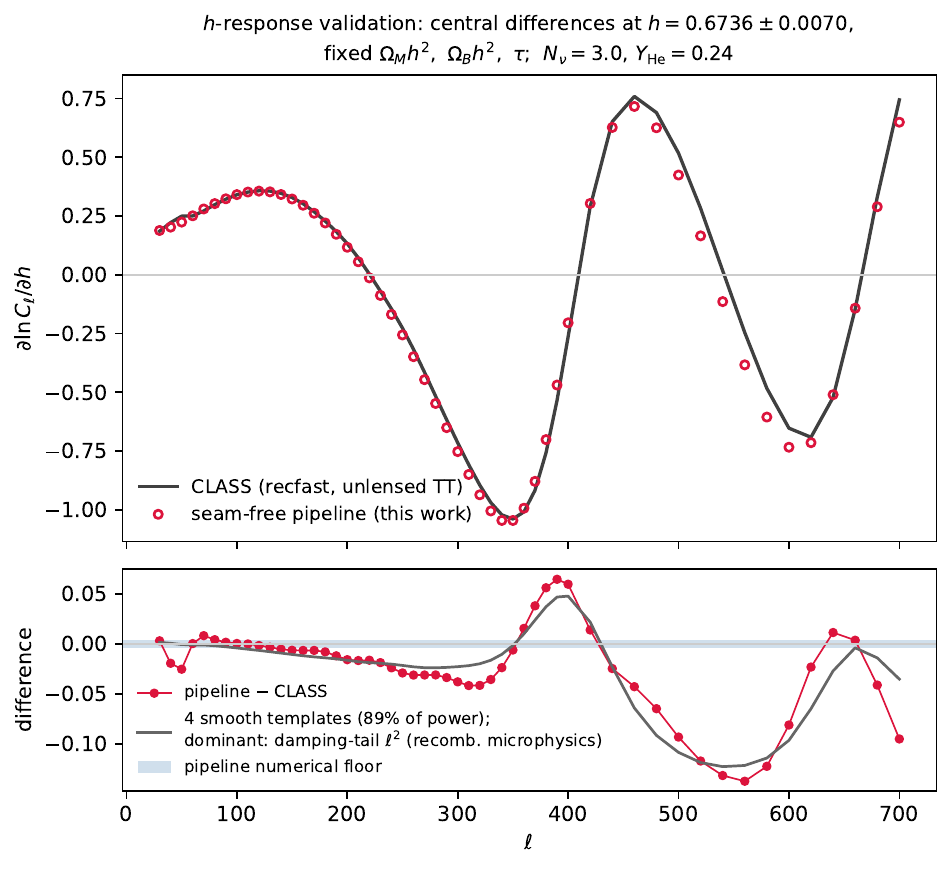}
\caption{Validation of the $h$-response. \emph{Top:} the fractional derivative
$\partial\ln C_\ell/\partial h$ at fixed $\Omega_M h^2$, $\Omega_B h^2$, $\tau$,
from the seam-free code (points) and \textsc{class} (line) under matched
conventions ($N_{\rm ur} = 3.0$, $Y_{\rm He} = 0.24$, unlensed TT).
\emph{Bottom:} the residual (points) with a four-template smooth model (line)
capturing $89\%$ of its power; the band marks the code's numerical floor.
Correlation $0.9965$; the residual tracks the steep slopes of the derivative ---
the static acoustic-phase offset of the absolute spectra ($\delta\ell \simeq +3$
at the first peak) propagated through differentiation --- and shows no
$\ell$-localized structure.}
  \label{fig:classval}
\end{figure}

\emph{(ii) The point-by-point derivative.} Across \(30 \le \ell \le 700\), the
two fractional derivatives match very closely: the two curves track each other
with a correlation coefficient of \(0.9965\). The typical difference between
them, their raw residual RMS, is \(9.4\%\) of the derivative's typical size, but
that number makes the disagreement sound worse than it really is. Almost all of
it comes from $\ell \gtrsim 400$, which carries $88\%$ of the residual power;
two-thirds of the total lies between $\ell = 480$ and $620$, around the second
acoustic peak (bottom panel of Fig.~\ref{fig:classval}). There, the derivative crosses zero steeply: even a
tiny shift in $\ell$ causes a big change in the derivative's value. If we
instead restrict to \(30 \le \ell \le 350\) --- the range where the acoustic
peaks are sharpest and most reliable --- the raw residual RMS falls to \(4.1\%\)
of scale. Checking whether the remaining difference can be explained by a simple
shift in $\ell$ (as if one calculation were slightly offset from the other along
the $\ell$ axis) reveals that a single shift of $\delta\ell = +3.3$ fits the
pattern well. After correcting for that shift, the difference falls to just
$1.8\%$. Neither code's numerical floor can account for this. Our code's
\(0.004\%\) noise floor propagates to roughly \(0.004\) on the fractional
derivative, which translates to about $0.8\%$ of the derivative's scale.
\textsc{class}'s own derivative moves by less than \(0.1\%\) of scale when its
recombination module is changed (see below) --- both well below the residual
$1.8\%$ we are seeing. The disagreement is therefore not a numerical artifact.

\emph{(iii) The residual comes from the input spectra, not the derivative.}

If we divide the residual by the local slope of the derivative, we get a clean,
consistent picture: the two curves are shifted relative to each other by a small
amount in $\ell$. Across the first acoustic peak the shift is about $+3.1$ to
$+3.4$ multipoles, growing to $+9$ to $+12$ near the zero crossing at
$\ell \simeq 550$. The important point is that this shift is \emph{already
there} in the absolute spectra, before any derivative is taken. Our pipeline
puts the first peak, first trough, second peak, and second trough at
$\ell = 218.5,\ 407.7,\ 532.0,\ 666.6$, while \textsc{class} puts them at
$220.5,\ 411.8,\ 535.5,\ 675.5$ --- an offset of roughly $0.7\%$ to $1.3\%$
across every feature. So when we differentiate with respect to $h$, we are not
creating a new disagreement; we are simply carrying this existing offset along.
That offset has a physical explanation, not a numerical one. Two documented
differences in the Stage~1 conventions push the acoustic features in opposite
directions: one convention shifts them to higher $\ell$ by about $+0.49\%$, and
the simplified recombination treatment shifts them to lower $\ell$ by
${\cal O}(1\%)$. The two partly offset each other, leaving the net shift we see.
The shape of the residual is the decisive clue. It is smooth in $\ell$ and
tracks the acoustic structure everywhere, rather than being concentrated in one
multipole range. A localized residual would point to a mistuned approximation
(like a tight-coupling switch being flipped at the wrong place). A smooth,
acoustic-wide offset instead points to a difference in the recombination phase
--- which is exactly what the comparison reveals.

One may wonder whether the result depends on which recombination code
\textsc{class} happens to use. It does not. Rerunning \textsc{class} with
\textsc{HyRec} instead of \textsc{recfast} changes its own fractional derivative
by only $0.1\%$ of its RMS scale, and both $d\ln\theta_s/dh = +0.2879$ and the
fitted displacement $\delta\ell = +3.3$ come out unchanged. So the displacement
is not an artifact of a particular recombination choice; it reflects the
deliberately simple Stage~1 recombination used here, relative to any modern
treatment. Nothing about \textsc{class} was adjusted to make the comparison
work. Two further consistency checks are worth mentioning. First, the
recombination redshift is independent of $h$ in both codes when the physical
densities are held fixed --- $z_L$ agrees to ten digits across the cached $h\pm$
runs, and $z_{\rm rec} = 1088.61$ is identical in all three \textsc{class} runs
at $0.6736 \pm 0.0070$. Second, the dependency stamps on the source caches are
identical across the modules, which guarantees that the derivative is taken
within one uniform code setup.

In short: our pipeline reproduces \textsc{class}'s $h$-response to $0.55\%$ on
the dominant acoustic channel and $0.9965$ in correlation. The only difference
is a smooth, inherited recombination-phase offset, and there is no seam in
either code at this level. Figure~\ref{fig:classval} shows the comparison and
the residual.

\subsection{The finite-temperature responses and what absorbs them}
\label{sec:responses}

Before computing the responses it is worth being exact about what
$\Omega_{\L_3}$ does and does not share with $\Omega_K$, since the two are
easily conflated and the difference is what both the Bayesian analysis of
Sec.~\re{sec:bayesian} and the response analysis in the present section hinge
on. The way $\Omega_{\L_3}$ enters $E(a)$ is not merely a matter of bookkeeping:
$H$ sets $\dd t=\dd a/(aH)$, so it sets the comoving distance $\chi$, the
distance $r_L$ to last scattering, and the angular-diameter distance $d_A$.
$\Omega_{\L_3}$ therefore \emph{does} shift distances --- numerically $r_L$
swings by $\pm0.73\%$ across $\Omega_{\L_3}=\mp0.015$ --- exactly as $\Omega_K$
does, and exactly as $h$ does. What $\Omega_K$ does in addition is of two kinds,
and both are geometrical in origin: only spatial curvature bends the
comoving-to-radial map, replacing $\chi$ by $\sinh$ or $\sin$ of it, and only
$\Omega_K$ modifies the perturbation equations. An $a^{-2}$ component that is
smooth --- that is, spatially homogeneous, with no perturbations of its own,
which is how $\Omega_{\L_3}$ is treated throughout (see
Sec.~\ref{sec:conclusion}) --- leaves that map flat in a spatially flat
universe, and leaves the perturbation equations untouched. The contrast with
$\Omega_K$ drawn in this section therefore holds within that approximation.

The geometric bending of the comoving-to-radial map is not a small
correction layered on top of the $E(a)$ effect of $\Omega_K$: it is larger,
and it acts in the opposite direction. Expanding the map for small curvature
gives, at last scattering,\footnote{The comoving angular-diameter distance is
$S_K(\chi)=\sinh(\sqrt{\Omega_K}\,H_0\chi)/(\sqrt{\Omega_K}\,H_0)$ for
$\Omega_K>0$, with $\sinh\to\sin$ and $\Omega_K\to|\Omega_K|$ for
$\Omega_K<0$. Expanding for small curvature gives
$S_K/\chi \simeq 1+\Omega_K(H_0\chi)^2/6$ in both cases. At last
scattering $H_0\chi\approx3.1$, so $(H_0\chi)^2/6\approx1.6$--$1.7$.}
\begin{equation}
\frac{S_K}{\chi}-1 \simeq \frac{\Omega_K (H_0\chi)^2}{6} \simeq +1.7\,\Omega_K .
\end{equation}
The same $a^{-2}$ term acting through $E(a)$ alone changes $r_L$ by a
fractional amount of about $-0.5$ per unit, as the $\pm0.73\%$ swing above
shows. Per unit parameter, then, $\Omega_K$ changes $d_A$ by about $+1.2$ in
fractional terms and $\Omega_{\L_3}$ by about $-0.5$. The opposite signs explain
the opposite correlations with $h$ in Sec.~\ref{sec:bayes_results}: at fixed
physical densities a larger $h$ shrinks $d_A$, so $h$ must move against
$\Omega_{\L_3}$ but along with $\Omega_K$. The two parameters are also not
interchangeable: the change that leaves $E(a)$ untouched,
$\delta\Omega_K=-\delta\Omega_{\L_3}$, still moves $d_A$ by about
$1.7\,\delta\Omega_K$ through the bending. That $\Omega_K$ is bounded, while
$\Omega_{\L_3}$ is not, requires more than $d_A$: curvature also acts on CMB
lensing and on the late-time ISW, and its steeper degeneracy with $h$ runs
into the limits on $h$ sooner.\footnote{None of this makes $\Omega_{\L_3}$ detectable. That $\Omega_{\L_3}$ is
distinct from $\Omega_K$ answers a different question: it shows that the
finite-temperature term is not curvature under another name, and it is what
justifies keeping both parameters free. Whether the data require a nonzero
$\Omega_{\L_3}$ is decided by its degeneracy with $h$, and that degeneracy is
what hides it. Over $\ell\ge30$ a shift in $h$ absorbs $99.9\%$ of the
$\Omega_{\L_3}$ signal (see below), and the unabsorbable remainder lies at low $\ell$,
where cosmic variance dominates. Being distinct from $\Omega_K$ changes none
of this.} The same logic separates $\Omega_{\L_2}$ from $\Omega_R$. The two enter $E(a)$
identically, so the expansion history alone cannot tell them apart. Radiation,
however, also perturbs the plasma and $\Omega_{\L_2}$ does not, and that is the
only handle that separates them. At the magnitudes the data allow, it is too
small to see.

\begin{figure}[H]
  \centering
  \includegraphics[width=\columnwidth]{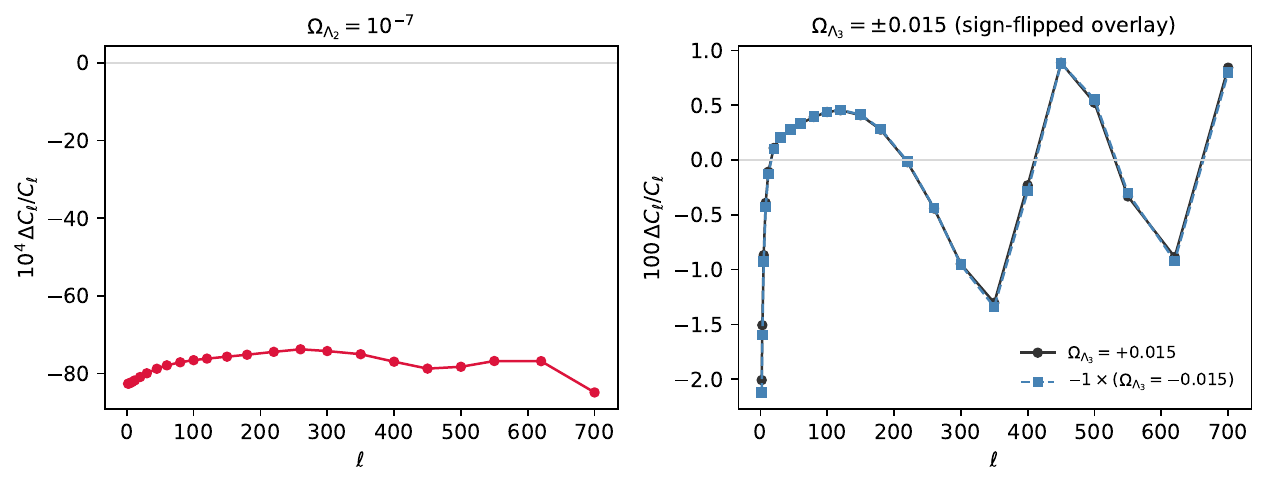}
\caption{Fractional response of the temperature spectrum. \emph{Left:}
$\Omega_{\Lambda_2} = 10^{-7}$ --- a nearly uniform suppression of about
$0.8\%$, varying by a few parts in $10^4$ across the acoustic peaks, i.e.,
effectively an amplitude shift. \emph{Right:} $\Omega_{\Lambda_3} = +0.015$
(circles) overlaid with the sign-flipped response at $-0.015$ (squares); the two
differ by about $7\%$, showing that the response is not exactly odd.}
  \label{fig:responses}
\end{figure}

Now that we have validated the derivative machinery, we compute the responses to
the two new parameters. The new parameters enter the calculation in only one
place --- through the expansion history $E(a)$ in equation \eqref{eq:friedmann}
--- so everything downstream (recombination timing, acoustic driving, distances,
and the ISW effect) inherits the modification consistently. We form the
fractional response
$\Delta C_\ell/C_\ell = C_\ell(\Omega_{\Lambda_2}, \Omega_{\Lambda_3}) / C_\ell(0,0) - 1$
using the same grid and settings, so the numerical noise floor from equation
\eqref{eq:floor} is the relevant error bar. Both responses lie well above this
floor. Figure~\ref{fig:responses} shows the results for representative values.

The two responses could hardly look more different. For
$\Omega_{\Lambda_2} = 10^{-7}$ --- about a tenth of a percent of the radiation
density --- the response is a nearly uniform suppression of about $0.8\%$,
varying by only a few parts in $10^4$ across the acoustic range. Runs at
$2\times10^{-7}$, $3\times10^{-7}$ and $-10^{-7}$ (not shown in
Fig.~\ref{fig:responses}) confirm that it is linear in the parameter to within
$1\%$ and symmetric under a change of sign to within $0.5\%$. A response this
flat is, for all practical purposes, just an overall amplitude change. One
caution: at $\Omega_{\Lambda_2} = 10^{-5}$, a value used in some early
explorations, the term is $11\%$ of the radiation density and acts like a change
of $N_{\rm eff}$ by $0.8$ --- far too large to be treated linearly. Keeping
$\Omega_{\Lambda_2} \ll \Omega_R$ is what makes the linear-response description
legitimate.

For $\Omega_{\Lambda_3} = \pm 0.015$, the response has structure and changes
sign: it suppresses $C_\ell$ at low $\ell$, crosses zero near the first peak,
and oscillates across the higher acoustic peaks. This is what a shifted expansion rate
does to the acoustic phase: it moves the sound horizon $r_L$. At this amplitude
the response is linear to about $3\%$: the response at
$\Omega_{\Lambda_3} = 0.015$ differs from twice that at $0.0075$ by $3\%$ RMS.
But it is not symmetric between positive and negative $\Omega_{\Lambda_3}$. If
we flip the sign of the response at $\Omega_{\Lambda_3} = -0.015$ and overlay it
on the response at $+0.015$, the two differ by about $7\%$. The difference grows
to $14\%$ at $|\Omega_{\Lambda_3}| = 0.030$. Both numbers lie above the floor.
This asymmetry is a genuine quadratic effect: the physical solutions at
$+|\Omega_{\Lambda_3}|$ and $-|\Omega_{\Lambda_3}|$ are distinct, not mirror
images.\footnote{The response contains a component that is even in
$\Omega_{\Lambda_3}$. This is not the same as the apparent bimodality seen in
the under-converged chains of Sec.~\ref{sec:bayesian}. The even component is a
genuine property of the response and persists once the chains converge. The
bimodality was a sampling artifact and disappears.}

\bigskip\noindent {\bf {Fisher analysis with the Planck covariance: which
absorbs what.}}
\label{sec:fisher}

We use the Fisher matrix, which is the standard shortcut for forecasting what a
data set can measure without having to run slow parameter-fitting chains. For a
Gaussian likelihood with a parameter-independent covariance, it takes the
form\footnote{This is the mean-dependent part of the general Fisher matrix for
Gaussian data, $F_{ij}=\tfrac12\,\mathrm{Tr}\bigl[\mathrm{Cov}^{-1}
\partial_i\mathrm{Cov}\,\mathrm{Cov}^{-1}\partial_j\mathrm{Cov}\bigr]
+\partial_i C^{\rm T}\,\mathrm{Cov}^{-1}\partial_j C$
\ci{Tegmark:1996bz}; the trace term vanishes when the covariance is held fixed,
as it is here.}
\begin{align}
F_{ij}=\sum_{bb'}\frac{\partial C_b}{\partial\theta_i}\,
\left(\mathrm{Cov}^{-1}\right)_{bb'}\frac{\partial C_{b'}}{\partial\theta_j},
\label{eq:fisher}
\end{align}
where the sum runs over the bandpowers $b$ (the power spectrum binned into
multipole ranges). In words: for each bandpower, we ask how much the spectrum
changes when we vary parameter $i$, and how much it changes when we vary
parameter $j$; we multiply those two changes together, weight each bandpower by
how precisely it is measured through the inverse covariance, and sum over all
bandpowers. The larger the resulting entries, the better the data constrain the
parameters. Inverting the Fisher matrix gives the forecast parameter covariance.
The square root of each diagonal entry, $[(F^{-1})_{ii}]^{1/2}$, is therefore
the expected $1\sigma$ error on $\theta_i$ after marginalizing over the other
parameters --- that is, the error once we account for the fact that all the
parameters are being fitted simultaneously. These are the $\sigma$ values quoted
in Table~\ref{tab:drags}. The same formalism answers a second question. Suppose
the data contain some extra signal $\Delta C_b$, but our model does not include
it. The fitted parameters will shift slightly to imitate that signal as best
they can, and to first order those shifts are
\begin{align}
\delta\theta_i=\sum_j (F^{-1})_{ij}\sum_{bb'}
\frac{\partial C_b}{\partial\theta_j}\left(\mathrm{Cov}^{-1}\right)_{bb'}\Delta C_{b'}.
\label{eq:fisherbias}
\end{align}
This is the standard Fisher-matrix estimate of the bias in parameters caused by
an unmodeled contribution to the data (see, e.g., \ci{Huterer:2004tr}). It
follows from fitting the standard model to the signal in the linear
approximation. Minimizing
\begin{equation*}
\chi^2=\sum_{bb'}\Bigl(\Delta C_b-\sum_j\partial_j C_b\,\delta\theta_j\Bigr)
(\mathrm{Cov}^{-1})_{bb'}\Bigl(\Delta C_{b'}-\sum_k\partial_k C_{b'}\,\delta\theta_k\Bigr),
\qquad \partial_i\equiv\frac{\partial}{\partial\theta_i},
\end{equation*}
with respect to $\delta\theta$ gives
$\sum_j F_{ij}\,\delta\theta_j=\sum_{bb'}\partial_i C_b\,(\mathrm{Cov}^{-1})_{bb'}\Delta C_{b'}$,
whose solution is \eqref{eq:fisherbias}.
The part of the signal that survives after these compensating shifts is the part
that no standard parameter can imitate.

The responses tell us about detectability once we weigh them against the
measurement errors and against how the standard parameters respond to the same
signals. We build the Fisher matrix from central-difference derivatives with
respect to $\{\Omega_M h^2, \Omega_B h^2, h, n_S\}$, computed by the same code
on the same grid and with the same settings, so that shared systematics cancel
in the differences.

The overall multiplicative amplitude of the spectrum,
$A \equiv N^2 e^{-2\tau_{\rm reio}}$ --- the primordial normalization together
with the reionization suppression, which enter the temperature spectrum only in
this combination --- is carried as a nuisance parameter $\ln A$, with a prior of
width $\sigma_{\ln A}$ that interpolates between a completely free amplitude
($\sigma_{\ln A} = \infty$) and a fixed one. This is distinct from the Planck
calibration parameter $y_{\rm cal}$ of Sec.~\ref{sec:bayes_setup}, which does
not enter here. The derivatives are binned with the exact \texttt{plik\_lite}
weights, and the Fisher matrix is assembled with the full $81 \times 81$
bandpower covariance --- the real Planck TT noise, not a simplified
model.\footnote{This means the analysis inherits \texttt{plik\_lite}'s domain,
$\ell \ge 30$; the information at $\ell < 30$, which the Bayesian analysis does
use through the low-$\ell$ likelihoods, is absent here. This matters, as we will
see.}

We then ask two questions about each of the two new parameters. Suppose one of
them is genuinely present in the universe, but an observer fits standard
$\Lambda$CDM to the data anyway. Two things follow. The fitted parameters are
pulled away from their true values, and the omitted signal leaves behind a
$\chi^2$ that those shifts partly remove. Let us call the first the \emph{drag}
on each standard parameter --- it is the Fisher bias of
Eq.~\eqref{eq:fisherbias} --- and the second the \emph{absorbed fraction}: the
fraction of the omitted signal's $\chi^2$ that the best standard fit removes.
Table~\ref{tab:drags} gives the drags; Figure~\ref{fig:sweep} gives the absorbed
fractions as a function of the amplitude prior.

\begin{table}[h]
\centering
\begin{tabular}{llcc}
\toprule
signal & amplitude $A$ & largest drags & absorbed \\
\midrule
$\Omega_{\Lambda_2} = 10^{-7}$ & free
  & (none above $0.05\sigma$) & $100.0\%$ \\
$\Omega_{\Lambda_2} = 10^{-7}$ & fixed
  & $\Omega_M h^2\!: -1.6\sigma$,\ \ $h\!: +0.7\sigma$,\ \ $n_S\!: +0.4\sigma$
  & $92.7\%$ \\
$\Omega_{\Lambda_3} = +0.015$ & free
  & $h\!: -0.012\ (-0.5\sigma)$ & $100.0\%$ \\
$\Omega_{\Lambda_3} = +0.015$ & fixed
  & $h\!: -0.013\ (-0.7\sigma)$ & $99.9\%$ \\
$\Omega_{\Lambda_3} = -0.015$ & free
  & $h\!: +0.013\ (+0.6\sigma)$ & $100.0\%$ \\
\bottomrule
\end{tabular}
\caption{Parameter shifts induced when a genuine finite-temperature component is
present but a standard \(\Lambda\)CDM model, together with the overall amplitude
\(A = N^2 e^{-2\tau_{\rm reio}}\), is fitted, using the Planck
\texttt{plik\_lite} TT covariance for \(\ell \ge 30\). In the ``amplitude''
column, ``free'' means \(\ln A\) is fitted with no prior and ``fixed'' means it
is held at its true value; these are the two ends of the sweep in
Fig.~\ref{fig:sweep}. ``Absorbed'' denotes the fraction of the signal's
\(\chi^2\) that is removed by the best-fit standard combination. The \(\sigma\)
values are the expected \(1\sigma\) uncertainties on the parameters, obtained
from a forecast that uses only the TT spectrum up to \(\ell=700\) and after
marginalizing over the other parameters. Each \(\sigma\) is computed with the
same treatment of \(A\) as its row; the values with \(A\) free are given in the
paragraph following Fig.~\ref{fig:sweep}. }
\label{tab:drags}
\end{table}

To read the table, note that ``amplitude $A$'' records whether the overall
normalization is allowed to float (free) or held fixed. This choice controls how
easily the standard fit can hide a new signal. The ``largest drags'' column
lists the biggest induced shifts of the standard parameters, in units of their
expected \(1\sigma\) errors; a small value means the parameters barely move,
while a value of order unity means the fit is pulled noticeably. The
``absorbed'' column reports how much of the new signal's \(\chi^2\) the best
standard fit removes: \(100\%\) means the signal can be hidden completely,
whereas a smaller number leaves a residual \(\chi^2\) that would reveal it. Two
patterns stand out. For \(\Omega_{\Lambda_2} = 10^{-7}\), the response is
essentially a uniform amplitude change, so when \(A\) is free the fit absorbs
the signal entirely and drags nothing; fixing \(A\) removes that escape route,
forcing a large drag on \(\Omega_M h^2\) and leaving \(7.3\%\) of the \(\chi^2\)
behind. For \(\Omega_{\Lambda_3} = \pm 0.015\), the drag on \(h\) stays below
\(1\sigma\) and absorption is essentially complete, so this signal is difficult
to detect through the standard parameters alone. The two amplitude-free
\(\Omega_{\Lambda_3}\) rows, which differ only in the sign of the signal, give
drags on \(h\) of \(-0.012\) and \(+0.013\); the small difference in magnitude
is another sign that the response is not exactly odd.

Three detailed observations can be made, in increasing order of importance.

First, everything about $\Omega_{\L_2}$ comes down to the amplitude. When the
amplitude is free, the signal is absorbed completely and no standard parameter
moves appreciably. As information about the amplitude is added from outside
(Fig.~\re{fig:sweep}), the signal begins to show through, and the shift appears
in the parameters that control the shape of the spectrum: $\Omega_M h^2$ is
pulled down, $h$ and $n_S$ up. With Planck's real errors, the amplitude has to
be known to better than $\sigma_{\ln A}\sim0.01$ before even $2\%$ of the signal
survives absorption. This is why the Bayesian analysis returns a genuine
two-sided bound rather than nothing at all: there the amplitude is not free, but
constrained by the data themselves through $A_s$ and through $\tau_{\rm reio}$
in the low-$\ell$ polarization, and pinned further by the calibration prior.

Second, everything about $\Omega_{\Lambda_3}$ comes down to $h$. A signal of
$+0.015$ shifts $h$ by $-0.012$ to $-0.013$, while no other parameter moves by
more than a few percent of its uncertainty. The result barely changes whether
the amplitude is free or fixed. This is the pipeline's version of the
$h$--$\Omega_{\Lambda_3}$ anticorrelation that the chains of
Sec.~\ref{sec:bayesian} display. One qualification belongs with it. The drag is
a central difference taken about a single fiducial point, so it measures the
slope of the degenerate direction at that point and nowhere else. The posterior
of Fig.~\ref{fig:triangle} shows that direction curving and steepening as
$\Omega_{\Lambda_3}$ grows. The two calculations agree on the degeneracy
wherever a linear description of it holds, and the sampling calculation carries
the additional information about how it bends. The independence from the
amplitude is exact to the precision of the calculation. The absorbed fraction of
the $\Omega_{\Lambda_3}$ signal over the full range $2\le\ell\le700$, under the
Planck-like diagonal noise model described below, is $97.30\%$ to four
significant figures whether the amplitude is free, constrained by a prior of
width $0.1$ or $0.01$, or held fixed. This contrasts with $\Omega_{\Lambda_2}$,
whose absorbed fraction falls from $100\%$ to $88\%$ over the same range.

Third --- and this is the mechanism behind the one-sided limit of
Sec.~\re{sec:bayes_priors} --- over the multipoles where Planck TT is precise,
the absorption is very nearly complete. Two calculations are involved, and it is
worth separating them. Using the \texttt{plik\_lite} covariance, which begins at
$\ell=30$, an adjustment of $h$ removes $99.9\%$ of the $\Omega_{\L_3}$ signal.
Repeating the calculation over the full range of multipoles, $2\le\ell\le700$,
under a Planck-like diagonal noise model that is cosmic-variance dominated at
low $\ell$, leaves $2.7\%$ that no standard parameter can imitate.

The difference between the two is the entire unabsorbable signal, and it sits
below $\ell=30$. That is exactly where a component scaling as $a^{-2}$ leaves
its mark, through the late-time expansion and the ISW effect, and exactly where
the errors are set by cosmic variance rather than by instrumental noise.
Building a better instrument cannot reduce them. So it is not that Planck failed
to detect an $\Omega_{\Lambda_3}$ signature that was there to be found. The
distinctive part of that signature lies where Planck has the least statistical
power. The sampling analysis reaches the same conclusion from the other
direction. Because $99.9\%$ of the signal can be absorbed, the extent of the
$\Omega_{\Lambda_3}$ posterior is set by whichever prior first truncates the
degenerate direction, not by the likelihood. There is no upper limit that does
not depend on the prior.

\begin{figure}[H]
  \centering
  \includegraphics[width=0.8\columnwidth]{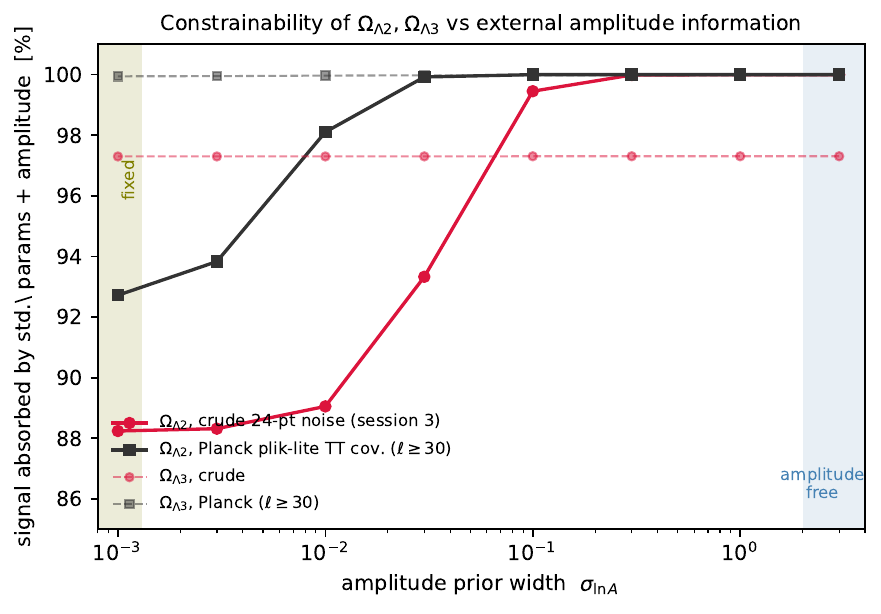}
\caption{The fraction of the $\Omega_{\L_2}$ signal (solid) and the
$\Omega_{\L_3}$ signal (dashed) that standard parameters can absorb, as a
function of how well the overall amplitude is known externally
($\sigma_{\ln A}$). Red: a simple single-channel noise model. Black: the real
\texttt{plik\_lite} covariance. Whether $\Omega_{\L_2}$ can be constrained is
entirely a question of external amplitude information; $\Omega_{\L_3}$ is
governed instead by its degeneracy with $h$, and hardly responds to the
amplitude prior at all.}
  \la{fig:sweep}
\end{figure}

For completeness, the marginalized forecasts underlying the $\sigma$'s of
Table~\re{tab:drags} are $\sigma(\Omega_M h^2)=0.0038$,
$\sigma(\Omega_B h^2)=0.00058$, $\sigma(h)=0.024$ and $\sigma(n_S)=0.017$ with
the amplitude left free. These are two to three times weaker than Planck's
\texttt{TT+lowE} constraints, as they should be, since they come from TT alone
at $\ell \le 700$. Table~\ref{tab:drags} reports two kinds of quantity, and they
respond differently to a uniform rescaling of the covariance. The absolute drags
and the absorbed fractions are ratios, so the rescaling cancels and both are
left unchanged. Drags quoted in units of \(\sigma\) are not ratios, and they
scale as the inverse square root of the rescaling. For that reason, the
\(\sigma\)-normalized drags should be read as an indication of how large an
effect is, not as a forecast of how significantly the effect could be detected.


\section{Conclusion}
\la{sec:conclusion}

Finite-temperature quantum-gravitational corrections to the cosmological
constant add two density parameters to the Friedmann equation,
$\Omega_{\L_2}\propto a^{-4}$ and $\Omega_{\L_3}\propto a^{-2}$, and they are
not a relabeling of radiation and curvature. A best-fit
point for them was located in Papers~I and II by deterministic means. In the present work we took up the two
questions that a best-fit point leaves open --- what the data bound, and why the
bound is what it is --- and answered them with a Bayesian MCMC analysis and a
CMB pipeline built from Weinberg's kinetic theory.

In brief, the results are as follows. The converged ten-parameter posterior
(Sec.~\ref{sec:bayesian}) places $\Omega_{\L_2}$ at $(-0.2\pm4.6)\times10^{-7}$
--- a clean two-sided null at the level of half a percent of the radiation
density --- while $\Omega_{\L_3}$ remains constrained from one side only, with
$\Omega_{\L_3}>-0.0092$ at $68\%$. The response analysis
(Sec.~\ref{sec:kinetic}) provides the mechanism behind both. The
$\Omega_{\L_2}$ signature is flat in $\ell$ to a few parts in $10^4$ across the
acoustic range: it is an amplitude, and it is bounded only because the data
themselves fix that amplitude, through $A_s$ and the low-$\ell$ EE determination
of $\tau_{\rm reio}$. The $\Omega_{\L_3}$ signature, by contrast, is structured
and sign-changing; yet over the multipoles where Planck TT is precise, it is
$99.9\%$ absorbable by a shift in $h$ of $-0.013$ per $+0.015$ of signal. The
part that no standard parameter can imitate --- a residual of a few percent ---
lives at $\ell\lesssim30$, where an $a^{-2}$ component acts through the
late-time expansion and the integrated Sachs--Wolfe effect, and where cosmic
variance is irreducible. Planck does not lack an $\Omega_{\L_3}$ signature to
find. It lacks statistical power exactly where the distinctive part of that
signature lives.

\subsection*{Sensitivity, detectability, and the status of Papers I and II}

This also resolves an apparent tension. The machine-learning analyses of
Papers~I and~II reported that the two parameters systematically improve the fit:
in Paper~I the surrogate regression attains $R^2 = 0.9988$ on held-out data,
against $0.9912$ for $\Lambda$CDM, with sharply reduced information criteria.
The Bayesian analysis, by contrast, finds that neither parameter is detected.
These two statements do not conflict, and the reason is not merely that the
methods differ in character.

The machine-learning analyses are surrogate-modeling exercises. They construct a
smooth map from the parameter vector to the Euclidean distance
\begin{equation}
  d(\boldsymbol{\theta})
  = \biggl[\sum_\ell
      \Bigl(C_\ell^{\textsc{class}}(\boldsymbol{\theta})
      - C_\ell^{\text{Planck}}\Bigr)^{2}
    \biggr]^{1/2},
  \label{eq:distance}
\end{equation}
where $C_\ell^{\text{Planck}}$ is the fixed, noiseless best-fit $\Lambda$CDM
theory curve of Ref.~\cite{Planck:2018vyg} --- the object identified in
Sec.~\ref{sec:theoryfile} as $C_\ell(\boldsymbol{\theta}_{\text{best}})$, an
ensemble average rather than a measurement. No covariance matrix, no cosmic
variance, no marginalization, and no likelihood enter Eq.~\eqref{eq:distance}. A
high $R^2$ therefore indicates only that the distance surface is smooth and that
the finite-temperature parameters move it in a way the regression can exploit. It
states that a signature is present in the curve; it does not state that the data
favor the extended model. The Bayesian analysis asks the complementary question
--- what remains compatible with the observed sky once every other parameter has
been integrated over --- and there the parameter scales are physical rather than
standardized: a shift of $10^{-7}$ in the Friedmann equation remains a shift of
$10^{-7}$.

\begin{table}[h]
  \centering
  \begin{tabular}{lcc}
    \toprule
     & ML analysis (Papers I, II) & Bayesian MCMC \\
    \midrule
    target          & distance to a fixed $C_\ell$ curve & likelihood of the observed data \\
    noise           & none                               & full covariance \\
    marginalization & none                               & over all parameters \\
    parameter scale & standardized                       & physical \\
    answer          & a signature is present             & the signature is not detectable \\
    \bottomrule
  \end{tabular}
  \caption{The two analyses differ in every substantive respect. Their outputs
  are therefore not the same kind of statement and should not be expected to
  agree numerically.}
  \label{tab:ml_vs_bayes}
\end{table}

Rather than attributing the discrepancy to noise and marginalization in general, the pipeline 
localizes it. For $\Omega_{\Lambda_2}$, the gap is the amplitude: the
parameter is absorbed completely while $\ln A$ remains free, and breaking that
absorption requires external amplitude information at the level of
$\sigma_{\ln A}\lesssim0.01$ under the real \texttt{plik\_lite} covariance. For
$\Omega_{\Lambda_3}$, the gap lies in the $h$ direction, and a Fisher drag
computed without any sampling reproduces the $h$--$\Omega_{\Lambda_3}$
anticorrelation seen in the chains --- a two-code cross-check of the degeneracy
structure, with one code sampling a posterior and the other differentiating a
spectrum. This pattern is familiar in cosmology, having appeared for early dark
energy \cite{Poulin:2018cxd,Smith:2020iux}, running vacuum
\cite{SolaPeracaula:2021gxi}, spatial curvature \cite{Planck:2018vyg}, and
neutrino mass \cite{Alam:2021inq}: a component that improves a fit in a
noise-free comparison, yet whose posterior is broad once degeneracies and
covariance are taken seriously.

\subsection*{What the pipeline established along the way}

The pipeline was built to answer a physics question, but some of what it
produced are statements about the kinetic derivation itself. Those are worth
separating from the parameter results.

The central design choice was to integrate the photon and polarization
hierarchies directly: no tight-coupling approximation, and no switch from an
approximate to an exact system. Standard codes place such a switch according to
criteria tuned against a fiducial cosmology. Here the entire subject is a
modified expansion history, and an error introduced at a switch is silent, so we
chose a slower calculation with no seam. The cost is several minutes per
spectrum; the return is that Silk damping, Landau damping, and the integrated
Sachs--Wolfe (ISW) contribution become \emph{outputs} of the Boltzmann dynamics
rather than inserted quantities. The $16/15$ of \W{6.4.25} emerges from the
polarization ladder and the slaved shear. The Landau damping length agrees with
\W{7.2.31} to $0.42\%$. The ISW term is not a small correction confined to low-$\ell$: it raises the
first acoustic peak by $40\%$ and the rise toward it by nearly twice as much.
Because the code is Weinberg's derivation made numerical, its agreement with
observation tests the derivation, not merely an implementation.

That agreement was quantified in two ways. Against the Planck 2018 best-fit
theory spectrum, the shape agrees to $1.8\%$ RMS over $\ell\ge30$ on the dense
thirty-one-point grid with a single fitted amplitude, and to $1.34\%$ on the
sparser tabulated multipoles; removing one overall rescaling of $\ell$ by
$0.8\%$ lowers the dense-grid figure to $0.9\%$. The reference spectrum is
lensed while our code computes the unlensed one. Over this range the difference
nowhere exceeds $1\%$, and correcting for it brings the rescaled residual down
to $0.6\%$: lensing is a small contaminant on the acoustic-scale offset, not
its cause. Against \textsc{class}, on the parameter response the
finite-temperature analysis actually uses, the dominant acoustic-scale channel
agrees to $0.55\%$ and the point-by-point fractional derivative correlates at
$0.9965$. The residual traces to a smooth, inherited recombination-phase offset,
not to anything generated by differentiation.

\subsection*{Outlook}

Both parameters are theoretically present and observationally undetected, and
the two analyses agree on what it would take to change that. For $\Omega_{\Lambda_2}$ the requirement is external amplitude information,
because the signature is an amplitude and nothing else. Improved measurements of
$\tau_{\rm reio}$ through large-scale polarization bear on it more directly than
any improvement in the acoustic peaks. For $\Omega_{\Lambda_3}$ the requirement is different in kind. Greater precision
at $\ell\ge30$ will not help: the degeneracy with $h$ is essentially exact
there. What is needed is either low-$\ell$ information, where cosmic variance
sets a floor no experiment can lower, or an external handle on $h$ from outside
the CMB, which would convert the degeneracy into a constraint. Polarization,
lensing, and the low-$\ell$--sensitive combinations they enable are the natural
next step. The sensitivity improvements expected from the Simons Observatory
\cite{SimonsObservatory:2018koc} and CMB-S4 \cite{CMB-S4:2016ple} will sharpen
the question even where they cannot settle it.

An external handle on $h$ is already available, and it does not require new CMB data. Late-time distance
measurements address both requirements, in different ways. Baryon acoustic
oscillations constrain the distance--redshift relation at a series of redshifts
below $z\simeq2$, where an $a^{-2}$ term and $h$ enter $E(a)$ with different
redshift dependence. In the CMB the same two quantities act almost entirely
through the single number $d_A$, which is why they are hard to separate there.
Current BAO compilations \cite{Alam:2021inq,DESI:2025zgx}, together with
supernova distances \cite{Brout:2022vxf}, should collapse the
$h$--$\Omega_{\Lambda_3}$ direction that dominates the present posterior and
turn the one-sided limit of Table~\ref{tab:mcmc} into a two-sided constraint.
One caveat carries over from Sec.~\ref{sec:responses}: such data do much less
for the $\Omega_{\Lambda_3}$--$\Omega_K$ direction, because curvature and the
finite-temperature term enter the expansion history through the same sum.
Separating those two still rests on the geometric handle described there.
A joint analysis that fixes $\Omega_K$ assumes what the geometric handle was
meant to determine; one that marginalizes over it does not. The distinction is
substantive.

For $\Omega_{\Lambda_2}$ the same data act through a different channel. A
component scaling as $a^{-4}$ behaves as extra radiation before recombination
and shifts the sound horizon at the drag epoch, the scale in whose units BAO
measures distances. This is the channel through which $N_{\rm eff}$ is
constrained, and it is independent of the amplitude channel offered by the
temperature spectrum. It also bears on the assumption, noted in
Sec.~\ref{sec:bayes_priors}, that $\Omega_R$ is fixed at its FIRAS value. A
chain with $N_{\rm eff}$ free alongside $\Omega_{\Lambda_2}$ would measure how
much of the limit comes from the data and how much from the assumption.
Taken together, these two lines would move the detectability question away from
the shape of the acoustic peaks, where the present paper finds absorption to be
essentially complete above $\ell=30$. The question would then rest on late-time
geometry and the drag scale. That is a different argument, and it deserves to be
made on its own terms.

\vspace*{.5in}

\ni {\bf Acknowledgements}\\

\ni This work used Anvil at Purdue University through allocation
PHY240277 from the Advanced Cyberinfrastructure Coordination
Ecosystem: Services \& Support (ACCESS) program~\cite{ACCESS},
which is supported by U.S. National Science Foundation grants
\#2138259, \#2138286, \#2138307, \#2137603, and \#2138296.

\newpage
\appendix

\renewcommand{\theequation}{A.\arabic{equation}}
\setcounter{equation}{0}

\section{Conventions}
\la{app:conv}

We follow \ci{Weinberg2008} throughout, with one difference of normalization.
Weinberg does not set the scale factor to unity today: as he notes below his
(1.1.12), the overall normalization of $a(t)$ enters the dynamical equations
only through the term $K/a^2(t)$, so it is free when $K=0$ but carries physical
meaning otherwise, and he keeps $a(t_0)$ explicit for that reason. The
calculations of Secs.~\re{sec:hydro} and~\re{sec:kinetic} are carried out for a
spatially flat background, where the normalization is free, and we set
$a(t_0)=1$; comoving wavenumbers are then present-day physical wavenumbers. This
applies to those two sections only. The chains of Sec.~\re{sec:bayesian} sample
$\Omega_K$ freely and use \textsc{class}'s own conventions, in which the
curvature enters the expansion rate, the $\chi\to S_K(\chi)$ map and the
perturbation mode functions alike.

Overdots denote differentiation with respect to cosmic time $t$, $a(t)$ is the
Robertson--Walker scale factor and $H=\dot a/a$. Comoving wavenumbers are
denoted $q$, and $k \eqv q/a$ is the corresponding physical wavenumber;
$\mu \eqv \hat q \cdot \hat p$ is the cosine between the wavevector and the
photon direction. The unperturbed radial coordinate is
$r(t) = \int_t^{t_0} \dd t'/a(t')$, and $r_L \eqv r(t_L)$ its value at last
scattering.

The Thomson collision rate is $\om_c(t) = n_e(t)\,\sigma_T$, with $n_e$ the
physical number density of free electrons, and the optical depth
$\tau(t) = \int_t^{t_0} \om_c(t')\, \dd t'$ is measured back from the present,
so that $e^{-\tau}$ increases toward $t_0$. The visibility function is
$P(t) = \om_c(t) e^{-\tau(t)}$, normalized to $\int P\, \dd t = 1$ and peaked
at $t_L$. The reionization optical depth $\tau_{\rm reio}$, which appears in
the spectrum normalization, is a separate quantity.

The baryon-to-photon ratio is $R = 3\bar\r_B/4\bar\r_\g$, which grows as $a$;
the acoustic frequency of the tightly coupled plasma is
$\om_0 = k/\sqrt{3(1+R)}$, and the tight-coupling expansion parameter is
$\eps \sim \om_0/\om_c = \om_0 t_\g$, with $t_\g = 1/\om_c$ the mean free time.
The damping rate $\G$ is the imaginary part of the WKB frequency. Perturbations
are worked in the synchronous gauge with metric amplitudes $A_q(t)$ and $B_q(t)$
in the conventions of \ci{Weinberg2008} \S6.1. Two shorthands for metric
combinations are ours rather than Weinberg's:
$Y \eqv a^2 \ddot B_q + a \dot a \dot B_q$, which appears in the form factor
\rf{eq:Fqt} and is used in Appendices~\re{app:sec3} and~\re{sec:mb}, and
$W \eqv q^2 \dot B_q = 3\dot A_q - 2\psi_q$, which follows from Weinberg's
definition of $\psi_q$ and is used in the construction of the form factors in
Sec.~\re{sec:pipeline}.

In the projection integrals $\nu \eqv \ell + \fr12$ and $\beta \eqv q r_L/\nu$
measures the wavenumber in units of the multipole's resonant value. Density
parameters $\Omega_i$ are fractions of the critical density today. In
Secs.~\re{sec:hydro} and~\re{sec:kinetic}, $\Omega_R$ includes three massless
neutrino species, $\Omega_R = 1.6813\, \Omega_\g$; the chains of
Sec.~\re{sec:bayesian} instead carry $N_{\rm ur}=2.0328$ massless species plus
one massive species, following the Planck baseline. Temperatures are in Kelvin
and spectra in $\mu$K$^2$, reported as
$\mathcal{D}_\ell \eqv \ell(\ell+1) C_\ell/2\pi$.

\renewcommand{\theequation}{B.\arabic{equation}}
\setcounter{equation}{0}

\section{Details of Section~\re{sec:hydro}}
\la{app:sec3}

In this appendix we provide the material set aside in Sec.~\re{sec:hydro}. It is
arranged in four parts: the definitions of Weinberg's kinetic variables together
with the Boltzmann hierarchy they satisfy, the steps of the tight-coupling
reduction, the slow modes and the transfer functions, and the large-$\ell$
projection with the evaluation of the form factors.

\subsection{Kinetic variables, the hierarchy, and the fluid moments}
\la{app:hierarchy}

The modes of $J_{ij}$ obey a Boltzmann equation inherited from that of
$\delta n^{ij}$. Introducing the momentum-space counterpart,
\bea
J_{ij}(\mathbf{q},\hat{p},t) &=& \alpha(\mathbf{q}) \Bigg\{
\frac{1}{2}\Big(\Delta_T^{(S)}(q,\hat{q}\cdot\hat{p},t) - \Delta_P^{(S)}(q,\hat{q}\cdot\hat{p},t)\Big)\Big(\delta_{ij} - \hat{p}_i \hat{p}_j\Big) \nn\\
&&+ \Delta_P^{(S)}(q,\hat{q}\cdot\hat{p},t)
\left[
\frac{\Big(\hat{q}_i - (\hat{q}\cdot\hat{p})\hat{p}_i\Big)\Big(\hat{q}_j - (\hat{q}\cdot\hat{p})\hat{p}_j\Big)}
{1 - (\hat{p}\cdot\hat{q})^2}
\right]
\Bigg\} \la{JijF}
\eea
where the subscripts $T$ and $P$ stand for temperature and polarization and
$\alpha(q)$ is a stochastic parameter, the Boltzmann equation
\bea
&&\frac{\partial J_{ij}(\mathbf{q},\hat{p},t)}{\partial t}
+ i\frac{\hat{p}\cdot\mathbf{q}}{a(t)} J_{ij}(\mathbf{q},\hat{p},t)
+ \alpha(\mathbf{q}) \left[\dot{A}_q(t) - (\mathbf{q}\cdot\hat{p})^2 \dot{B}_q(t)\right] \left(\delta_{ij} - \hat{p}_i \hat{p}_j\right) \nn\\
&=& -\omega_c(t) J_{ij}(\mathbf{q},\hat{p},t)
+ \frac{3\omega_c(t)}{8\pi} \int d^2\hat{p}_1 \Big[
J_{ij}(\mathbf{q},\hat{p}_1,t)
- \hat{p}_i \hat{p}_k J_{kj}(\mathbf{q},\hat{p}_1,t)
- \hat{p}_j \hat{p}_k J_{ik}(\mathbf{q},\hat{p}_1,t) \nn\\
&&+ \hat{p}_i \hat{p}_j \hat{p}_k \hat{p}_l J_{kl}(\mathbf{q},\hat{p}_1,t)
\Big]
 + \frac{2\omega_c(t)}{a(t)} \left(\delta_{ij} - \hat{p}_i \hat{p}_j\right) \hat{p}_k\, \delta u_{Bk}(\mathbf{q},t)
\eea
leads to
\begin{align}
\dot\Delta^{(S)}_P+\frac{iq\mu}{a}\,\Delta^{(S)}_P
&=-\omega_c\,\Delta^{(S)}_P+\tfrac34\,\omega_c\,(1-\mu^2)\,\Pi(q,t),
\label{eq:boltzP}\\
\dot\Delta^{(S)}_T+\frac{iq\mu}{a}\,\Delta^{(S)}_T
&=-\omega_c\,\Delta^{(S)}_T
-2\dot A_q+2q^2\mu^2\dot B_q
\nonumber\\
&\quad
+3\omega_c\,\Phi(q,t)+\tfrac34\,\omega_c(1-\mu^2)\,\Pi(q,t)
+\frac{4iq\mu\,\omega_c}{a}\,\delta u_{Bq}(t),
\label{eq:boltzT}
\end{align}
with $\Phi$ and $\Pi$ defined by
\bea
\int \frac{d^2\hat{p}}{4\pi} J_{ij}(\mathbf{q},\hat{p},t)
= \alpha(\mathbf{q}) \left[
\delta_{ij} \Phi(q,t) + \frac{1}{2} \hat{q}_i \hat{q}_j \Pi(q,t)
\right],
\eea
acting as sources. The partial-wave expansion
of the temperature and polarization amplitudes, with
$\mu\equiv\hat q\cdot\hat p$,
\begin{equation}
\Delta^{(S)}_{T,P}(q,\mu,t)=\sum_{\ell=0}^{\infty}
i^{-\ell}(2\ell+1)\,P_\ell(\mu)\,\Delta^{(S)}_{T\!,P,\ell}(q,t),
\label{eq:pw}
\end{equation}
then turns these into the coupled hierarchy
\begin{align}
\dot\Delta^{(S)}_{P,\ell}
+\frac{q}{a(2\ell+1)}\Bigl[(\ell+1)\DP{\ell+1}-\ell\,\DP{\ell-1}\Bigr]
&=-\omega_c\,\DP{\ell}
+\tfrac12\,\omega_c\,\Pi\Bigl(\delta_{\ell 0}+\tfrac{\delta_{\ell2}}{5}\Bigr),
\label{eq:hierP}\\[1mm]
\dot\Delta^{(S)}_{T,\ell}
+\frac{q}{a(2\ell+1)}\Bigl[(\ell+1)\DT{\ell+1}-\ell\,\DT{\ell-1}\Bigr]
&=-2\dot A_q\,\delta_{\ell 0}
+2q^2\dot B_q\Bigl(\tfrac{\delta_{\ell0}}{3}-\tfrac{2\delta_{\ell2}}{15}\Bigr)
-\omega_c\,\DT{\ell}
\nonumber\\
&\hspace{-1in}
+\omega_c\Bigl(3\Phi+\tfrac12\Pi\Bigr)\delta_{\ell0}
+\tfrac1{10}\,\omega_c\,\Pi\,\delta_{\ell2}
-\frac{4}{3}\,\frac{q}{a}\,\omega_c\,\delta u_{Bq}\,\delta_{\ell1}.
\label{eq:hierT}
\end{align}
The sources $\Phi$ and $\Pi$ in terms of the lower multipoles are
\rf{eq:sources}.\footnote{Three errata to \ci{Weinberg2008} affecting these
equations have been taken into account throughout.} One identity is worth
displaying, because it is a check on the whole construction and because it
reappears in the tight-coupling expansion. From \rf{eq:sources},
\bea
3\Phi = \DT{0} - \ft12 \Pi
\qquad \Longrightarrow \qquad
-\om_c \DT{0} + \om_c\left(3\Phi + \ft12 \Pi\right) = 0 ,
\la{eq:phiid}
\eea
so that Thomson scattering drops out of the $\ell = 0$ line of the hierarchy:
scattering conserves photon number and energy, as it must. This is also what
turns the scattering source of $F$ into \rf{eq:Fsource}. The remaining
ingredients are the photon fluid moments
\begin{equation}
\delta p_{\gamma q}=\frac{\bar\rho_\gamma}{3}\bigl(\DT{0}+\DT{2}\bigr),
\qquad
q^2\pi^S_{\gamma q}=\bar\rho_\gamma\,\DT{2},
\qquad
\delta\rho_{\gamma q}=\bar\rho_\gamma\,\DT{0},
\qquad
\frac{q}{a}\,\delta u_{\gamma q}=-\tfrac34\,\DT{1},
\label{eq:moments}
\end{equation}
and the photon--baryon momentum balance
\begin{equation}
\delta p_{\gamma q}-q^2\pi^S_{\gamma q}
+\bigl[\partial_0+3H\bigr]
\Bigl[\bar\rho_B\,\delta u_{Bq}+\tfrac43\bar\rho_\gamma\,\delta u_{\gamma q}\Bigr]=0 .
\label{eq:mom}
\end{equation}
With \eqref{eq:moments}, and using
$[\partial_0+3H](\bar\rho_B\,\delta u_B)=\bar\rho_B\,\delta\dot u_B$ (since
$\bar\rho_B\propto a^{-3}$) and
$[\partial_0+3H](\tfrac43\bar\rho_\gamma\,\delta u_\gamma) =\tfrac43\bar\rho_\gamma(\delta\dot u_\gamma-H\delta u_\gamma)$
(since $\bar\rho_\gamma\propto a^{-4}$), Eq.~\eqref{eq:mom} divided by
$\tfrac43\bar\rho_\gamma$ reads
\begin{equation}
\tfrac14\DT{0}-\tfrac12\DT{2}
+\delta\dot u_{\gamma q}+R\,\delta\dot u_{Bq}
-H\,\delta u_{\gamma q}=0
\label{eq:momclean}
\end{equation}
with $R = 3\bar\r_B/4\bar\r_\g$ the baryon-to-photon ratio.

\subsection{The tight-coupling reduction}
\la{app:tcgroup}

Collected here are the three steps of the $1/\om_c$ expansion of
Sec.~\re{sec:tc} that were quoted there without proof: where the numerical
coefficient of the slaved shear comes from, what the $\ell=0$ line of the
hierarchy becomes under the WKB ansatz, and how the photon--baryon slip leads to
the dispersion relation.

\subsubsection*{The shear coefficient and the treatment of Thomson scattering}
\la{app:shear}

The numerical coefficient in the slaved shear \rf{eq:shear} is where the
treatment of Thomson scattering matters:
\begin{center}
\small
\begin{tabular}{@{}lccc@{}}
\toprule
treatment of $\ell=2$ collisions & relax.\ factor &
$\DT{2}\big/\bigl[\tfrac{2q}{5a\omega_c}\DT{1}\bigr]$ &
shear coefficient in $\Gamma$ \\
\midrule
full [Eqs.~\eqref{eq:Pi}--\eqref{eq:shear}] &
$\tfrac34$ & $\tfrac43$ & $16/15$\\[1mm]
no polarization ($\DPm\equiv0$, so $\Pi=\DT{2}$) &
$\tfrac9{10}$ & $\tfrac{10}{9}$ & $8/9$\\[1mm]
isotropic scattering (drop $\tfrac1{10}\omega_c\Pi\,\delta_{\ell2}$) &
$1$ & $1$ & $4/5$\\
\bottomrule
\end{tabular}
\end{center}

The last column gives the damping rate~\eqref{eq:Gamma}, which is derived below.
Its factor of $4/5$ is exactly $\frac34 \times \frac{16}{15}$, and this matches
a footnote to Weinberg's Eq.~(6.4.25). There Weinberg notes that earlier work,
before Kaiser, obtained a photon shear viscosity equal to $3/4$ of the correct
value. The coefficient $16/15$ is Kaiser's result for the shear viscosity,
$\eta = \frac{16}{15} \bar{\rho}_\gamma t_\gamma$, given in Weinberg's
Eq.~(6.4.24). This $16/15$ has a purely kinetic origin: it comes from combining
anisotropic Thomson scattering with the polarization feedback contained in
\eqref{eq:Pi}. This is the only place where polarization effects survive into
the final form of Weinberg's Eq.~(7.2.41), entering there through the damping
scale $\ell_D$.

\subsubsection*{The $\ell=0$ line under the WKB ansatz}
\la{app:ell0}

The $\ell=0$ line of \eqref{eq:hierT} reads
\begin{equation}
\dot{\Delta}_{T,0} + \frac{q}{a} \Delta_{T,1}
= -2\dot{A}_q + \frac{2}{3} q^2 \dot{B}_q - \omega_c \Delta_{T,0} + \omega_c \left( 3\Phi + \frac{1}{2}\Pi \right),
\end{equation}
which can be rewritten as
\begin{equation}
\dot{\Delta}_{T,0} + \frac{q}{a} \Delta_{T,1}
= -2\dot{A}_q + \frac{2}{3} q^2 \dot{B}_q + \omega_c \left( 3\Phi + \frac{1}{2}\Pi - \Delta_{T,0} \right).
\end{equation}
The last term vanishes by \eqref{eq:phiid}, and the metric terms are dropped for
the fast mode. Replacing the time derivative by $-i\om$ then gives \rf{eq:e0}.
The sign in the WKB ansatz is a convention tied to the direction of time
evolution: the full solution is a linear combination of
$\exp(-i\int\omega\,dt')$ and $\exp(+i\int\omega\,dt')$, and for a single
propagating mode the initial conditions select one of the two.

\subsubsection*{The slip and the WKB dispersion relation}
\la{app:slipwkb}

The $\ell=1$ line of \eqref{eq:hierT}, together with the relation
$(q/a)\,\delta u_{\gamma q}= -\frac34 \Delta_{T,1}^{(S)}$ from
\eqref{eq:moments}, gives
\begin{equation}
\dot\Delta_{T,1} + \frac{q}{3a}\left(2\Delta_{T,2} - \Delta_{T,0}\right)
= -\frac{4}{3}\,\frac{q}{a}\,\omega_c\,
\left(\delta u_{Bq} - \delta u_{\gamma q}\right).
\label{eq:ell1}
\end{equation}
The right-hand side is the momentum exchanged between photons and baryons by
scattering, and it is proportional to the \emph{difference} of their velocities,
as momentum conservation requires. In the tight-coupling regime that difference,
the slip, is of order $\epsilon$. Solving for it,
\begin{equation}
\delta u_{Bq} - \delta u_{\gamma q}
= -\frac{3a}{4q\,\omega_c}
\left[\dot\Delta_{T,1} + \frac{q}{3a}\left(2\Delta_{T,2} - \Delta_{T,0}\right)\right],
\label{eq:slip}
\end{equation}
so the slip too is slaved to the photon multipoles and suppressed by the large
collision rate: the photons and baryons move almost together, and any residual
relative velocity is quickly erased by scattering. Inserting \eqref{eq:e0} into
\eqref{eq:slip} and dropping the $O(\eps^2)$ shear term leads to
\rf{eq:slipval}, and feeding the result into the momentum balance
\eqref{eq:momclean} leads to the dispersion relation \rf{eq:disp}. The details
are as follows. With \eqref{eq:e0} inserted and the $O(\eps^2)$ shear term
dropped, the bracket of \eqref{eq:slip} becomes
\begin{equation}
\dot\Delta_{T,1}-\frac{k}{3}\DT{0}
=\Bigl[-i\omega\cdot\frac{i\omega}{k}-\frac{k}{3}\Bigr]\DT{0}
=\frac{1}{k}\Bigl[\omega^2-\frac{k^2}{3}\Bigr]\DT{0}
\simeq-\frac{kR}{3(1+R)}\,\DT{0},
\end{equation}
using $\omega^2\simeq\omega_0^2\equiv k^2/[3(1+R)]$ at leading order. Hence
\begin{equation}
\delta u_{Bq}-\delta u_{\gamma q}
=\frac{R}{4(1+R)\,\omega_c}\,\DT{0}+O(\eps^2).
\end{equation}
Now insert everything into the momentum balance \eqref{eq:momclean} (dropping
the $H$ term):
\begin{align}
0&=\tfrac14\DT{0}-\tfrac12\DT{2}
+(1+R)\,\delta\dot u_{\gamma q}
+R\,\frac{\dd}{\dd t}\bigl(\delta u_{Bq}-\delta u_{\gamma q}\bigr)
\nonumber\\
&=\Bigl[\tfrac14
-\frac{4i\omega}{15\,\omega_c}
-\frac{3\omega^2(1+R)}{4k^2}
-\frac{i\omega R^2}{4(1+R)\,\omega_c}\Bigr]\DT{0},
\end{align}
where the second term used
$\tfrac12\DT{2}=\tfrac12\cdot\tfrac{8k}{15\omega_c}\DT{1} =\tfrac{4i\omega}{15\omega_c}\DT{0}$
from \eqref{eq:shear} and \eqref{eq:e0}, and the third and fourth used
\eqref{eq:e0} and \eqref{eq:slipval}. Multiplying by $4k^2/3$, one gets
\begin{equation}
\omega^2(1+R)
=\frac{k^2}{3}
-\,i\,\omega\,\frac{k^2}{\omega_c}
\Bigl[\frac{16}{45}+\frac{R^2}{3(1+R)}\Bigr].
\end{equation}

\subsection{The slow modes and the transfer functions}
\la{app:slowgroup}

In this subsection we discuss the following three items: what the slow modes are, the fitting formulas for the three transfer
functions together with the solutions they give at last scattering, and a remark
on the exponent of the fit for $\Delta(\kappa)$.

\subsubsection*{The slow modes and the growth suppression}
\la{app:slow}

The fast modes are acoustic waves in the photon--baryon fluid, and they
oscillate on the acoustic timescale. The slow modes are different in kind. They
are the particular solutions of the Einstein--Boltzmann system driven by the
metric perturbations, they evolve on the much longer Hubble timescale $H^{-1}$,
and their time derivatives are of order $H$ times the amplitude rather than
$\omega$ times the amplitude, with $\omega\gg H$. Physically they are the
gravitational condensations of the non-relativistic matter, which grow once that
matter comes to dominate the energy density and which eventually become galaxies
and large-scale structure.

The key quantity that characterizes the slow modes is the transfer function
$\mathcal{T}(\kappa)$, introduced by Weinberg in his Chapter 6. It encapsulates
the suppression of the slow-mode amplitude on small scales (large wavenumbers
$q$), which arises because these modes entered the horizon during the
radiation-dominated era, when pressure forces prevented them from growing. For
modes that entered the horizon during the matter-dominated era, there is no such
suppression, and the transfer function approaches unity. In terms of the
rescaled wavenumber $\kappa \equiv \sqrt{2}\,q/q_{\mathrm{EQ}}$, where
$q_{\mathrm{EQ}}$ is the wavenumber entering the horizon at matter--radiation
equality, the two limits are recorded in \eqref{eq:limits}.

The slow-mode solution for the dark matter density perturbation $\delta_{Dq}$ in
the matter-dominated era is Weinberg's (6.5.15),
\begin{equation}
\delta_{Dq}(t) = \frac{9 q^2 t^2}{10 a^2} \mathcal{R}_q^o \, \mathcal{T}(\kappa),
\end{equation}
where $\mathcal{R}_q^o$ is the primordial curvature perturbation, and the
gravitational potential $\psi_q$ is his (6.5.16),
\begin{equation}
\psi_q(t) = -\frac{3 q^2 t}{5 a^2} \mathcal{R}_q^o \, \mathcal{T}(\kappa).
\end{equation}
Both are proportional to the transfer function, which therefore encodes the
scale-dependent growth history. The slow modes also contribute to the photon
temperature perturbation. Their contribution is subdominant to the fast acoustic
oscillations for modes inside the horizon, but for modes still outside the
horizon at last scattering --- large angular scales, small $\ell$ --- they
dominate and produce the Sachs--Wolfe plateau. This is why the transfer function
enters the form factor $\bar F(q)$, as the term proportional to
$3 \mathcal{T}(\kappa) R_L$ in Weinberg's (7.2.36) and in our \rf{eq:Fbar}.

\subsubsection*{Transfer functions and the solutions at last scattering}
\la{app:transfer}

The solution of the slow system, interpolating between the long- and
short-wavelength regimes, is parameterized by the transfer functions
$\TT(\kappa)$, $\SS(\kappa)$, $\Delta(\kappa)$,
\bea
{\cal S}(\k) &\equiv& \left(\frac{  5^{1/2} (0.1657 \k)^6 +(0.5116 \k)^4+(1.209 \k)^2+1}{(0.1657 \k)^6+(0.4249 \k)^4+(0.9459 \k)^2+1}\right)^2 \nn\\
{\cal T}(\k) &\equiv& \frac{\log [(0.124 \k)^2+1]}{(0.124 \k)^2} \sqrt{ \frac{(0.2197 \k)^6+(0.4452 \k)^4+(1.257 \k)^2+1}{(0.3927 \k)^6+(0.8568 \k)^4+(1.606 \k)^2+1}}\nn\\
\D(\k) &\equiv&\frac{(0.2578 \k)^6+(0.5986 \k)^4+(1.1547 \k)^2}{(0.2204 \k)^8+(0.4581 \k)^6+(0.8707 \k)^4+(1.723 \k)^2+1}
\eea
with limits
\begin{equation}
\kappa\ll1:\ \ \TT,\SS\to1,\ \Delta\to\frac{2\kappa}{\sqrt3};
\qquad
\kappa\gg1:\ \ \TT\to\frac{45}{2\kappa^2}
\Bigl[-\frac72+\gamma+\ln\frac{4\kappa}{\sqrt3}\Bigr],\ \ \SS\to5,\ \
\Delta\to0,
\label{eq:limits}
\end{equation}
with the baryon-corrected generalization of \S 6.5:
\begin{align}
\psi_q(t_L)&=-\frac{3q^2t_L\,\Ro\,\TT(\kappa)}{5\,a_L^2},
\label{eq:psiL}\\
\delta_{\gamma q}(t_L)&=\frac{3\Ro}{5}\Bigl[
\TT(\kappa)\,(1+3R_L)
-(1+R_L)^{-1/4}\,e^{-\int_0^{t_L}\Gamma\,\dd t}\,\SS(\kappa)\,
\cos\Theta_L\Bigr],
\label{eq:dgL}\\
\delta u_{\gamma q}(t_L)&=\frac{3\Ro}{5}\Bigl[
-t_L\,\TT(\kappa)
+\frac{a_L}{\sqrt3\,q\,(1+R_L)^{3/4}}\,
e^{-\int_0^{t_L}\Gamma\,\dd t}\,\SS(\kappa)\,\sin\Theta_L\Bigr],
\label{eq:duL}
\end{align}
where $\Theta_L\equiv\Theta(t_L)=q\,d_H/a_L+\Delta(\kappa)$ and $d_H$ is the
acoustic horizon \W{7.2.39}.

\subsubsection*{The exponent of the $\Delta(\kappa)$ fit}
\la{app:delta}

The peak positions returned by \eqref{eq:final} depend sensitively on the
power to which the fit for $\Delta(\kappa)$ is raised, and we record the
three possibilities here. As written in Sec.~\ref{sec:hydro}, with no
exponent, the fit gives the values quoted there. Raised to the power
$1/2$, the same fit gives $219.5$, $541.0$, $820.6$, $1133.2$ and
$1425.8$ --- within one multipole of Weinberg's Table~7.1 at every peak.
Raised to the power $1/4$, as in the uncorrected printing, it gives
$203$, $521$, $798$, $1110$ and $1402$. The power $1/2$ is also the only
one of the three that reproduces \W{6.5.10}: the numerator of the fit
begins at $(1.1547\kappa)^2$ and the denominator at $1$, so the square
root yields $\Delta\to1.1547\kappa=2\kappa/\sqrt3$, while the behavior
$\Delta\to0$ at large $\kappa$ is unchanged. We record this without changing the exponent, since the fit was checked
against Table~6.1 in the form printed in Sec.~\ref{sec:hydro}, and that check would have to be
repeated before the exponent could be changed. We note in passing that
changing the exponent shifts all five peaks by nearly the same amount,
so the near-constancy of the offset does not by itself rule out $\Delta$
as its source.

\subsection{The large-$\ell$ projection and the form factors}
\la{app:projgroup}

What remains is the projection from wavenumber to multipole and the evaluation
of the three form factors: the time derivative of the spherical Bessel phase,
the Debye asymptotics together with the average over the rapid phase, and the
visibility averages that produce $\bar F$, $\bar G$ and $\bar H$.

\subsubsection*{Derivative of the spherical Bessel phase}
\la{app:phase}

In the asymptotic expansion of the spherical Bessel functions, the phase
\(\Phi(t)\) is
\begin{equation}
\Phi(t) \equiv \nu \left( \tan b - b \right),
\end{equation}
where \(\nu \equiv \ell + 1/2\) and \(\cos b \equiv \nu / \rho\), with
\(\rho \equiv q r(t)\). Here \(r(t)=\int_t^{t_0} dt'/a(t')\) is the comoving
radial coordinate of a photon at time \(t\), so that \(\dot{r}(t) = -1/a(t)\)
and
\begin{equation}
\dot{\rho} = q \dot{r} = -\frac{q}{a}.
\end{equation}
Differentiating \(\cos b = \nu \rho^{-1}\) with respect to time gives
\(-\sin b \, \dot{b} = -\nu \rho^{-2} \dot{\rho}\), hence
\begin{equation}
\dot{b} = \frac{\nu}{\rho^2 \sin b} \dot{\rho}
= -\frac{q \cos^2 b}{a \nu \sin b},
\end{equation}
where the second equality uses \(\dot{\rho} = -q/a\) and
\(\rho = \nu / \cos b\). Differentiating the phase,
\begin{equation}
\dot{\Phi} = \nu \left( \sec^2 b \, \dot{b} - \dot{b} \right)
      = \nu \frac{\sin^2 b}{\cos^2 b} \dot{b}
      = \nu \frac{\sin^2 b}{\cos^2 b}
            \left( -\frac{q \cos^2 b}{a \nu \sin b} \right)
          = -\frac{q \sin b}{a},
\end{equation}
which is the result quoted in Sec.~\re{sec:hydro}. The negative sign reflects
the fact that the comoving distance \(r(t)\) decreases as the photon propagates
toward the observer, and the factor \(\sin b\) arises from the geometric
relation \(\sin b = \sqrt{1 - \nu^2/(q^2 r^2)}\). The same expression follows
directly from the WKB phase
\(\Phi = q \int_t^{t_0} dt' / a(t') \, \sqrt{1 - \nu^2/(q^2 r^2(t'))}\), which
confirms the consistency of the derivation. The result governs the rapid
oscillation of the line-of-sight integrand at large $\ell$.

\subsubsection*{Debye asymptotics and the phase average}
\la{app:debye}

For $\ell\gg1$, with $\nu\equiv\ell+\tfrac12$ and $\cos b\equiv\nu/\rho$
($0\le b\le\pi/2$), the uniform asymptotic formula reads,
for $\rho>\nu$ with $|\nu^2-\rho^2|\gg\nu^{4/3}$,
\begin{equation}
j_\ell(\rho)\;\to\;\frac{\cos b}{\nu\sqrt{\sin b}}\,
\cos\bigl[\varphi(\rho)\bigr],
\qquad
\varphi(\rho)\equiv\nu(\tan b-b)-\tfrac{\pi}{4},
\label{eq:debye0}
\end{equation}
and $j_\ell(\rho)\to0$ for $\rho<\nu$. Since $\varphi'(\rho)=\sin b$ while the
amplitude varies on the scale $\rho$ itself, each $\rho$-derivative acts at
leading order on the phase only:
\begin{equation}
j'_\ell(\rho)\;\to\;-\sin b\;\frac{\cos b}{\nu\sqrt{\sin b}}\,
\sin\varphi,
\qquad
j''_\ell(\rho)\;\to\;-\sin^2\! b\;\frac{\cos b}{\nu\sqrt{\sin b}}\,
\cos\varphi .
\label{eq:debye12}
\end{equation}
The first of these is \W{7.2.15}; the second is its immediate extension, needed
because we keep $\bar H$. (As with \W{7.2.15}, letting the derivative act on the
amplitude would introduce a spurious divergence at $\rho\to\nu$, where
\eqref{eq:debye0} itself fails; in the final formula the $j''$ term carries an
explicit factor $\sin^2 b\propto\beta^2-1$ that suppresses that region.)
Inserting the localized line-of-sight expression \eqref{eq:thin} together with
the Debye asymptotic forms \eqref{eq:debye12} into \eqref{eq:7210}, the
integrand becomes
\begin{equation}
\frac{\cos^2 b}{\nu^2\sin b}\,
\Bigl\langle
\Bigl[\bigl(\bar F-\sin^2\!b\,\bar H\bigr)\cos\varphi
-\sin b\,\bar G\,\sin\varphi\Bigr]^2\Bigr\rangle
=\frac{\cos^2 b}{2\nu^2\sin b}\,
\Bigl[\bigl(\bar F-\sin^2\!b\,\bar H\bigr)^2+\sin^2\!b\;\bar G^2\Bigr],
\label{eq:avg}
\end{equation}
where the rapid oscillation of the phase $\varphi$ over the $q$ integration has
been used to replace the oscillating trigonometric functions by their averages,
$\langle\cos^2\varphi\rangle = \langle\sin^2\varphi\rangle = 1/2$ and
$\langle\cos\varphi\,\sin\varphi\rangle = 0$.

\subsubsection*{Evaluation of the visibility-averaged form factors}
\la{app:formfactors}

Here we carry out the three evaluations quoted in Sec.~\re{sec:hydro}.
Throughout, $P(t)=\om_c e^{-\tau}$ is the visibility function and
$\bar t_\gamma \equiv 1/\omega_c(t_L)$.

\paragraph{$\bar{\textbf{\textit{F}}}$:}
Starting from $F(q,t)$ in \rf{eq:Fqt} and using the scattering-source identity
\rf{eq:Fsource}, the term proportional to $P(t)\,\frac{1}{4}\Delta_{T,0}$
combines with the boundary term arising from the metric contributions after
integration by parts. These two pieces assemble exactly as in Weinberg's
transition from (7.1.16) to (7.2.18), yielding the combination
\begin{equation}
\frac{1}{3}\delta_{\gamma q}(t_L)+\frac{a_L^2\psi_q(t_L)}{3q^2t_L},
\end{equation}
which is valid in the matter-dominated era under the relations $\dot A_q=0$,
$\dot B_q=-2\psi_q/q^2$, and $\ddot B_q=2\psi_q/(3tq^2)$. Inserting the explicit
solutions for $\psi_q(t_L)$ and $\delta_{\gamma q}(t_L)$ from \rf{eq:psiL} and
\rf{eq:dgL}, the slow pieces combine as
\begin{equation}
\frac{\Ro}{5}\TT(\kappa)(1+3R_L)-\frac{\Ro}{5}\TT(\kappa)
= \frac{3\Ro}{5}\TT(\kappa)R_L,
\end{equation}
while the fast piece acquires the Landau damping factor from the visibility
averaging. The ingredient not present in Weinberg's original derivation is the
explicit polarization source $P(t)\,\Pi/16$ appearing in \rf{eq:Fsource}. To
evaluate it, we use the tight-coupling relations \rf{eq:Pi}, \rf{eq:shear} and
\rf{eq:e0} applied to the real fast-mode solution \rf{eq:fast}:
\begin{align}
\Delta_{T,1} &= -\frac{1}{k}\,\dot{\Delta}_{T,0}^{\rm fast}
\simeq -\frac{\omega_0}{k}\,A\,\sin\Theta, \\
\Pi &= \frac{5}{2}\cdot\frac{8k}{15\omega_c}\,\Delta_{T,1}
= -\frac{4\omega_0}{3\omega_c}\,A\,\sin\Theta,
\label{eq:Pifast}
\end{align}
where $A$ is the bare monopole amplitude of the fast-mode solution. The factor
$1/\omega_c$ in $\Pi$ cancels exactly against the $\omega_c$ inside the
visibility function, so the visibility integral of $P\,\Pi/16$ is finite and
localized at last scattering. It is important to distinguish between the bare
monopole amplitude $A$ and the observed oscillation amplitude $A_{\rm osc}$ that
appears in the assembled combination $\frac{1}{4}\Delta_{T,0}-\frac{1}{2}Y$. The
acoustic term in $\bar F$ is normalized to $A_{\rm osc}=A/4$, where the factor
$1/4$ comes from the product of the $1/3$ of Weinberg's (7.2.18) and the $4/3$
of the dictionary relating $\Delta_{T,0}$ to $\delta_{\gamma q}$. Using this,
\begin{equation}
\frac{1}{16}\Pi
= \frac{1}{16}\cdot\frac{4}{3}\,\omega_L\bar t_\gamma\,A
= \frac{1}{12}\,\omega_L\bar t_\gamma\,A
= \frac{1}{3}\,\omega_L\bar t_\gamma\,A_{\rm osc},
\end{equation}
and, evaluating at the visibility peak,
\begin{equation}
\frac{1}{16}\Pi
= \frac{\Ro}{5}(1+R_L)^{-1/4}\,\SS(\kappa)\,e^{-q^2 d_D^2/a_L^2}
\,\frac{\omega_L\bar t_\gamma}{3}\,\sin\Theta_L.
\end{equation}
Adding this to the no-polarization result gives \rf{eq:Fbar}.

\paragraph{$\bar{\textbf{\textit{H}}}$:}
Directly from \rf{eq:Hqt} and the tight-coupling expression for $\Pi$ in
\rf{eq:Pifast}, and using the same dictionary $A_{\rm osc}=A/4$,
\begin{align}
\frac{3}{16}\Pi
&= \frac{3}{16}\cdot\frac{4}{3}\,\omega_L\bar t_\gamma\,A
= \frac{1}{4}\,\omega_L\bar t_\gamma\,A
= \omega_L\bar t_\gamma\,A_{\rm osc}.
\end{align}
Averaging over the visibility function and evaluating at the peak $t_L$, with
the damping and phase localization already accounted for, gives \rf{eq:Hbar}.
The overall minus sign is inherited from the fact that $A_{\rm osc}$ is negative
in the convention used for the fast-mode solution,
$A_{\rm osc} = -\frac{\Ro}{5}(1+R_L)^{-1/4}\SS(\kappa)e^{-q^2d_D^2/a_L^2}$.
Comparing \rf{eq:Hbar} with the polarization term in \rf{eq:Fbar}, the ratio
$\bar H/\bar F_{\rm pol}=3$ is fixed entirely by the source coefficients $3/16$
and $1/16$, since both quantities average the same underlying polarization
source $\Pi$.

\paragraph{$\bar{\textbf{\textit{G}}}$:}
In the tight-coupling limit the baryon and photon velocity potentials agree up
to corrections of order $\epsilon$,
$\delta u_{Bq} = \delta u_{\gamma q} + O(\epsilon)$. The slip \eqref{eq:slipval}
contributes an $O(\epsilon)$ term proportional to $\cos\Theta$, in phase with
the monopole; we drop it at the same order as the polarization correction
$\eps_P$, since it is subdominant for the large-$\ell$ power spectrum. Using the
matter-dominated relation $\dot B_q = -2\psi_q/q^2$, the combination inside the
brackets of \rf{eq:Gqt} becomes
\begin{equation}
\frac{\delta u_{Bq}}{a} + \frac{a \dot B_q}{2}
= \frac{\delta u_{\gamma q}}{a} - \frac{a \psi_q}{q^2} + O(\epsilon).
\end{equation}
When the solutions for $\psi_q$ and $\delta u_{\gamma q}$ from \eqref{eq:psiL}
and \eqref{eq:duL} are inserted, the slow parts of these two terms cancel
identically, because both carry the same transfer function $\mathcal{T}(\kappa)$
with matched coefficients. This is Weinberg's observation below (7.2.24). What
remains is the pure Doppler oscillation \rf{eq:Gbar}, in which the factor
$(1+R_L)^{-3/4}$ comes from the WKB amplitude of the fast mode and the damping
factor carries both Silk and Landau damping.

\renewcommand{\theequation}{C.\arabic{equation}}
\setcounter{equation}{0}

\section{Initial conditions, the residual gauge, and consistency with Weinberg's adiabatic mode}
\la{app:ic}

We collect here the treatment of the initial data used by the code of
Section~\ref{sec:kinetic}. The first part covers how the growing mode is imposed
and how the residual gauge freedom of synchronous gauge is fixed. The second
part presents the constraint algebra, which shows that this choice is consistent
with Weinberg's adiabatic mode.

We start the integration at an initial time $a_i$, chosen deep in the radiation
era and well outside the horizon, with $q/(aH) \lesssim 0.1$. In practice we set
\begin{equation}
a_i = \min\!\left(10^{-5},\ \frac{2\times10^{-7}}{q}\right),
\qquad q \text{ in Mpc}^{-1}.
\end{equation}
The two entries serve different purposes, and it is worth seeing where
each comes from. The second entry enforces a horizon condition. During the radiation era
the expansion rate behaves as $H \propto a^{-2}$, so the comoving Hubble
scale is $aH = H_0\sqrt{\Omega_R}\,a^{-1}$ and therefore
\begin{equation}
\frac{q}{aH} = \frac{q\,a}{H_0\sqrt{\Omega_R}} .
\end{equation}
With $H_0\sqrt{\Omega_R} \simeq 2.2\times10^{-6}\ \mathrm{Mpc}^{-1}$,
choosing $a_i = 2\times10^{-7}/q$ gives
$q/(aH) \simeq 0.1$: every mode is started just outside the horizon,
which is the natural point at which to begin integrating it. The first entry caps this start for long-wavelength modes. Without the
cap, small $q$ would push $a_i$ to arbitrarily early times. The cap holds
those modes at $a_i = 10^{-5}$, which is still deep in the radiation era:
since $a_{\rm EQ} \approx 2.9\times10^{-4}$, this corresponds to
$a/a_{\rm EQ} \simeq 3\%$, so matter is only a small correction to the
radiation-era growing mode. Taken together, the prescription starts each mode at
$q/(aH) \lesssim 0.1$: the horizon condition sets the start for
short-wavelength modes, and the cap sets it for long-wavelength ones.
The initial conditions are taken
from the adiabatic growing mode, as given in Weinberg's equations
(6.2.27)--(6.2.29):
\begin{align}
\delta &= \frac{q^2 t^2}{a^2} \mathcal{R}_q^o, \\
\psi_q &= -\frac{q^2 t}{a^2} \mathcal{R}_q^o, \\
\delta u &= -\frac{2}{9} \frac{q^2 t^3}{a^2} \mathcal{R}_q^o,
\label{eq:ic_growing}
\end{align}
which are the super-horizon growing-mode power laws for the radiation-dominated
era, where $a \propto t^{1/2}$. In this era, $\delta \propto q^2 t$ while
$\psi_q$ is constant. The overall amplitude $\mathcal{R}_q^o$ cancels out when
we construct the power spectrum $C_\ell$, so its value is irrelevant. These
initial conditions are distributed among the different species according to the
adiabatic condition: each species carries the same fractional density
perturbation, so $\delta_D = \delta_B = \delta_\gamma = \delta_\nu = \delta$,
with no relative entropy perturbations. Photons and baryons share a common
velocity, $\delta u_B = \delta u$, because they are tightly coupled at the
initial time. Using the normalization $\Delta_{T,0} = \frac{4}{3}\delta_\gamma$
from Weinberg (6.2.8) and the dictionary from (6.1.31)--(6.1.34), we obtain
$\Delta_{T,0} = J_{\nu,0} = \frac{4}{3}\delta$ and
$\Delta_{T,1} = J_{\nu,1} = -\frac{4}{3}(q/a)\delta u$. All higher photon
multipoles and all polarization multipoles are set to zero, since they are
suppressed by factors of $1/\omega_c$ deep in the tight-coupling regime. The
higher neutrino multipoles are also set to zero, because they are only populated
by free streaming, which has not yet had time to operate.

Weinberg provides the initial conditions for the physical combinations $\delta$,
$\psi_q$, and $\delta u$, but he does not specify the synchronous gauge metric
functions $A_q$ and $B_q$, which contain the residual gauge freedom. We fix this
freedom by imposing
\begin{equation}
A_q(a_i) = 0, \qquad q^2 B_q(a_i) = 0,
\label{eq:ic_gauge}
\end{equation}
while initializing $a^2\psi_q$ from equation \eqref{eq:ic_growing} and evolving
the metric through Weinberg's equations (6.1.70)--(6.1.73). These are not
arbitrary zeros. For the adiabatic growing mode, the matter data determine the
metric \emph{rates} through the momentum constraint (6.1.72) and the relation
$q^2\dot B_q = 3\dot A_q - 2\psi_q$ from (6.1.73), while the metric
\emph{values} $A_q$ and $q^2B_q$ are precisely the integration constants that
the residual gauge transformation shifts. In the radiation era, one can check
analytically that
$\dot A_q \propto \bar\rho_\gamma \delta u_{\gamma q} \propto t^{-2} \cdot t^2$
is constant, so $A_q$ grows linearly from zero exactly as the growing mode
requires. A single residual gauge function suffices to impose both conditions of
\eqref{eq:ic_gauge} at once. The two are not independent: the Einstein
constraints place them in the compatible ratio; see the constraint algebra below.

Numerically, when we load the growing-mode matter data and evaluate the
source terms on the right-hand side of the metric equations of motion, we find that the metric rates $|\dot A_q/H|$ and
$|q^2\dot B_q/H|$ are below $10^{-5}$ of the mode scale $|\psi_q|$ across the
range $q = 0.002$ to $0.1 \, \mathrm{Mpc}^{-1}$, with no transient at the start of the forward integration: the
metric rates do not drift, overshoot, or settle over time. This
means that our hand-set zeros lie on the growing-mode attractor. The computed
spectrum is insensitive to a constant shift $A_q \to A_q + c$ at the $10^{-6}$
level, as expected for a gauge choice. However, the intermediate metric
quantities $A_q(a)$ and $B_q(a)$ do depend on the gauge choice, so any
comparison of transfer functions --- as opposed to the observable $C_\ell$ ---
against a code that uses a different gauge requires that we match
\eqref{eq:ic_gauge} explicitly. The cold-dark-matter rest frame
$\theta_{\rm CDM} = 0$ used by Ma \& Bertschinger \ci{MaBertschinger1995}, which
\textsc{class} inherits, is a different gauge choice (see Section~\ref{sec:mb}).

Two subleading omissions should be noted. First, we set the neutrino quadrupole
to zero in the initial conditions, whereas the exact adiabatic mode carries a
small $J_{\nu,2} = O((qt)^2)$ that sources the neutrino anisotropic stress and
slightly corrects the initial gravitational potential. Retaining only the
monopole and dipole is standard and accurate at $a_i = 10^{-5}$, but it is a
known place where the leading-order adiabatic conditions are incomplete at the
sub-percent level. The omission is confined to the initial data: the neutrino
hierarchy is evolved with free streaming from $a_i$ onward, so the quadrupole is
regenerated within a few $e$-folds of the start. Second, the initial data from
\eqref{eq:ic_growing}--\eqref{eq:ic_gauge} are imposed directly on the evolution
equations; checking that they also satisfy the constraint from Weinberg (6.1.70)
at $a_i$, so that the solver starts on the constraint surface rather than
relaxing onto it, is a worthwhile check when sub-percent accuracy is required.

\bigskip\noindent\emph{Consistency of the gauge choice.}
Here we verify that the initial conditions
\begin{align}
A_q(a_i) &= 0, \label{eq:Ainit}\\
q^2 B_q(a_i) &= 0, \label{eq:Binit}
\end{align}
imposed on the synchronous-gauge metric perturbations at some early time $a_i$,
are fully consistent with the initial conditions on the matter fields that
Weinberg imposes for the dominant (adiabatic) mode.

\medskip\noindent\emph{Weinberg's initial conditions for the adiabatic mode.} In
the radiation-dominated era ($a \propto \sqrt{t}$), Weinberg identifies the
growing adiabatic mode (Mode 1) with the initial conditions:
\begin{align}
\delta_{\gamma q} = \delta_{Bq} = \delta_{\nu q} = \delta_{Dq} &= \frac{q^2 t^2}{a^2} \mathcal{R}_q^o, \label{eq:delta_init}\\
\psi_q &= -\frac{t q^2}{a^2} \mathcal{R}_q^o, \label{eq:psi_init}\\
\delta u_{\gamma q} = \delta u_{\nu q} &= -\frac{2 t^3 q^2}{9 a^2} \mathcal{R}_q^o. \label{eq:vel_init}
\end{align}
Here $\mathcal{R}_q^o$ is the (constant) primordial curvature perturbation, and
$\psi_q$ is the synchronous-gauge metric combination
\begin{equation}
\psi_q \equiv \frac{1}{2}\left(3\dot{A}_q - q^2 \dot{B}_q\right). 
\end{equation}

\medskip\noindent\emph{What the constraints fix, and what they leave free.} The
$0i$ component of the Einstein equations in synchronous gauge reads
\begin{equation}
\dot{A}_q = 8\pi G \sum_\alpha \bar{\rho}_\alpha \delta u_{\alpha q},
\label{eq:momentum}
\end{equation}
where the sum runs over all species with non-negligible velocity perturbations.
During radiation domination the universe is dominated by photons and neutrinos,
so
\begin{equation}
\dot{A}_q \simeq 8\pi G \left( \frac{4}{3}\bar{\rho}_\gamma \delta u_{\gamma q} + \frac{4}{3}\bar{\rho}_\nu \delta u_{\nu q} \right).
\end{equation}
Using $\bar{\rho}_\gamma \propto a^{-4} \propto t^{-2}$ and
$\delta u_{\gamma q} \propto t^3/a^2$ from \eqref{eq:vel_init}, and
$a^2 \propto t$, we find that $\dot{A}_q \propto t/a^2$ is constant in the
radiation era, so
\begin{equation}
A_q(t) = A_q(a_i) + \dot{A}_q(a_i)(t - t_i).
\end{equation}
Setting $A_q(a_i) = 0$ is therefore perfectly consistent: the metric
perturbation grows linearly from zero, as expected for the growing adiabatic
mode. The trace-free part of the $ij$ Einstein equation, which for a perfect
fluid ($\pi_q^S = 0$) gives 
\begin{equation}
A_q = a^2 \ddot{B}_q + 3a\dot{a}\dot{B}_q, \label{eq:anisotropic}
\end{equation}
does not constrain $B_q(a_i)$ either; it relates $A_q$ to the derivatives of
$B_q$. Both metric values are thus integration constants. Once they are set to
zero, the rates are determined: the definition of $\psi_q$ gives
\begin{equation}
q^2 \dot{B}_q(a_i) = 3\dot{A}_q(a_i) - 2\psi_q(a_i),
\end{equation}
with $\dot{A}_q(a_i)$ fixed by the momentum constraint \eqref{eq:momentum} and
$\psi_q(a_i)$ by \eqref{eq:psi_init}, while \eqref{eq:anisotropic} then fixes
$\ddot{B}_q(a_i)$. Neither zero is an additional assumption about the matter.

\medskip\noindent\emph{Residual gauge freedom.} Synchronous gauge retains a
residual gauge freedom:
\begin{align}
\Delta A &= -\frac{2\dot{a}}{a}\tau(\mathbf{x}), \label{eq:resA}\\
\Delta B &= -2\tau(\mathbf{x}) \int a^{-2}(t)\,dt. \label{eq:resB}
\end{align}
A single residual function $\tau(\mathbf{x})$ suffices to impose both
$A_q(a_i) = 0$ and $q^2 B_q(a_i) = 0$, although two conditions might appear to
require two functions. The reason is that for the adiabatic growing mode the
Einstein constraints above place the initial metric data in precisely the ratio
that \eqref{eq:resA} and \eqref{eq:resB} relate: the two conditions are not
independent, but compatibly linked by the constraints. The choice therefore
involves no loss of generality. The initial conditions \eqref{eq:Ainit}--\eqref{eq:Binit} are in this sense
fully consistent with Weinberg's initial conditions for the adiabatic mode:
they fix only the residual gauge, not the physical state, so they cannot excite
a spurious mode or bias the computed $C_\ell$. This is what makes hand-set
zeros acceptable. They differ from the choice made in \textsc{class}, which
fixes the residual gauge in the cold-dark-matter rest frame; the two agree on
$C_\ell$ but not on intermediate metric quantities, as noted above.

\renewcommand{\theequation}{D.\arabic{equation}}
\setcounter{equation}{0}

\section{Relation of Section~\re{sec:kinetic} to the Ma--Bertschinger formulation}
\la{sec:mb}

The standard reference for numerical Boltzmann cosmology is Ma \& Bertschinger
\ci{MaBertschinger1995} (see \cite{Dodelson2021} for a review), and the codes in general use (\textsc{cmbfast},
\textsc{camb}, \textsc{class}) descend from its variables and conventions. It is
therefore worth stating precisely how the present calculation, built on
Weinberg's Chapter~7, relates to that formulation, and in particular where the
two make different technical choices.

\vspace{.1in}
\ni {\bf Line-of-sight versus direct hierarchy integration}

The two approaches differ at the outermost level in how the free streaming from
last scattering to the present is performed. Ma \& Bertschinger integrate the
full coupled Boltzmann hierarchy in time, propagating every temperature
multipole $\Delta^{(S)}_{T,\ell}$ up to a truncation order $\ell_{\max}$ that
must be comparable to the largest angular multipole of interest. For a spectrum
resolved to $\ell\sim1500$ this is a system of order $10^3$ coupled ordinary
differential equations per wavenumber. The present calculation instead uses the line-of-sight form
\W{7.2.10} of the multipole integral, our \rf{eq:7210}, in which the spherical
Bessel functions carry out the free streaming analytically. This is the method
introduced by Seljak \& Zaldarriaga \ci{Seljak:1996is} the year after Ma \&
Bertschinger, and adopted by all subsequent Boltzmann codes. Its practical
consequence is that the \emph{sources} $F,G,H$ require only a small number of
low multipoles, evaluated over the narrow support of the visibility function,
rather than the full high-$\ell$ hierarchy propagated to today. In the present
implementation the photon and polarization hierarchies are truncated at
$\ell_{\max}=10$ (Sec.~\re{sec:pipeline}); before decoupling the collision term
keeps the higher multipoles exponentially suppressed, so the sources are already
converged well below this, and the angular structure of the observed spectrum is
supplied entirely by the Bessel functions.

\vspace{.1in}
\ni {\bf Gauge and dynamical variables}

The principal content of Ma \& Bertschinger is a systematic treatment of scalar
perturbations in both the synchronous and conformal Newtonian gauges, carried
out in parallel, along with an explicit dictionary relating the two. Their
synchronous metric perturbations are \(h\) and \(\eta\), and their fluid
variables are the density contrast \(\delta\), the velocity divergence
\(\theta\), and the shear \(\sigma\).

Weinberg's variables are not the same. The metric perturbations are $A_q$ and
$B_q$ \W{6.1.70}--\W{6.1.73}, the Newtonian-like potential is $\psi_q$, and the
fluid velocities are carried as a velocity \emph{potential} $\delta u$ rather
than a divergence $\theta=(q^2/a)\,\delta u$. The translation is straightforward
at the level of the fluid equations, but the metric sector has no term-by-term
correspondence. In particular, the integrated Sachs--Wolfe source \W{7.1.9}
appears in Weinberg's variables as the combination
\bea
\frac{\dd}{\dd t}\Bigl[A_q+a^2\ddot B_q+a\dot a\dot B_q\Bigr]
\equiv\frac{\dd}{\dd t}\bigl[A_q+Y\bigr],
\qquad Y\equiv a^2\ddot B_q+a\dot a\dot B_q,
\la{eq:mb_isw}
\eea
whereas in the Ma--Bertschinger synchronous variables the same physical effect
is expressed through $\dot h$ and $\dot\eta$. A feature of the Weinberg form
worth noting is that, at late times, $A_q$ is very nearly constant while the
entire dark-energy-driven decay of the gravitational potential is carried by
$Y$: with $\Omega_\Lambda$ switched off, the change in $A_q+Y$ between $a=0.1$
and the present drops by more than two orders of magnitude. The late integrated
Sachs--Wolfe effect therefore lives, in these variables, almost entirely in the
second-derivative metric combination rather than in the potential $A_q$ itself.

A related point concerns the residual gauge freedom of the synchronous gauge. Ma
\& Bertschinger fix it explicitly by working in the rest frame of the cold dark
matter, $\theta_{\rm CDM}=0$; Weinberg does not phrase the corresponding choice
as a gauge condition at all, and in the present implementation it is fixed by
the initial data \rf{eq:ic_gauge}. The two prescriptions need not coincide, with
the consequence for transfer-function comparisons noted in \S\re{app:ic}.

\vspace{.1in}
\ni {\bf Time variable}

Ma \& Bertschinger use conformal time $\tau$, with derivatives $'=\dd/\dd\tau$;
Weinberg uses cosmic time $t$, with $\dot{}=\dd/\dd t$; the present code
integrates in $\ln a$. These are equivalent, but the powers of $a$ that appear
differ throughout. The two treatments serve different purposes, and
this is the reason for adopting Weinberg's here. Ma \& Bertschinger is a
reference: it presents the perturbation equations in both gauges in a form ready
for direct numerical integration. Weinberg's Chapter~7 is a derivation, which
obtains the line-of-sight form factors from the collision term and then carries
out the analytic reduction of Sec.~\re{sec:hydro}. The consequence for what the
present calculation can establish is the third of the properties listed at the
start of Section~\re{sec:kinetic}: because the pipeline implements Weinberg's
derivation rather than Ma \& Bertschinger's equations, its agreement with that
derivation --- the emergent $16/15$ and the Landau damping length --- is a test
of the derivation itself.

\end{document}